\documentclass[twocolumn,rmp,nofootinbib, floatfix]{revtex4}
\usepackage{ latexsym,amssymb, amsmath,amsfonts,epsfig, epic, eepic, subfigure}
\newcommand{\z}{\langle z\rangle}
\def\theequation{\arabic{section}.\arabic{equation}}

\begin{document}

\preprint{hep-th/0703289}
\title{Conifolds and Geometric Transitions}
\author{Rhiannon Gwyn \& Anke Knauf}
\affiliation{McGill University\\
 Department of Physics\\
 3600 Rue Universit\'e\\
 Montr\'eal, Qu\'ebec, H3A 2T8\\
 Canada}
\email{gwynr, knauf@physics.mcgill.ca}

\date{\today}

\begin{abstract}
Conifold geometries have recieved a lot of attention in string theory and string-inspired cosmology recently, in particular the Klebanov--Strassler background that is known as the ``warped throat''. It is our intention in this article to give a pedagogical explanation for the singularity resolution in this geometry and emphasise its connection to geometric transitions. The first part focuses on the gauge theory dual to the Klebanov--Strassler background, which we also explain from a T--dual intersecting branes scenario. We then make the connection to the Gopakumar--Vafa conjecture for open/closed string duality and summarise a series of papers verifying this
model on the supergravity level.

An appendix provides
extensive background material about conifold geometries. We pay
special attention to their complex structures and re-evaluate the
supersymmetry conditions on the background flux in constructions
with fractional D3-branes on the singular (Klebanov--Tseytlin) and
resolved (Pando Zayas--Tseytlin) conifolds. We agree with earlier
results that only the singular solution allows a supersymmetric
flux, but point out the importance of using the correct complex
structure to reach this conclusion.
\end{abstract}

\maketitle

%\setcounter{tocdepth}{2}
%\tableofcontents
%%%%%%%%%%%%%%%%%%%%%%%%%%%%%%%%%%%%%%%%%%%%%%%%%%%%%%%%%%%%%%%%%%%%%%%%%%%%

\setcounter{equation}{0}
\section{Introduction}
The geometric transition between conifold geometries is an example of a string
theory duality between compactifications on different geometrical
backgrounds. Initial arguments came from  two different angles: a
generalisation of AdS/CFT via the \textcite{ks} model and independently as a duality between open and closed
topological strings by \textcite{gopakumar}. A common
embedding has since been found in IIB, IIA and M-theory \cite{vafa,
cachazo, flop, 0105066, 0110050} and the geometric transition has been
confirmed on the supergravity level \cite{gtone, realm, gttwo, dipole, thesis}. It
is the intention of this article to give a comprehensive overview of
the different ideas underlying geometric transitions and review
briefly the lengthy supergravity calculations of the latter references.

\begin{center}
\begin{figure*}[htp]
\begin{center}
\mbox{\begin{minipage}{14.5 cm}
\begin{center}
\includegraphics[scale=0.45]{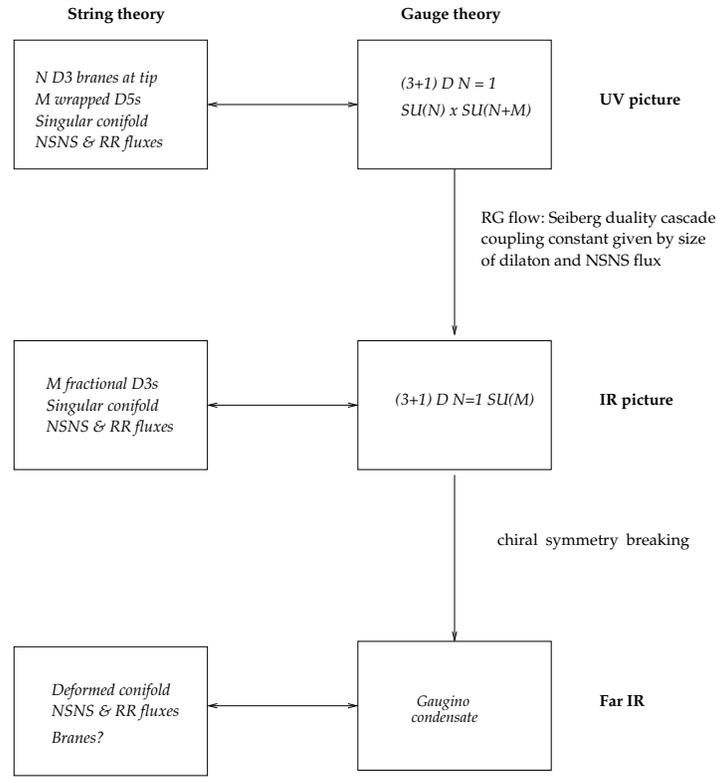}
\caption{The Klebanov-Strassler model}
\label{KSsum}
\end{center}
%\vspace*{0.9cm}
\end{minipage}
 }
 \end{center}
\end{figure*}
\end{center}

\begin{center}
\begin{figure*}[htp]
\begin{center}
\mbox{\begin{minipage}{14.5 cm}
\begin{center}
\includegraphics[scale=0.45]{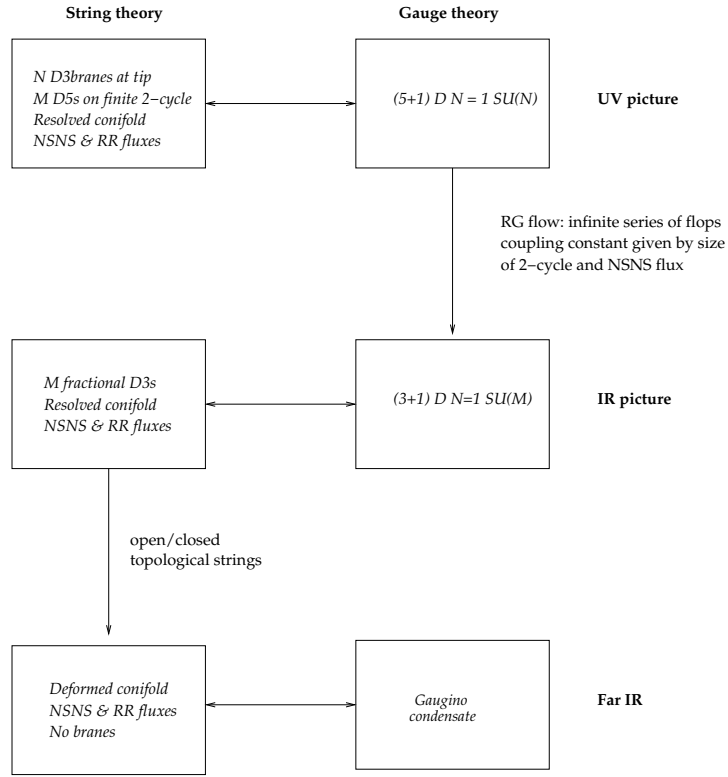}
\caption{Vafa's model}
\label{Vsum}
\end{center}
%\vskip 1 cm
\end{minipage}
 }
 \end{center}
\end{figure*}
\end{center}

The model of Klebanov and Strassler is based on a series of papers
\cite{kw, kt, 9808075, 9911096} generalising the AdS/CFT
correspondence \cite{Maldacena:1997re, Gubser:1998bc,
Witten:1998qj}. Instead of the $\mathcal{N}=4$ superconformal field
theory one obtains from considering $AdS_5\times S^5$, theories with
less supersymmetry can be found by taking $AdS_5\times M^5$, where
$M^5$ is some five-dimensional manifold. One can break conformal
invariance by introducing fractional D3-branes instead of (only)
D3-branes. These are objects that wrap compact cycles in the
internal manifold and therefore appear effectively three
dimensional. Once conformal invariance is broken, the gauge theory
exhibits a running coupling. The coupling constant is related to the
NS-NS B-field in the string theory dual. One approaches the far IR limit of the
gauge theory as the radial co-ordinate in the
supergravity dual approaches zero. The manifold $M_5$ considered in
this model is the base of a conifold, so there is a singularity at
$r=0$. This does not mean that the far IR limit of the gauge theory
is not well defined. On the contrary, knowledge about the
strong-coupling behaviour of the dual Super--Yang--Mills theory led
Klebanov and Strassler to the following remarkable conclusion: since
SYM exhibits gaugino condensation and chiral symmetry breaking
(which breaks the $U(1)$ symmetry down to $\mathbb{Z}_2$) in the far
IR, the dual string theory background has to be modified for $r\to
0$, in order to reflect this symmetry property. The singularity is
smoothed out, giving a manifold which looks like the conifold at
large radial distances, but approaches a finite three-sphere at the
tip of the cone. This manifold is called the ``deformed conifold''
and has precisely the required symmetry property, i.e. it is only
invariant under $\mathbb{Z}_2$, where the (singular) conifold was
invariant under a full $U(1)$. We summarise this picture in Figure
\ref{KSsum}.

\textcite{gopakumar} also considered conifold
geometries, but they were interested in topological string
amplitudes. They showed that the open A-model on the deformed
conifold (with a blown-up $S^3$) agrees with the closed A-model on
the resolved conifold (with a blown-up $S^2$) on the level of
topological string partition functions. One has to identify the
correct parameters from each theory: roughly speaking the size of
the three-cycle in the deformed geometry (its complex structure
modulus) is identified with the size of the two-cycle in the
resolved geometry (its K\"ahler modulus). Via mirror symmetry, the
same can be said for the B-model, but here the roles of deformed and
resolved conifold are exchanged. The connection to the KS model
becomes apparent if one embeds this B-model into IIB superstring
theory, as done by \textcite{vafa}. Before the geometric
transition, D5-branes wrap the non-vanishing two-cycle in the
resolved conifold and appear as fractional D3-branes that carry an
$\mathcal{N}=1$ $SU(N)$ SYM. 

This seems to be precisely the picture
one finds at the ``bottom of the duality cascade'' in the KS model:
for the model with $N$ D3 \em and \em $M$ fractional D3-branes the gauge
group is $SU(N)\times SU(N+M)$ and the RG flow is such that one of
the theories flows towards strong and the other towards weak
coupling. This translates via Seiberg duality to an $SU(N-M)\times
SU(N)$ theory, where the two gauge group factors now exhibit the
opposite running coupling behaviour. Ergo, one can follow a
``cascade'' of such Seiberg dualities, where in each step the gauge
group factor drops by $M$. If $N$ is a multiple of $M$ (we will discuss the more general case in section \ref{ks}) 
all the regular D3-branes will ``cascade away'' and the gauge group becomes $SU(M)$,
like in the Vafa set-up. The
only difference is that Vafa considers a resolved conifold, whereas
KS started with the singular version. Nevertheless, the picture they
both find in the IR is very similar: Vafa also argues that in the
large $M$ limit the string theory background is rather given by a
deformed conifold. He gives an even more concise description of this
picture: since the topological string argument was based on an
open/closed duality, there is no equivalent for the D-branes in the
dual closed theory. The geometric transition conjectured by Vafa is
therefore a duality between a background with D-branes (on which the
gauge degrees of freedom propagate) and a background with only
fluxes (where a geometrical parameter enters into the flux-generated
superpotential to give the correct confining IR behaviour). We
therefore conclude that the D-branes in the KS model should also
disappear once the singular conifold is traded against its
deformation -- see Figure \ref{Vsum} for comparison.

\begin{figure}
\centering
\includegraphics[scale=0.4]{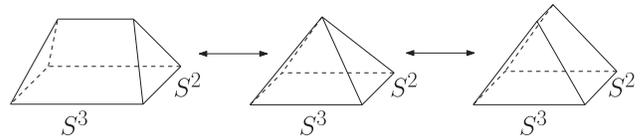}
\caption{The conifold transition. All three geometries share the base $S^2\times S^3$, but the $S^3$ of the deformed conifold (left) remains finite. In the conifold transition it can be shrunk to zero size to give the singular conifold (centre), from which blowing up the $S^2$ gives the resolved conifold (right).}
\label{contrans}
\end{figure}

All the transitions discussed so far take either the singular or resolved conifold to the deformed conifold by blowing up a non-trivial two-cycle. Thus what we have really been discussing is the ``conifold transition".  As all conifold geometries are cones over $S^2\times S^3$, this can be pictured as depicted in Figure \ref{contrans}.

Generalisations could be imagined for other manifolds that allow for
non-trivial two- and three-cycles. In fact, when trying to confirm
Vafa's picture on the supergravity level, we immediately encountered
generalisations of the conifold \cite{gtone}. These come about
because under T-duality the original geometry attains a twisting of
its fibration structure by the B-field. The resulting manifold is
therefore non-K\"ahler and we will review its construction in
Section \ref{nsmirror}. The B-field is a key ingredient in the Vafa
and KS model, as its radial dependence is what gives rise to the
running coupling \cite{9911096}. It cannot be avoided when
introducing fractional D3-branes. Therefore, either the IIA or IIB
embedding of the geometric transition will have such a B-field and
its mirror (or T-dual) will be non-K\"ahler.

The outline of this article is as follows. In Section \ref{evid} we
will review the Klebanov--Strassler and the Gopakumar--Vafa models.
Section \ref{KS} explains in great detail the gauge theory duals of
regular and fractional D3-branes, as well as the duality cascade and
the singularity resolution via chiral symmetry breaking. The
discussion of Gopakumar and Vafa's model, Section \ref{vafa}, starts
with a short review of topological string theory and states their
conjecture. References are provided for the detailed calculations,
as they alone would fill an entire article. We do, however, discuss
the embedding of the open/closed duality into superstring theory and
review the derivation of the flux-generated superpotential in
Section \ref{vafasuper}. The IIB and IIA pictures can be connected
to an M-theory background in which the geometric transition
manifests itself as a flop; this is what we call Vafa's duality
chain in Section \ref{chainvafa}. Both IIB models, the KS and the
Vafa model, have an intuitive description in T-dual IIA theory,
where the conifold background turns into a pair of NS5-branes. This
picture is useful for observing the gauge theory construction and
the unification of the geometric transition with the cascading
solution in M-theory \cite{0105066, 0106040, 0110050}. We review the
arguments in Section \ref{bc} before we finally turn to summarising
the supergravity analysis of the duality chain in Section
\ref{chainour}. We walk the reader through the main steps, as there
appear some non-trivial issues along the way, but we try to be as
non-technical as possible.

An extensive appendix summarises known
facts about conifold geometries. In Appendix \ref{cs} we give the
choice of complex structures that make all three conifold metrics
Ricci flat and K\"ahler and we use this knowledge in Appendix
\ref{fluxes} to evaluate the supersymmetry requirements for known
supergravity solutions for fractional D3-branes on conifold
geometries. We agree with earlier results of \textcite{cvetgiblupo} that
the KS model on the singular conifold preserves supersymmetry,
whereas the \textcite{pt} solution for D5-branes
wrapped on the resolution of the conifold does not.

\setcounter{equation}{0}

\section{Evidence for geometric transitions}\label{evid}

\subsection{Gauge theory argument from Klebanov--Strassler}\label{KS}

In the Klebanov-Strassler model \cite{ks} a configuration of $N$
D3-branes and $M$ fractional D3-branes on a singular conifold geometry
is considered. The D3-branes sit at the singular point of the
conifold, while the fractional branes arise from wrapping $M$
D5-branes on the vanishing two-cycle of the conifold. The gauge
theory living on the branes is non-conformal, and in the IR is given
by an $SU(M)$ theory which exhibits chiral symmetry breaking and
gaugino condensation, suggesting that the correct dual of the gauge
theory in the IR limit is a deformed conifold. In this section we
review the argument for this duality from the Klebanov-Strassler
model. We begin by constructing the gauge theory of the
Klebanov-Witten model in which no fractional branes are present and
the gauge theory is conformal, and then proceed to the non-conformal
case corresponding to the presence of wrapped D5-branes.

%\begin{figure}[htp]
%\centering
%\subfigure[The Klebanov-Witten Model] % caption for subfigure a
%{
 %\label{fig:KW}
 %\includegraphics[scale=0.5]{gwyn_fig04a.eps}
%}
%\vspace*{0.5cm}

%\subfigure[The Klebanov-Strassler Model] % caption for subfigure b
%{
%\label{fig:KS}
%\includegraphics[scale=0.5]{gwyn_fig04b.eps}
%}
%\caption{Comparing the Klebanov-Witten and Klebanov-Strassler set-ups.}
%\label{fig:KWKS} % caption for the whole figure
%\end{figure}
\begin{figure}[htp]
\centering
\includegraphics[scale=0.5]{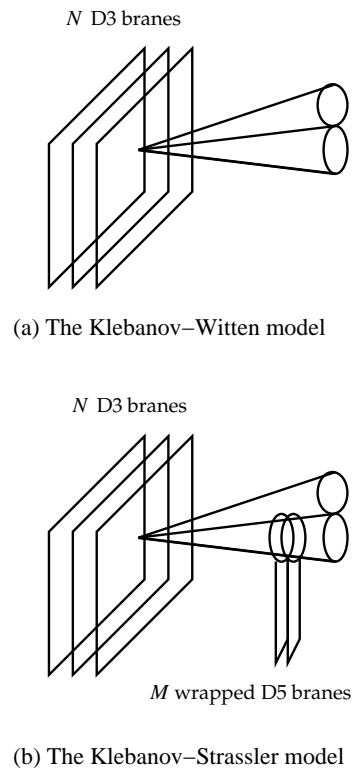}
\caption{Comparing the Klebanov-Witten and Klebanov-Strassler set-ups.}
\label{fig:KWKS}
\end{figure}

\subsubsection{The Klebanov--Witten Model}\label{KW}

First consider the Klebanov--Witten model \cite{kw} in which a stack
of D3-branes is placed at the tip of a conifold (see Figure
\ref{fig:KWKS} (a)). As in the original scenario of the AdS/CFT
conjecture, we expect a duality between the gauge theory living on
the branes and the gravity theory, found by taking the near-horizon
limit. The near-horizon geometry in this case is  $AdS_5 \times
T^{1,1}$ where $T^{1,1}$ is the base of the conifold, a
Sasaki-Einstein manifold. The reader not familiar with conifold
geometries might wish to consult Appendix A for more details. It is
the coset space $(SU(2) \times SU(2))/U(1)$ and has topology
$S^2\times S^3$ (as depicted in Figure \ref{contrans}). The metric
of $T^{1,1}$ was found by \textcite{candelas} and is given by
 \begin{eqnarray}
\label{T11}
 d\Sigma_{T^{1,1}}^2 & = & \frac{1}{9} ( d \psi + \cos \theta_1 d \phi_1 + \cos \theta_2 d \phi_2)^2 \\
 \nonumber &&+ \frac{1}{6} \sum_{i=1,2}(d \theta_i^2 + \sin^2 \theta_i d\phi_i^2),
 \end{eqnarray}
 so that the metric on the (singular) conifold is
 \begin{eqnarray}
ds^2 & = & dr^2 + r^2 d \Sigma_{T^{1,1}}^2.
 \end{eqnarray}
The conifold is a non-compact Calabi-Yau. Note that although the conifold is singular, the supergravity solution for the configuration of $N$ D3-branes at the tip is given by 
\begin{eqnarray}
ds^2 & = & H^{-\frac{1}{2}} (r) ds_{0123}^2 + H^{\frac{1}{2}}(r) \left ( dr^2 + r^2d\Sigma_{T^{1,1}}^2 \right )
\end{eqnarray}
and is nonsingular everywhere since $H(r) = 1 + \frac{L^4}{r^4}$ with \(L^4 = 4 \pi g_S N(\alpha')^2$.

We would like to study the gauge theory living on the D3-branes. To do this we will use the symmetries of the conifold to find a convenient set of co-ordinates that can be promoted to fields. We will thus construct a gauge theory with exactly the conifold symmetries, and find the correct gauge theory on the branes by adding first one D-brane and then generalising to the case where there is a stack of $N$ D3-branes on the conifold tip. We begin by rewriting the defining equation for the conifold (\ref{eq-con}),
\begin{eqnarray}
\label{con}
z_1^2 + z_2^2 + z_3^2 + z_4^2 & = & 0, \quad z_i \in \mathbb{C}^4
\end{eqnarray}
as
\begin{eqnarray}
\label{conmat}
\det Z_{ij} &= &0,
\end{eqnarray}
where \( Z_{ij} = \frac{1}{\sqrt{2}} \sum_n \sigma^n_{ij}z_n\), with $\sigma^n$ the Pauli matrices for $n=1, 2, 3$ and $\sigma^4 = \imath${\bf 1}. The defining equation for the conifold is now $\det Z = 0$, where we have a choice of co-ordinates\footnote{The main references for this section are \cite{kw}, \cite{ks} and \cite{strassler}. A recent useful reference is \cite{arvind}.}
\begin{eqnarray}
\label{coord}
Z \:= \: \left ( \begin{array}{cc} z_3 + \imath z_4 & z_1 - \imath z_2 \\ z_1 + \imath z_2 & - z_3 + \imath z_4  \end{array} \right ) \:=\:\left ( \begin{array} {cc} A_1B_1 & A_1B_2 \\ A_2B_1 & A_2 B_2 \end{array} \right ) 
\end{eqnarray}

As explained in more detail in Appendix \ref{appsingconi}, the
conifold is invariant under $SO(4) \approx SU(2) \times SU(2)$ where
$SO(4)$ acts on the $z_i$ in (\ref{con}) and the $SU(2)$s act on the
$i$ and $j$ indices in (\ref{conmat}), respectively. In addition,
there is a $U(1)$ symmetry under which all the $z_i$ in (\ref{con})
are rotated by the same phase. It is easy to see that the metric
(\ref{T11}) possesses the same symmetries. Each $SU(2)$ acts on one
$\{\theta_i, \phi_i, \psi_i \}$ (where \( \psi\) in (\ref{T11}) is
given by  \( \psi_1 - \psi_2 \)) while the $U(1)$ symmetry corresponds
to invariance under shifts in $\psi$. To find the gauge theory we
will consider D3-branes on this space and study the low-energy field
theory of the modes localised on the brane. The moduli space of
vacua of this theory should be exactly $N$ copies of the manifold -
in this case the conifold - modulo the action of the permutation group (since the D3-branes are identical). The global symmetry group of the gauge
theory living on the D3-branes will therefore be $SU(2)\times SU(2)
\times U(1)$.

The co-ordinates $A_i$ and $B_j$, $i,j = 1,2$ give a useful parametrisation of the conifold. The $A_i$ are rotated into each other under one $SU(2)$ and the $B_j$ transform under the other. As they stand, these co-ordinates represent 8 real degrees of freedom. However, in describing the conifold using $A_1, A_2, B_1$ and $B_2$ we have invariance under
\begin{eqnarray}
A_i &\rightarrow& \lambda A_i, \hspace{2cm} \lambda \in \mathbb{C}\\
\nonumber B_j & \rightarrow & \frac{1}{\lambda} B_j.
\end{eqnarray}
We fix the magnitude and phase of $\lambda$ separately. To fix the magnitude, impose
\begin{eqnarray}
\label{constraint}
|A_1|^2 + |A_2|^2 - |B_1|^2 - |B_2|^2 & = & 0.
\end{eqnarray}
This removes one degree of freedom.\footnote{It's then further dividing by the scale invariance that one obtains the base, see \eqref{eq-int}.} To fix the phase, we have to divide by $U(1)$, which means we make the identifications
\begin{eqnarray}
\label{U1}
A_i & \sim & e^{\imath \alpha} A_i,\\ \nonumber
B_j & \sim & e^{-\imath \alpha} B_j,
\end{eqnarray}
leaving 6 degrees of freedom. In other words no more conditions are required in order for these co-ordinates to give a complete description of the conifold. Thus we have arrived at a description of the conifold given by the 4 co-ordinates $A_i, B_j$ and the magnitude constraint (\ref{constraint}), subject to the identifications (\ref{U1}). This description is equivalent to the equation of the conifold given by (\ref{conmat}).

Next we turn to the gauge theory. We can formulate the field theory in terms of $A_i$ and $B_i$ because they have the symmetries we want; we promote them to chiral superfields.  Consider then an ${\cal N} = 1$ supersymmetric theory with $U(1)$ gauge group and chiral superfields $\tilde A_i, \tilde B_j$ with $i,j =1,2$, such that the $\tilde A_i$ have charge 1 and the $\tilde B_j$ have charge -1.  The superpotential is given by
\begin{eqnarray}\label{superpotential}
 W & = & \lambda  \det (\tilde A_i \tilde B_j)\\ \nonumber  
   & = & \lambda  (\tilde A_1 \tilde B_1 \tilde A_2 \tilde B_2 - \tilde A_1 \tilde B_2 \tilde A_2 \tilde B_1 ),
 \end{eqnarray}
 which preserves $SU(2) \times SU(2) \times U(1)_R$. The D-term condition for supersymmetry is given by $D=0$, where
\begin{eqnarray}\label{dterm}
 D &= &- \frac{1}{2} \sum_i \left [ q_{\tilde A_i} \tilde A_i^*\tilde A_i  +  q_{\tilde B_i} \tilde B_i^* \tilde B_i  \right ]+ \xi,
\end{eqnarray}
$\xi$ being the coefficient of the Fayet-Iliopoulos term and $q_i$
being the $U(1)$ charge of the relevant field. When $\xi$ is
zero\footnote{In fact $\xi \neq 0$ corresponds to a resolved
conifold, as discussed in more detail in \textcite{arvind}.} this is
exactly our conifold constraint (\ref{constraint}). The moduli space
of vacua is found by furthermore dividing by the gauge group $U(1)$
which is equivalent to imposing (\ref{U1}).  Now we claim that for a
single D3-brane on a conifold, the gauge theory whose moduli space
exactly matches the conifold is in fact given by a gauge group $U(1) \times U(1)$, so the fields now have charges
$(1,-1)$ and $(-1,1)$, respectively, under the two $U(1)$ groups. The
two D-term conditions (one for each $U(1)$) both yield the relation
(\ref{constraint}). The superpotential \eqref{superpotential} is
unchanged and zero, so there are no F-term conditions
\[ F_i = - \frac{ \partial W}{\partial \tilde A_i} = 0, \]
and this theory's moduli space is the conifold.

When we place instead of a single brane a stack of $N$ D3-branes on the conifold, the gauge group becomes\footnote{Actually it is $U(N) \times U(N)$ but the $U(1)$ factors decouple in the IR.} $SU(N) \times SU(N)$ so the superpotential no longer vanishes. It is given by
 \begin{eqnarray}
 \label{kwsup}
 W & = & \lambda tr \det (\tilde A_i \tilde B_j),
 \end{eqnarray}
where the trace is now necessary because the superfields carry a gauge index for each $U(N)$ and should therefore be treated as matrices.  The $\tilde A$ and $\tilde B$ fields are now bifundamental fields transforming in the $(N, \overline N)$ and $(\overline N, N)$ representations of the gauge groups, respectively. Together with the D-term conditions and the gauge invariance they give a description of the moduli space which matches the description of the conifold in terms of $A$ and $B$ arrived at above. The F-term equations
 \begin{eqnarray}
\tilde B_1 \tilde A_i \tilde B_2 - \tilde B_2 \tilde A_i \tilde B_1 & = & 0,\\
\nonumber \tilde A_1 \tilde B_j \tilde A_2 - \tilde A_2 \tilde B_j \tilde A_1 & = & 0,
\end{eqnarray}
together with the D-term conditions \eqref{dterm}, can be shown to
have a solution if and only if $\tilde A$ and $\tilde B$ can be
simultaneously diagonalised \cite{strassler}. In this case the
superpotential vanishes, giving exactly (\ref{conmat}). The theory
flows to a non-trivial infrared fixed point \cite{kw}. We have seen
how the symmetry group $SU(2) \times SU(2) \times U(1)$ acts on
$A_i, B_j$.  The $U(1)$ global symmetry of the conifold manifests as
what is called a $U(1)_R$ symmetry in the gauge theory, under which
\begin{eqnarray}
(A_i, B_j) & \longrightarrow & e^{R\alpha} (A_i, B_j).
\end{eqnarray}
The R-charges of the fields $A_i$ and $B_j$ are found by imposing
conformal invariance\footnote{Setting the $\beta$-functions equal to
zero yields anomalous dimensions -1/2 for the fields, and one can
make use of the relation \(dim {\cal{O}} = 1 + \frac{1}{2}
\gamma_{\cal{O}} = \frac{3 R_{\cal{O}}}{2}\) for an operator
$\cal{O}$ to solve for the R-charges.} and are equal to $1/2$. This
leads to an R-charge of 2 for the superpotential, as required. The
extra $U(1)$ symmetry expressed in (\ref{U1}) is referred to in the
gauge theory as baryonic symmetry.

\subsubsection{The duality cascade}
\label{ks}

Having discussed the Klebanov-Witten model in detail, we are now in a position to study the Klebanov-Strassler model \cite{ks}. As shown in Figure \ref{fig:KWKS}(b), the difference between the two set-ups is that in the configuration considered by Klebanov and Strassler there are $M$ D5-branes wrapping the vanishing 2-cycle of the conifold. These are effectively D3-branes with fractional charge, called fractional branes. Their effect is to render the dual gauge theory non-conformal, with many interesting consequences.

This set-up was studied by \textcite{9808075} and \textcite{9911096}, where the running of the gauge coupling was found. An exact solution, including backreactions, was set out in \textcite{kt}, but the details of the duality cascade, chiral symmetry breaking and confinement were elucidated in \textcite{ks}, which we follow closely here. A comprehensive review is given by \textcite{strassler}.

Consider first the gravity theory. The fractional branes are magnetic sources for the three-form field strength ($F_3$ is the Hodge dual of $F_7 = dC_6$)
\begin{eqnarray}
F_3 & = & M \omega_3,
\end{eqnarray}
where
\begin{eqnarray}
\label{omega3}
\omega_3 & = & \frac{1}{2} d \psi \wedge ( \sin \theta_1 d \theta_1 \wedge d \phi_1 - \sin \theta_2 d \theta_2 \wedge d \phi_2) \\ \nonumber
&&+ \frac{1}{2} d \phi_1 \wedge d \phi_2 \wedge ( \cos \theta_1 \sin \theta_2 d \theta_2 + \sin \theta_1 \cos \theta_2 d \theta_1)
\end{eqnarray}
is the 3-form dual to the non-trivial 3-cycle of the conifold. In
addition there are $N$ units of five-form flux due to the D3-branes,
but the total five-form RR field strength also has a contribution
from the NS-NS B-field, which is necessarily present in the
supergravity solution:
\begin{eqnarray}
\label{run}
\tilde F_5 & = & d C_4 + B_2 \wedge F_3,\\
B_2 & = & 3 g_s M \omega_2 \ln \left (\frac{r}{r_0} \right ).
\end{eqnarray}
Here $\omega_2 = \frac{1}{2} \left ( \sin \theta_1 d \theta_1 \wedge d \phi_1 - \sin \theta_2 d \theta_2 \wedge d \phi_2 \right )$ is the 2-form on the 2-cycle wrapped by the D5s and $r_0$ is some UV scale.

From the gauge theory point of view, the presence of the $M$ fractional branes changes the gauge group\footnote{See \textcite{9808075, 9911096} and \textcite{kt}. An easy way to understand this generalisation is given in the brane construction discussion in Section \ref{bc}.} from $SU(N) \times SU(N)$ to $SU(N+M) \times SU(N)$. The field content is still given by four chiral superfields $A_i$ and $B_j$, while the superpotential is given by (\ref{kwsup}).
However, the relative gauge coupling depends on $B_2$ and therefore
on $r$ \cite{kt, 9911096}:
\begin{eqnarray}
\frac{1}{g_1^2} - \frac{1}{g_2^2} & \sim & \frac{1}{g_s} \left [ \int_{S^2} B_2 - \frac{1}{2} \right ].
\end{eqnarray}
According to the usual AdS/CFT dictionary, $r$ maps to the RG scale in the dual gauge theory. Thus this theory is no longer conformal.

The gauge couplings $\frac{1}{g_1^2}$ and $\frac{1}{g_2^2}$ flow in
opposite directions. Facing a divergence in one we can continue by
performing a Seiberg duality transformation \cite{9411149}, under
which we obtain an $SU(N) \times SU(N-M)$ theory which resembles
closely the theory we started with. We can see this by noting the
running of the five-form flux under which the D3-branes are charged
(\ref{run}). Since $C_4$ is oriented along the worldvolume of the
D3-branes in the 0123-directions, we can write the self-dual
five-form
\begin{eqnarray*}
\tilde F_5 & = & \mathcal{F}_5 + \star \mathcal{F}_5
\end{eqnarray*}
in terms of
\begin{eqnarray*}
{\cal {F}}_5 & = & N(r)  \text{vol} (T^{1,1})\,.
\end{eqnarray*}
Here we used $\omega_2 \wedge \omega_3 \sim \mathrm{vol} (T^{1,1})$
and defined
\begin{eqnarray*}
N(r) & = & N + \frac{3}{2 \pi} g_s M^2 \ln \left (\frac{r}{r_0} \right ).
\end{eqnarray*}
So the number of colours in the theory has become a scale-dependent quantity. $N(r)$ will decrease in units of $M$, i.e. the running gives rise to a flow under which $SU(N+M) \times SU(N)$ $\rightarrow$ $SU(N) \times SU(N-M)$. This process will continue until the gauge group is $SU(2M) \times SU(M)$ or just $SU(M)$, corresponding to a situation in which only the $M$ fractional D3-branes remain -- the five-form flux has decreased to zero indicating that the $N$ D3-branes have ``cascaded away". The process is called a duality cascade, since the $SU(N)$ and $SU(N-M)$ theories are related by a Seiberg duality \cite{9411149}.

\subsubsection{Chiral symmetry breaking and deformation of the conifold}
\label{chisb}

When $N- k M$ approaches zero we should do a more careful analysis; the cascade must stop because negative $N$ is physically nonsensical. At sufficiently small $r$ the solution becomes singular. By studying the far IR of the gauge theory, Klebanov and Strassler argued that this singularity is removed by the IR dynamics, via a gluino condensate which breaks the anomaly-free $\mathbb{Z}_{2M}$ R-symmetry in the theory to $\mathbb{Z}_2$. The expectation value acquired by the gaugino condensate maps to the deformation parameter of the conifold: $\mu$ in (\ref{eq-def}). Thus the singularity is removed in the gravity picture by blowing up the $S^3$ of $T^{1,1}$.

Although the metric of the KS set-up has a continuous $U(1)_R$ symmetry, the full SUGRA solution is only invariant under a $Z_{2M}$ subgroup of this, under which
\begin{eqnarray}
(A_i, B_j) & \rightarrow & (A_i, B_j) e^{\frac{2 \pi \imath n}{4M}}
\end{eqnarray}
and the superpotential rotates by $e^{\frac{2 \pi \imath n}{M}}$.

The theory has a moduli space with $M$ independent branches in the IR. To see this, we probe the space with a single additional D3-brane. Since in the far IR the five-form field strength has cascaded to zero and only the $M$ fractional D3-branes remain, the gauge group is $SU(M+1) \times SU(1)$ or $SU(M+1)$. The fields are $A_i$ and $B_j$, $i,j = 1,2$, in the $M+1$ and $\overline{M+1}$ representations, and the superpotential is of the form of (\ref{kwsup}). We can write this in terms of the gauge invariant $N_{ij} = A_i B_j$, which one can think of as a meson.

At low energy the theory is described by the Affleck--Dine--Seiberg superpotential \cite{affleck}
\begin{eqnarray}\nonumber
W_L & = & \lambda N_{ij} N_{kl} \epsilon^{ik} \epsilon^{jl} + (M-1)\left [ \frac{2 \Lambda^{3M+1}}{N_{ij} N_{kl} \epsilon ^{ik} \epsilon^{jl}}\right ] ^{\frac{1}{M-1}}
\end{eqnarray}
to which the only solutions for a supersymmetric vacuum are
\begin{eqnarray}
\left ( N_{ij} N_{kl} \epsilon^{ik} \epsilon ^{jl}\right ) ^M & = & \frac{2 \Lambda^{3M+1}}{\lambda^{M-1}}.
\end{eqnarray}
In other words, the theory in the far IR has a moduli space with $M$ independent branches. The $\mathbb{Z}_{2M}$ R-symmetry permutes these branches, rotating $N_{ij} N_{kl} \epsilon^{ik} \epsilon ^{jl} $ by $e^{2 \pi \imath M}$, so the R-symmetry is spontaneously broken from $\mathbb{Z}_{2M}$  to  $\mathbb{Z}_2$. Thus the gaugino condensate is responsible for the chiral symmetry breaking. To see how this symmetry breaking smooths out the singularity in the IR, we should remind ourselves of Section \ref{KW} in which we promoted our co-ordinates to chiral superfields. We can now go the other way, using $N_{ij}$ to find a co-ordinate description of the geometry in terms of the gauge theory fields. In the classical theory, $\det N_{ij} = 0$, which should be compared to (\ref{conmat}). In this case the probe brane is moving on the singular conifold. However, as we have seen, in the Klebanov-Strassler model the field theory in the IR gives
\begin{eqnarray}
\det N_{ij} & = & \left (\frac{\Lambda^{3M+1}}{(2 \lambda)^{M-1}} \right ) ^{\frac{1}{M}}.
\end{eqnarray}
Thus the probe brane moves not on the singular but on the deformed conifold, where the deformation is given by the gaugino condensate responsible for chiral symmetry breaking. It is the deformed conifold which gives the correct moduli for the field theory and which is the correct background geometry for the SUGRA solution that is dual to the gauge theory in the far IR.

It has been pointed out by \textcite{ghk} that the IR of the KS background should really be thought of as the next--to--last step of the cascade. Studying the field theory dual to this $SU(2M)\times  SU(M)$ one finds that baryonic operators, instead of mesonic ones as for mobile D3--branes, acquire a vev. This background is therefore said to lie on the baryonic branch.

%%%%%%%%
\subsection{Open/Closed duality}\label{vafa}

A different perspective on geometric transitions comes from topological string theory. \textcite{gopakumar} conjectured a duality between an open and a closed topological string theory that live on different backgrounds. The connection to the Klebanov--Strassler model became apparent after \textcite{vafa} embedded this duality into superstring theory. We will only review the arguments and refer the reader to the original work \cite{gopakumar, vafa, bcov} or the reviews \cite{neitzke, koshkin} for details. The purpose of Section \ref{topol} is to define topological string amplitudes and explain the difference between open and closed topological string theories. The reader familiar with topological string theory may want to skip ahead to the Gopakumar--Vafa conjecture explained in Section \ref{conjecture}.

\subsubsection{Topological Sigma Models and String Theory}\label{topol}

Throughout this section we will restrict ourselves to the case $H_3=0$.  We will closely follow the review \textcite{neitzke}, see e.g. \textcite{wittentopol, topol, eguchi} and \textcite{marino} for details.

String theory is intrinsically linked to sigma models. We can view string theory as the description of a two-dimensional worldsheet $\Sigma$ propagating through a 10-dimensional target space $M$. The sigma model that describes this theory deals with maps $\phi: \Sigma\to M$. These maps can be promoted to chiral superfields $\Phi$ that have $\phi$ in their lowest component and obey the 2d sigma model action
\begin{equation}\label{sigmaaction}
S\,=\,-\frac{1}{4}\,\int d\tau\,d\sigma\,d^2\theta\,\left(g_{ij}(\Phi)+
B_{ij}(\Phi)\right)\, D^\alpha\Phi^i\,D_\alpha\Phi^j\,,
\end{equation}
with indices $i,j=1,...,d$ parametrising the target space and
\begin{equation}\label{defD}
D_\alpha \,=\, \frac{\partial}{\partial\overline{\theta}^\alpha}
+i\rho^\mu\theta_\alpha\partial_\mu\,,
\end{equation}
where $\rho^\mu$ is a 2d $\gamma$-matrix and $\theta_\alpha$ a two-component Grassmann-valued spinor. Chiral superfields are defined by $\overline{D}^\alpha \Phi^i=0$. Written in terms of $\mathcal{N}=1$ superfields, this action has explicit $\mathcal{N}=1$ supersymmetry. In the case $H=dB=0$ it  has further non-manifest supersymmetry if and only if the target space is K\"ahler \cite{zumino}.

Considering a sigma model that does contains not only chiral but also
twisted chiral superfields, one can find additional supersymmetry with $H\ne 0$ 
even if the target is {\sl not} K\"ahler. This was proposed by
\textcite{gates} more than twenty years ago and has recently found an
embedding in generalised complex geometry \cite{hitch}. It turns out
that the target manifolds in this model define a (twisted)
generalised K\"ahler structure \cite{gualtieri}. Stimulated by this
new mathematical language,  there has been tremendous progress in
defining generalised (topological) sigma models \cite{lindstrom, lindstrocek, lindstrpersson,
kapuli, pestun}, but we will not discuss this direction any further.

Topological string theory integrates not only over all maps $\phi$ but also over all metrics on $\Sigma$; this is often called a sigma model coupled to two-dimensional gravity. Classically, the sigma model action depends only on the conformal class of the metric, so the integral over metrics reduces to an integral over conformal (or complex) structures on $\Sigma$.

The sigma model \eqref{sigmaaction} with K\"ahler target can be made topological by a procedure called ``twisting'' \cite{wittentopol}, which basically shifts the spin of all operators by 1/2 their R-charge. There are two conserved supercurrents for the two worldsheet supersymmetries that are nilpotent
\begin{equation}
(G^\pm)^2\,=\,0\,,
\end{equation}
so one might be tempted to use these as BRST operators and build cohomologies of states. But they have spin 3/2. The twist shifts their spin by half their R-charge to obtain spin 1 operators:
\begin{equation}
S_{new}\,=\,S_{old}+\frac{1}{2}\,q,
\end{equation}
where $q$ is the $U(1)$ R-charge of the operator in question. Classically, the theory has a vector $U(1)_V$ symmetry and an axial $U(1)_A$ symmetry. Twisting by $U(1)_V$ gives the so-called A-model; twisting by $U(1)_A$ the B-model.  The $U(1)_A$ might suffer from an anomaly unless $c_1(M)=0$, which leads to the requirement that the target must be a Calabi--Yau manifold for the B-model. One could now define $Q=G^+$ or $Q=G^-$ and use this nilpotent operator as a BRST operator, i.e. restrict one's attention to observables which are annihilated by $Q$.

Before doing so let us note a special feature of $\mathcal{N}=(2,2)$ supersymmetry. Since left and right movers basically decouple, we can split any of the operators $G^\pm$ into two commuting copies, one for left and one for right movers. In terms of complex co-ordinates let us denote the left movers as holomorphic $G^\pm$ and the right movers as antiholomorphic $\overline{G^\pm}$. This makes the (2,2) supersymmetry more apparent. Now twisting can be defined for left and right movers independently and we obtain in principle four models, depending on which we choose as BRST operators:
\begin{eqnarray}\nonumber
\mbox{A model}:\; & &(G^+,\,\overline{G}^+)\,,\qquad
\mbox{B model}:\; (G^+,\,\overline{G}^-),\\ \nonumber
\mbox{\={A} model}:\; & & (G^-,\,\overline{G}^-) \,,\qquad
\mbox{\={B} model}:\;  (G^-,\,\overline{G}^+)\,.
\end{eqnarray}
Of these four models, only two are actually independent, since the correlators for A (B) and for \={A} (\={B}) are related by complex conjugation. So we will ignore \={A} and \={B} in the following.

Starting with this set-up, one can now discuss observables in topological theories. It turns out that $Q+\overline{Q}$ in the A-model reduces to the differential operator $d=\partial+\overline{\partial}$ on $M$, i.e. the states of the theory lie in the de Rham cohomology. A ``physical state'' constraint requires states to be in $H^{(1,1)}(M)$ only, which corresponds to deformations of the K\"ahler structure on $M$. One can also show that correlators are independent of the complex structure modulus of $M$, since the corresponding operators are $Q$-exact (they decouple from the computation of string amplitudes).

In the B-model the relevant cohomology is that of $\overline{\partial}$ with values in $\Lambda^*(TM)$, i.e. the observables are (0,1)-forms with values in the tangent bundle $TM$. These correspond to complex structure deformations. One can also show that in this case correlation functions are independent of K\"ahler moduli. So each of the two topological models depends only on half the moduli,
\begin{eqnarray}\nonumber
\mbox{A model on}\, M: & & \mbox{depends on K\"ahler moduli of}\,M;\\ \nonumber
 \mbox{B model on}\, M: & & \mbox{depends on complex structure}\\ \nonumber
 			& & \mbox{moduli of}\,M\,.
\end{eqnarray}
In this sense both models describe topological theories, because they only depend on the topology of the target, not its metric. It can also be shown that the relevant path integral $\int e^{-S}$ simplifies tremendously compared to ordinary field theories. It localises on $Q$-invariant configurations. These are simply constant maps $\phi:\Sigma\to M$ with $d\phi=0$ for the B-model and holomorphic maps $\overline{\partial}\phi=0$ for the A-model. In this sense the B-model is simpler than the A-model, because the string worldsheet ``reduces to a point'' on $M$; its correlation functions are those of a field theory on $M$. They compute quantities determined by the periods of the holomorphic 3-form $\Omega^{(3,0)}$, which are sensitive to complex structure deformations.

The holomorphic maps in the A-model are called ``worldsheet instantons''. Each worldsheet instanton is weighted by
\begin{equation*}
\exp\left(\int_C (J+i B)\right),
\end{equation*}
where $t=J+i B\;\in H^2(M,\mathbb{C})$ is the complexified K\"ahler parameter and $C$ is the image of the string worldsheet in $M$. Summing over all instantons makes this theory more complicated than the B-model, but only in the sense that it is not local on $M$ and does not straightforwardly reduce to a field theory on $M$. In summary, the A-model moduli are complexified volumes of 2-cycles, while the B-model moduli are the periods of $\Omega$.

Let us now talk about the relation of these topologically twisted sigma models to string theory. As mentioned before, string theory sums not only over all possible maps $\phi:\,\Sigma\to M$, as discused for the sigma models above, but also over all possible metrics on $\Sigma$. The latter actually reduces to a sum over the moduli space of genus $g$ Riemann surfaces. The topological string free energy is then defined as a sum over all genera
\begin{equation}
\mathcal{F}\,=\,\sum_{g=0}^\infty g_s^{2-2g}\,F_g,
\end{equation}
with the string coupling $g_s$ and $F_g$ being the amplitude for a fixed genus $g$. The string partition function is given by $\mathcal{Z}=\exp \mathcal{F}$.

The interesting quantities for the topological string theory are therefore the genus $g$ partition functions. Already at genus zero one finds a lot of interesting information about $M$. In the A-model the genus zero free energy turns out to be
\begin{equation}
F_0\,=\,\int_M J\wedge J\wedge J + \;\mbox{instanton corrections}\,.
\end{equation}
The first term corresponds to the classical contribution of the worldsheet theory: it gives the leading order contribution in which the string worldsheet just reduces to a point.
The instanton term contains a sum over all homology classes $H_2(M,\mathbb{Z})$ of the image of the worldsheet, each weighted by the complexified area, and a sum over ``multi-wrappings'' in which the map $\Sigma\to M$ is not one-to-one.

To define the genus zero free energy in the B-model requires a little more effort. We already noted that the relevant moduli are periods of $\Omega\in H^3(M,\mathbb{C})$. This cohomology can be decomposed as
\begin{equation}
H^3\,=\,H^{3,0}\oplus H^{2,1} \oplus H^{1,2} \oplus H^{0,3}\,.
\end{equation}
For a Calabi--Yau threefold the Hodge numbers are given by $h^{3,0}=h^{0,3}=1$, because there is one unique holomorphic 3-form, and $h^{2,1}=h^{1,2}$. Therefore, $H^3(M,\mathbb{C})$ has real dimension $2\,h^{1,2}+2$. It is customary to choose a symplectic basis of 3-cycles $A^i$ and $B_j$ with intersection numbers
\begin{equation}
A^i \cap A^j=0\,,\quad B_i\cap B_j=0\,,\quad A^i\cap B_j=\delta^i_j\,,
\end{equation}
with $i,j=1,...,h^{1,2}+1$.
One can then define homogeneous co-ordinates on the moduli space of complex structure deformations by
\begin{equation}
X^i\,:=\,\int_{A^i}\,\Omega\,.
\end{equation}
This gives $h^{1,2}+1$ complex co-ordinates, although the moduli space only has dimension $h^{1,2}$. This overcounting is due to the fact that $\Omega$ is only unique up to overall rescaling, so the same is true for the co-ordinates defined this way. Therefore they carry the name ``homogeneous co-ordinates''.
There are also $h^{1,2}+1$ periods over B-cycles
\begin{equation}
\hat{F}_i\,:=\,\int_{B^i}\,\Omega\,.
\end{equation}
Due to the relation between A and B cycles, there must be a relation between the periods. In other words, we can express $\hat{F}_i$ as a function of $X^j$.
\begin{equation}
\hat{F}_i\,=\,\hat{F}_i(X^j)\,.
\end{equation}
One can prove that these satisfy an integrability condition,
\begin{equation}
\frac{\partial}{\partial X^i}\,\hat{F}_j\,=\,\frac{\partial}{\partial X^j}\,\hat{F}_i,
\end{equation}
which allows us to define a new function $F$ via
\begin{equation}
\hat{F}_i\,=\,\frac{\partial}{\partial X^i}\,F\,.
\end{equation}
This function is actually nothing but the genus zero free energy of the B-model.
It is given by the simple formula
\begin{equation}
F\,=\,\frac{1}{2}\,X_i \hat{F}^i\,.
\end{equation}

In general, the sum over all worldsheet configurations is too hard to carry out explicitly. There are nevertheless some tools that enable one to calculate topological string amplitudes. For example, mirror symmetry between the A- and B-models can be used to compute amplitudes in the model of choice (usually the B-model since it does not obtain instanton corrections) and then extrapolating the result to the mirror theory. We will be more interested in a duality between open and closed strings, which enables one in principle to calculate the free energy at all genera for a particular class of non-compact geometries --- e.g. conifolds. To describe open topological strings we need to explain what we mean by topological branes that appear as boundaries of $\Sigma$.

A D-brane corresponds to a boundary condition for $\Sigma$ that is BRST-invariant. In the A-model this implies that the boundary should be mapped to a Lagrangian submanifold\footnote{This means $L$ has half the dimension of $M$ and the K\"ahler form restricted to $L$ vanishes or in other words it is an isotropic submanifold of maximal dimension (half of that of the ambient Calabi--Yau). However, a more generic condition that allows for a non--flat gauge field on the brane has been derived in \cite{kapuli1}.} $L$ of $M$. If we allow open strings to end on $L$, we say that the D-branes are wrapped on $L$. Having a stack of $N$ D-branes on $L$ corresponds to including a weighting factor $N$ for each boundary.

D-branes carry gauge theories in physical string theory (we will use ``physical'' for the target space perspective to distinguish it from toplogical string theory). The same is true for topological branes. In the A-model it turns out that one can actually compute the exact string field theory, which is again a topological theory: $U(N)$ Chern--Simons theory \cite{witten, wittened}. Its action in terms of the $U(N)$ gauge connection $A$ is given by
\begin{equation}
S\,=\,\int_L \,Tr\left(A\wedge dA+\frac{2}{3}\,A\wedge A\wedge A\right)\,.
\end{equation}
This action might still obtain instanton corrections, but \textcite{witten}
showed that in the special case where $L=S^3$ there
are none. This is fascinating, because the $S^3$ in the deformed
conifold (which is also $T^*S^3$) is such a Langrangian submanifold.

In physical superstring theory, D-branes are sources for RR fluxes. So under what quantity are topological branes charged? The only fluxes available are the K\"ahler 2-form $J$ and the holomorphic 3-form $\Omega$. Wrapping a topological brane on a Lagrangian subspace $L$ of $M$ (in the A-model) creates a flux through a 2-cycle $C$ which  ``links'' L. This link means that $C=\partial S$ for some 3-cycle S that intersects L once, so $C$ is homologically trivial in $M$, although it becomes nontrivial if considered as a cycle in $M\setminus L$. This implies that $\int_C J=0$ since $J$ is closed and $C$ trivial.

Wrapping $N$ branes on L has the effect of creating a K\"ahler flux through $C$
\begin{equation}
\int_C J\,=\,N g_s\,,
\end{equation}
because the branes act as a $\delta$-function source for the two-form, i.e. $J$ is no longer closed on L, but $dJ=Ng_s\,\delta(L)$. Similarly, a B-model brane on a holomorphic 2-cycle $Y$ induces a flux of $\Omega$ through the 3-cycle linking $Y$. In principle we could also wrap branes on 0, 4 or 6-cycles in the B-model, but there is no field candidate those branes could be charged under. This suggests a privileged role for 2-cycles.

\subsubsection{The Gopakumar--Vafa Conjecture}\label{conjecture}

After all these preliminaries we are now ready to explain the
geometric transition on conifolds. This is a duality between open
and closed topological strings (it has been shown that they give
rise to the same string partition function) which has profound
physical consequences. The dual gauge theory from the open string
sector is $\mathcal{N}=1$ SYM in d=4. The IR dynamics of this gauge
theory can be obtained either from the open or from the closed
string sector. In this sense, both string theory backgrounds are
dual; they compute the same superpotential because they have the
same topological string partition function. The key to this duality
in the gauge theory is to identify parameters from the open string
theory with parameters from the closed string theory. In the IR it
will be the gluino condensate which is identified either with the
K\"ahler or complex structure modulus of the closed or open string
theory background.

The geometric transition in question \cite{gopakumar} considers the A-model on the deformed conifold $T^*S^3$. As noted by \textcite{witten}, the exact partition function of this theory is simply given by $U(N)$ Chern--Simons theory. The {\it closed} A-model on this geometry is trivial, because it has no K\"ahler moduli. But the $T^*S^3$ contains a Lagrangian 3-cycle $L=S^3$ on which we can wrap branes in the open A-model. This creates a flux $Ng_s$ of $J$ through the 2-cycle C which links L, in this case $C=S^2$. So it is natural to conjecture that this background is dual to a background with only flux through this 2-cycle. The resolved conifold is the logical candidate for this dual background as it looks asymptotically like the deformed conifold, but has a finite $S^2$ at the tip of the cone. This led Gopakumar and Vafa to the following \vspace{5mm}

{\parindent=0.0mm
{\sl Conjecture: The open A model on the deformed conifold $T^*S^3$ with $N$ branes wrapping the $S^3$ is dual to the closed A model on the resolved conifold $\mathcal{O}(-1)\oplus\mathcal{O}(-1)\to \mathbb{CP}^1$, where the size of the $\mathbb{CP}^1$ is determined by $t=Ng_s$.}} \vspace{5mm}

There are no branes anymore in the dual geometry; there is simply no 3-cycle which they could wrap. The passage from one geometry to the other is called a ``geometric transition'' or ``conifold transition'' in this case. The agreement of the partition function on both sides was shown by \textcite{gopakumar} for arbitrary 't Hooft coupling $\lambda=N g_s$ and to all orders in $1/N$. In this sense, this duality is an example of a large $N$ duality as suggested by 't Hooft: for large $N$ holes in the Riemann surface of Feynman diagrams are ``filled in'' or ``condensed'', where one takes $N\to\infty$ with $g_s$ fixed. The authors \textcite{gopakumar} matched the free energy $F_g$ at every genus $g$ via the identification of the 't Hooft coupling
\begin{equation}\label{identifythooft}
i\lambda\,=\,Ng_s\quad\mbox{(open)} \quad\leftrightarrow\quad
i\lambda\,=\,t \quad\mbox{(closed)}\,,
\end{equation}
where $t$ is the complexified K\"ahler parameter of the $S^2$ in the resolved conifold and the identification of the 't Hooft coupling for open strings is dictated by the Chern--Simons theory on $S^3$.

Beyond that, it was also shown that the coupling to gravity (to the metric\footnote{It might seem contradictory that there can be a coupling to the metric when we are speaking about topological models. The classical Chern--Simons action is indeed independent of the background, but at the quantum level such a coupling can arise. On the closed side there are possible IR divergences, anomalies for non-compact manifolds, that depend on the boundary metric of these manifolds.}) and the Wilson loops take the same form for the open and closed theories. The two topological string theories described here correspond to the different limits $\lambda\to 0$ and $\lambda\to\infty$, but they are conjectured to describe the same string theory (with the same small $g_s$) only on different geometries.

\subsection{The Vafa Model}\label{vafasuper}
\subsubsection{Embedding Gopakumar--Vafa in Superstrings and Superpotential}\label{vafaembed}

This scenario has an embedding in ``physical'' type IIA string theory. Starting with $N$ D6-branes on the $S^3$ of the deformed conifold we find a dual background with flux through the $S^2$ of the resolved conifold. The Calabi--Yau breaks 3/4 of the supersymmetry (which leaves 8 supercharges), therefore the theory on the worldvolume of the branes has $\mathcal{N}=1$ supersymmetry (the branes break another half). There is a $U(N)$ gauge theory on the branes (in the low-energy limit of the string theory the $U(1)$ factor decouples and we have effectively $SU(N)$). As described in the last section, these wrapped branes create flux and therefore a superpotential.\footnote{We will set the string coupling $g_s=1$ throughout most of this section.} This superpotential is computed from topological strings, but we need a gauge theory parameter in which to express it. The relevant superfield for $\mathcal{N}=1$ $SU(N)$ is $S$, the chiral superfield with a gaugino bilinear in its bottom component. We want to express the free energy $F_g$ in terms of $S$. Since there will be contributions from worldsheets with boundaries, we can arrange this into a sum over holes $h$:
\begin{equation}
F_g(S)\,=\,\sum_{h=0}^\infty F_{g,h}\,S^h\,.
\end{equation}
It turns out that the genus zero term computes the pure gauge theory, i.e. pure SYM. Higher genera are related to gravitational corrections.

As discussed above, the open topological string theory is given by Chern-Simons on $T^*S^3$, which has no K\"ahler modulus. The superpotential created by the open topological amplitude of genus zero was found by \textcite{vafa} to be
\begin{equation}\label{wopen}
W^{\rm open}\,=\,\int d^2\theta\,\frac{\partial F_0^\text{open}(S)}{\partial S}
+\alpha \,S+\beta,
\end{equation}
with $\alpha,\beta=\,$constant and $\alpha\,S$ being the explicit annulus contribution ($h=2$).

Although the topological model is not sensitive to any flux through a 4- or 6- cycle, in the superstring theory the corresponding RR forms $F_4$ and $F_6$ can be turned on. On the closed string side this corresponds to a superpotential
\begin{eqnarray}\nonumber
W^{\rm closed} &=& \int F_2\wedge (J+\imath B)\wedge (J+\imath B)\\
  & &  +i\int F_4\wedge (J+\imath B) +\int F_6\,,
\end{eqnarray}
where $(J+\imath B)$ is the complexified K\"ahler class, whose
period over the basis 2-cycle gives the complex K\"ahler modulus $t$
(of the resolved conifold). According to \textcite{vafa} the topological
string amplitude is not modified by these fluxes. The genus zero
topological string amplitude $F_0$ determines the size of the
4-cycle to be $\frac{\partial F_0}{\partial t}$ \cite{vafa}. If we
have $M, L$ and $P$ units of 2-, 4- and 6-form flux, respectively,
the superpotential yields after integration
\begin{equation}
W^{\rm closed}\,=\,M\,\frac{\partial F_0}{\partial t}+itL+P\,.
\end{equation}
Note that requiring $W=0$ and $\partial_t W=0$ fixes $P$ and $L$ in terms of $M$ and $t$. $M$ is of course fixed by the number of branes in the open string theory.

This looks very similar to the superpotential for the open theory \eqref{wopen}. We have already discussed that the topological string amplitudes agree
\begin{equation}
F^\text{open} \,=\,F^\text{closed}
\end{equation}
if one identifies the relevant parameters as in \eqref{identifythooft}. To map the superpotentials onto each other we have to identify $S$ with $t$ and $\alpha, \beta$ with the flux quantum numbers $iL, P$. It is clear from the gauge theory side that $\alpha$ (or $L$) is related to a shift in the bare coupling of the gauge theory. In particular, to agree with the bare coupling to all orders we require $iL=V/g_s$, where $V$ is the volume of the $S^3$ that the branes are wrapped on. This gives an interesting relation between the size $V$ of the blown-up $S^3$ (open) and the size $t$ of the blown-up $S^2$ (closed):
\begin{equation}
\left(e^t-1\right)^N\,=\,\mbox{const}\cdot e^{-V}\,.
\end{equation}
This indicates that for small $N$ ($N/V\ll 1$) the description with D-branes wrapped on $S^3$ is good (since $t\to 0$), whereas for large $N$ the description with blown-up $S^2$ is good (since $V\to-\infty$ does not make sense). It should be clear from our discussion that after the $S^3$ has shrunk to zero size there cannot be any D6-branes in the background, but RR fluxes are turned on.

Let us finish this section with the explicit derivation of the Veneziano--Yankielowicz superpotential in type IIB \cite{cachazo}, from the perspective of the closed string theory. This theory is mirror dual to the IIA scenario discussed above,\footnote{The same result could of course be obtained in IIA, but the IIB treatment is not complicated by instantons.} so the open string theory lives on a resolved conifold background and the closed theory with fluxes lives on the deformed conifold described by
\begin{equation}
f \,=\, \sum_{i=1}^4 z_i^2 - \mu^2 \,=\, 0.
\end{equation}
On the deformed conifold there is only one compact A-cycle (the $S^3$) with N units of RR flux and one non-compact B-cycle with $\alpha$ units of NS-NS flux. (This is the same set-up as in the KS model where $B_2$ also threads through the non-compact cycle.) We therefore define the superfield S as the period of the A-cycle and its dual period $\hat F$:
\begin{equation}
S \,=\, \int_A \Omega\,,\qquad\qquad \hat F \,=\, \int_B \Omega\,.
\end{equation}
The holomorphic 3-form is given by
\begin{eqnarray}\nonumber
\Omega &=& \frac{dz_1\wedge dz_2\wedge dz_3\wedge dz_4}{df} \,=\, \frac{dz_1\wedge dz_2\wedge dz_3}{2z_4}\\
  &=& \frac{dz_1\wedge dz_2\wedge dz_3}{2\sqrt{\mu^2-z_1^2-z_2^2-z_3^2}} \,.
\end{eqnarray}
Viewing the 3-cycles A and B as 2-spheres (spanned by a real subspace of $z_2, z_3$) fibred over $z_1$, one can integrate $\Omega$ over $S^2$, resulting in a one-form
\begin{equation}
\int_{S^2} \Omega\,=\,\frac{1}{2\pi i}\,\sqrt{z_1^2-\mu^2}\,dz_1\,.
\end{equation}
The compact A-cycle, projected to the $z_1$ plane, becomes an interval $(-\mu,\mu)$, so that the A-period results in
\begin{equation}
  S \,=\, \frac{1}{2\pi\imath}\int_{-\sqrt{\mu}}^{\sqrt{\mu}}dz_1\,\sqrt{z_1^2-\mu^2}\,=\,\frac{\mu}{4}.
\end{equation}
The non-compact B-cycle is projected to $(\mu,\infty)$. We therefore introduce a cutoff $\Lambda_0$ such that
\begin{eqnarray}\nonumber
\hat F &=& \frac{1}{2\pi i}\,\int_\mu^{\Lambda_0^{3/2}} dz_1\,\sqrt{z_1^2-\mu^2}\\
   &=& \frac{1}{2\pi i}\,\left(\frac{\Lambda_0^3}{2} -S+ S\ln\frac{S}{\Lambda_0^3}\right) +
   \mathcal{O}\left(\frac{1}{\Lambda_0}\right).
\end{eqnarray}
With $N$ units of flux through the A cycle and $\alpha$ units of flux through the B-cycle the flux-generated superpotential \cite{vafa}
\begin{equation}
W_{eff} \,=\, -2\pi i\sum_i(N_i \hat F_i  + \alpha_i S_i)
\end{equation}
becomes
\begin{equation}\label{superpot}
W_{eff} \,=\, N\left(S\ln\Lambda_0^3 +S-\ln S\right) -2\pi i\alpha S\,.
\end{equation}
The first and last terms can be combined by replacing $\Lambda_0$ with the {\sl physical} scale of the theory. This is because $\alpha$ is related to the bare coupling of the four-dimensional $\mathcal{N}$=1 $SU(N)$ gauge theory (as discussed in Section \ref{ks}) via $2\pi i\alpha = 8\pi^2/g_0^2$, but the coupling of $\mathcal{N}$=1 SYM exhibits a logarithmic running
\begin{equation}
\frac{8\pi^2}{g^2(\Lambda_0)}\,=\,3N\ln\Lambda_0 + \,{\rm constant}\,.
\end{equation}
The term $3N\ln\Lambda_0-2\imath \pi\alpha$ is precisely what shows
up as the coefficient of $S$ in \eqref{superpot}. $\alpha$ can
therefore be absorbed into $\Lambda_0$ by introducing the physical
scale $\Lambda$ and the superpotential becomes
\begin{equation}
W_{eff} \,=\, NS\left(1-\ln\frac{S}{\Lambda^3}\right)\,,
\end{equation}
which is precisely the Veneziano--Yankielowicz superpotential \cite{vy} suggested for four-dimensional $\mathcal{N}=1$ $SU(N)$ Super--Yang--Mills, where $S$ plays the role of the glueball field. The vacuum of the theory exhibits all the known phenomena of gaugino condensation, chiral symmetry breaking and domain walls. This is a remarkable result and the first example where string theory produces the correct superpotential of a gauge theory.

\begin{center}
\begin{figure*}[htp]
\begin{center}
\mbox{\begin{minipage}{14.5 cm}
\begin{center}
\includegraphics[scale=1]{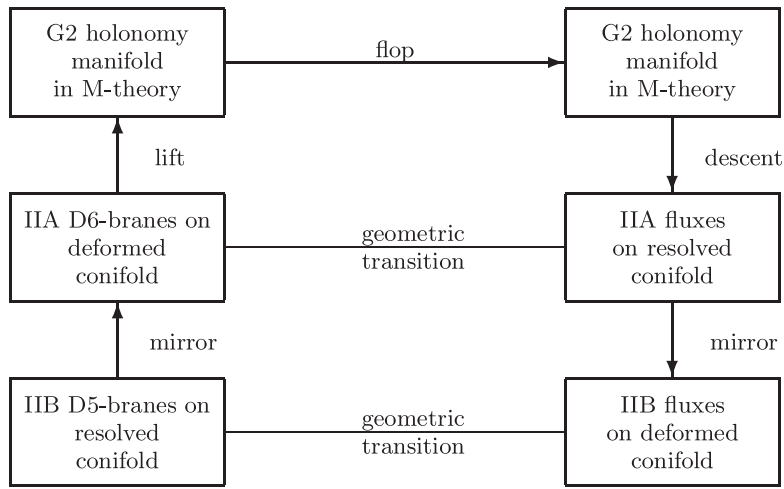}
\caption{Vafa's duality chain. By following the arrows through a series of mirror symmetries and a flop transition we can verify the geometric transition as conjectured for IIA and IIB.}
\label{vafachain}
\end{center}
\end{minipage}
 }
 \end{center}
\end{figure*}
\end{center}

The discussion of the IR limit deserves a word of caution. We argued earlier, that both the KS and the Vafa model reach an $SU(M)$ gauge theory.
This suggests that the open string background in Vafa's scenario is actually nothing but the IR limit of the KS cascade, which then in the far IR is dual to the closed string background. This is not quite accurate, as the UV limit of Vafa's scenario does \emph{not} coincide with the UV limit of KS. Instead, it approaches a (5+1)-dimensional theory. 
The UV limit of Vafa's scenario should rather correspond to a \textcite{mn} (MN) type of solution.
However, the MN UV--limit does not quite fit into interpolating
scenarios known as the KS baryonic branch \cite{ghk,papatseyt}, which is a one--parameter family of backgrounds obtained by deformations of the original KS solution \cite{barybranchone, barybranchtwo, barybranchthree}. These $SU(3)$ structure backgrounds are all expected to have a dual field theory description with vevs of baryonic operators turned on. The MN background, however, does not share this property \cite{ghk}. In terms of the interpolating solution of \textcite{barybranchone}, it can only be reached in an infinite limit of the parameter space and can therefore not be interpreted as on the same ``branch'' as the KS solution.
\footnote{We thank I. Klebanov for discussion on this point.}  
Although we have illustrated various similarities between the KS cascade and Vafa's geometric transition, the reader should keep this subtle distinction in the UV behaviour in mind.

\subsubsection{Vafa's duality chain}\label{chainvafa}

Let us summarise the superstring picture of the conifold transition: In type IIA we start with $N$ D6-branes on the $S^3$ of the deformed geometry and find as its dual $N$ units of 2-form flux through the $S^2$ of the resolved conifold. In the mirror type IIB, $N$ D5-branes wrapping the $S^2$ of the resolved conifold are dual (in the large $N$ limit) to a background without D-branes but with the 3-form flux turned on. The geometry after transition is given by the deformed conifold with blown-up $S^3$. In both cases we have to identify the complex structure modulus of the deformed conifold with the K\"ahler modulus of the resolved conifold or, roughly speaking, the size of the $S^3$ with the size of the $S^2$.

The type IIA scenario can be lifted to $M$-theory where the deformed and resolved conifolds are related by a flop transition. In seven dimensions both manifolds stem from a manifold with $G_2$-holonomy and symmetry $SU(2)\times SU(2)\times U(1)$. Topologically, the manifold in question is equivalent to a cone over $S^3\times \widetilde{S}^3$ that has a $U(1)$ fibre on which one can reduce to d=6. One can either reduce on a fibre that belongs to an $S^3$ of vanishing size (this yields a six-dimensional manifold with blown-up $\widetilde{S}^3$, the deformed conifold) or on a fibre that belongs to an $\widetilde{S}^3$ of finite size (this gives a finite size $\widetilde{S}^2$ in six dimensions, the resolved geometry).\footnote{Furthermore modding out by a $\mathbb{Z}_N$ in both cases gives a singularity corresponding to $N$ D6-branes or a non-singular solution with $N$ units of flux, respectively \cite{flop}.} In other words, both scenarios are related by an exchange of the finite-size $\widetilde{S}^3$ with the vanishing $S^3$, which is called a ``flop transition''.

A cone over $S^3\times \widetilde{S}^3$ is given by $\mathbb{R}^+\times S^3\times \widetilde{S}^3$ which is equivalent to $\mathbb{R}^4\times \widetilde{S}^3$. The topology of this manifold can be viewed as \cite{flop}
\begin{equation}
(u_1^2+u_2^2+u_3^2+u_4^2)-(v_1^2+v_2^2+v_3^2+v_4^2)\,=\,V\,,
\end{equation}
with $u_i,v_i\in \mathbb{R}$.
For $V>0$ the blown-up $\widetilde{S}^3$ is described by $u_i$, and the $v_i$ correspond to $\mathbb{R}^4$. For $V<0$ their roles are exchanged. The flop transition can then be viewed as a sign flip in $V$ or as an exchange of the two $S^3$. In the presence of $G$-flux in M-theory the volume $V$ can be complexified to $V+iG$, so that even for the transition point $V=0$ the singularity is avoided.

Using the arguments from this section, one can follow a ``duality chain'', as depicted in Figure \ref{vafachain}, which leads from D-branes in IIB through mirror symmetry to IIA, then via an M-theory flop to the closed string side in IIA and back to IIB via another mirror symmetry. Precisely this chain was followed by \cite{gtone}, \cite{realm}, \cite{gttwo}, \cite{thesis} and will be reviewed in Section \ref{chainour}.

\subsection{Brane constructions}
\label{bc}
%\subsubsection{Brane Constructions and the Gauge Group}
We saw in Section \ref{KW} that finding gauge theories on singular geometries like the conifold can be difficult. A more intuitive way to arrive at these gauge theories is to T-dualise the type IIB picture of branes on a conifold along a direction perpendicular to the branes. As we shall see, this gives type IIA configurations of branes suspended between orthogonal NS5-branes, from which the gauge theory living on the branes can be easily deduced as in the Hanany-Witten set-up \cite{hananywitten}. In addition, lifting these IIA configurations to M-theory allows one to reproduce the dualities conjectured via gauge or topological string theory arguments in the KS and Vafa set-ups, respectively. This was done by \textcite{0105066} for the geometric transition (see also \textcite{0106040, Hori:1997ab}), and in \textcite{0110050} for Klebanov-Strassler (following \textcite{9811139} and \textcite{9904131}). In both cases use was made of Witten's MQCD (an M-brane description of QCD) methods \cite{Witten:1997sc, Witten:1997ep} based on lifting to M-theory configurations of D4-branes suspended between NS5-branes.

In this section we discuss first the brane configuration dual to string theory on conifold geometries and then in turn the relevant brane configurations dual to both the Klebanov-Strassler and the geometric transition (as embedded in type IIB) set-ups. In the brane configuration picture, and even more so in the pictures lifted to M-theory which allow us to track the duality in each case, we see the similarities between the two arguments. The brane configuration picture is useful for
understanding several aspects of the theories in a different way,
and for highlighting deep similarities between the two scenarios.
However it has some limitations which should be kept in mind and
which we shall mention as they arise.

\subsubsection{The T-dual of a conifold}
\label{tdual}

As shown in \textcite{BSV} and \textcite{HanUranga}, a conifold can be described by two degenerating tori varying over a $\mathbb{P}^1$ base. When two T-dualities are performed, one over a cycle of each torus, the conifold gives rise to a pair of orthogonal NS5-branes. A configuration of orthogonal NS5-branes is also found upon performing a single T-duality along the $U(1)$ fibration of the conifold, as shown explicitly in \textcite{9811139}. To see this, consider the conifold equation in the form given in (\ref{eq-res}):
\begin{eqnarray}
xy - uv & = & 0.
\end{eqnarray}
We can define co-ordinates in such a way that $x=x^4 + \imath x^5$
and $y = x^8 + \imath x^9$. Here we map $\theta_1, \phi_1$ to $x^4,
x^5$ and $\theta_2, \phi_2$ to $x^8, x^9$. Following the literature,
we will also make the identifications $x^6 = \psi$ and $x^7 = r$ for
the rest of this section. We will perform the T-duality along
$x^6$. Our reason for adopting this change of coordinates is that in the
dual brane picture $x^4, x^5, x^8$ and $x^9$ are no longer compact
directions.

Upon T-dualising, the NS5-branes will arise on the degeneration loci of the fibration \cite{BSV, uranga}. The fibre degenerates at the conifold singularity, where $x = y = u = v=0$. A convenient choice of co-ordinates allows us to take the fibre as degenerating to two intersecting complex lines given by
\begin{eqnarray}
xy & = & 0.
\end{eqnarray}
In the T-dual picture this curve then defines the intersection of the two NS5-branes: one extends along $x$ ($y = u = v = 0$) and the other along $y$ ($x=u=v=0$). For the deformed conifold,
\begin{eqnarray}
xy - uv = \mu^2,
\end{eqnarray}
$xy$ can no longer be zero at the tip but is given by
\begin{eqnarray}
\label{conlift}
xy & = & \mu^2.
\end{eqnarray}
This gives the intersection curve of the NS5-branes in the
corresponding T-dual picture. In both cases $x^6$ and $x^7$
parametrise possible separations between the NS5-branes. If it is
present an NS-NS B-field on a vanishing 2-cycle in the conifold
picture maps to separation in $x^6$ of the NS5-branes
\cite{9803232}, while a finite $S^2$ implies a separation in $x^7$
given by the resolution parameter.

In \textcite{9811139, 9904131} and \textcite{uranga} this T-duality was exploited for
the construction of a type IIA brane configuration dual to the
Klebanov-Witten and Klebanov-Strassler set-ups, further studied in
\textcite{0106040, 0110050} where Witten's MQCD methods were used to
track the conjectured geometric transition via the M-theory
description. This was also applied to Vafa's geometric transition
(see also \textcite{0105066}). A discussion of the general approach is
given in \textcite{9803232}, where the references for several key
results are collected.

\subsubsection{The Klebanov-Strassler set-up via brane configurations}

We begin with the Klebanov-Witten set-up and then proceed to the Klebanov-Strassler scenario and discuss what becomes of the $M$ fractional D3-branes under the T-duality.

\begin{figure}
\centering
\includegraphics[scale=0.3]{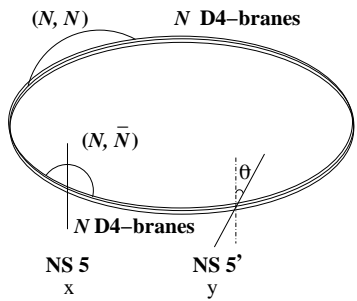}
\caption{The IIA brane configuration dual to the Klebanov-Witten set-up. }
\label{braneb} % caption for the whole figure
\end{figure}

In Section \ref{KW} we argued that while an ${\cal{N}} = 1$ supersymmetric theory with $U(1)$ gauge group reproduced the parametrisation of the conifold given by the fields $A_i, B_j$ and the equations (\ref{constraint}) and (\ref{U1}), the gauge theory whose moduli space corresponded to a single D3-brane on the conifold had gauge group $U(1) \times U(1)$ (which generalised to $U(N) \times U(N)$ for $N$ D3-branes). The appearance of two gauge group factors can be understood by T-dualising the conifold set-up in a direction perpendicular to the branes. Under a T-duality along $\psi$ or $x^6$ a conifold will give rise to a pair of NS5-branes \cite{9811139}.\footnote{The explicit calculation at the supergravity level presented in \textcite{9811139} takes the T-dual of the NS5-brane configuration described and does not exactly reproduce a conifold but something similar, where $x^4, x^5$ and $x^8, x^9$ remain 2-planes instead of the required fibred 2-spheres. This means that the $SU(2)\times SU(2)$ global symmetry is not directly visible in the brane construction. Other arguments, coming from the gauge theory and the fact that this symmetry is regained in the M-theory lift, nevertheless support the geometrical interpretation given here \cite{9811139}.} The presence of D3-branes as in the Klebanov-Witten set-up yields a configuration of D4-branes stretching between these NS5-branes, along the T-dual direction $x^6$, as shown in Figure \ref{braneb}.

We label one NS-brane NS5 and the other NS5'. In the figure, the 0123 and radial directions are suppressed. The branes are orthogonal: NS5 extends in $x$ and NS5' extends in $y$. The NS5-branes are much heavier than the D4-branes because they are infinite in all of their worldvolume directions and so they can be treated classically \cite{hananywitten, Witten:1997sc}. We study the gauge theory on the D4-branes where the NS5-branes are considered as fixed and the positions of the D4-branes parametrise the moduli space of the theory. The D4-branes have only four infinite directions, so the field theory living on them is effectively 3+1-dimensional.  It is the same as that living on the D3-branes at the tip of the conifold in the IIB picture.

As should be clear from Fig. \ref{braneb}, the gauge theory is
easily deduced. The configuration preserves 4 supercharges, and
therefore describes an ${\cal N} = 1 $ supersymmetric gauge theory
on the D4-branes. Gauge fields transforming as $(N,N)$ or $(\bar N,
\bar N)$ in the adjoint representation correspond to open strings
ending on the D4-branes between the NS5-branes, while matter fields
transforming in the bifundamental representations $(N, \bar N)$ or
$(\bar N, N)$ correspond to strings stretching between D4-branes on
opposite sides of an NS5-brane. The separation of the NS5-branes in
$x^6$-direction is given by the NS-NS two-form, and is not specified
by the geometry.\footnote{As expected, the separation of the
NS5-branes is related to the coupling(s) of the dual field theory.}
Thus generically  the $N$ D4-branes are broken into two segments by
the NS5-branes, and the gauge group is $U(N) \times U(N)$, as
claimed in Section \ref{KW}.

For  ${\cal N} = 1$ supersymmetry the NS5-branes must be perpendicular. If NS5' is rotated in the $(x, y)$-plane so that they are parallel, the gauge theory on the D4-branes has ${\cal N} = 2$ supersymmetry. In this case, the D4-branes can move in $x$ (the scalars corresponding to their fluctuations in these directions fit into the adjoint representation of ${\cal{N}}=2$). This is no longer true once there is a relative rotation of the NS5-branes, since any separation of the D4-branes in $x$ would lead to twisting of the D4-branes that would break supersymmetry completely. When the NS5-branes are perpendicular, there are no moduli for motion of the D4-branes in the directions of the NS5-branes. Furthermore, the angle between the NS5-branes serves as a SUSY-breaking parameter. In fact, this angle ($\theta$) is related to the mass of the adjoint chiral multiplets by \( \mu = \tan \theta\)  \cite{barbon, Witten:1997sc, Hori:1997ab}. Thus the adjoints can be integrated out of the superpotential when the NS5-branes are perpendicular, and agreement with Klebanov-Witten is also found at the superpotential level \cite{9811139}.  On the other hand the D4-branes on either side of each NS5-brane cannot be split from each other, so the bifundamental fields remain massless.  This T-duality therefore allows us to rederive the gauge theory dual of the Klebanov-Witten set-up.

\begin{figure}[htp]
\centering
\includegraphics[scale=0.3]{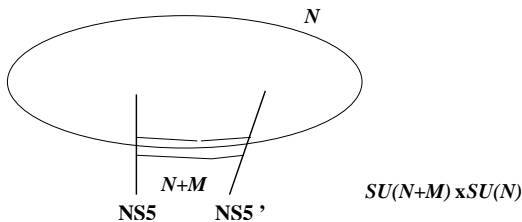}
\caption{The IIA brane configuration dual to the Klebanov-Strassler set-up.}
\label{KSdual}
\end{figure}

To study the brane configuration dual to the Klebanov-Strassler
model we need to know how a fractional D3-brane will transform under
a T-duality along a direction perpendicular to it. In a
five-dimensional spacetime, D3-branes and fractional D3-branes are
domain walls \cite{9807164, 9808075}. There is a jump in the 5-form
flux when one crosses them, and this is related in AdS
compactifications to the number of branes on which the gauge theory
lives. Thus crossing a domain wall corresponds to increasing or
decreasing the rank of the gauge group in the dual gauge theory. It
is easy to see in the brane configuration set-up dual to the
Klebanov-Witten scenario that addition of a single D3 brane will
increase the rank of the gauge group from $SU(N) \times SU(N)$ to
$SU(N+1) \times SU(N+1)$. However, we will soon see that the $M$
wrapped branes of the KS model map to D4-branes extended only along
part of $x^6$ between NS5 and NS5' \cite{9803232, 9904131}. Thus
fractional D3-branes function as \em exotic \em domain walls, the
crossing of which will take the gauge group of the dual field theory
from $SU(N) \times SU(N)$ to $SU(N+1) \times SU(N)$. With the
presence of $M$ wrapped D5-branes, the brane configuration shown in
Figure \ref{KSdual} results.

The action of the T-duality on the $M$ wrapped D5-branes of the KS
model can be deduced by noting that the $S^2$ they wrap is the
difference in the sense of homology between the two $S^2$s which
form the base of the $U(1)$ fibration giving $T^{1,1}$, i.e. $S^2 =
S^2_1 - S^2_2$, where the first $S^2$ is parametrised by $\theta_1$
and $\phi_1$ and the second is parametrised by $\theta_2\,,\phi_2$.
The cohomology $H^2$ is given by two-forms which are closed but not
exact. Candidate representatives of $H^2$ are \( \sin \theta_1 d
\theta_1 \wedge d\phi_1 \pm \sin \theta_2 d \theta_2 \wedge d \phi_2 \), both of
which live only on the two $S^2$ factors in $T^{1,1}$ and are
independent of the $U(1)$ fibre. One might think that they are exact,
since they can be written $ d(\cos \theta_1 d \phi_1 \pm \cos
\theta_2 d \phi_2)$, but these expressions are not globally defined
because $\phi_i$ is not well defined at the poles. As argued in
\textcite{9904131}, the term with the plus sign is actually exact, since
it can be written as \( d(d\psi + \cos \theta_1 d \phi_1 + \cos
\theta_2 d\phi_2) \sim d e^\psi\), where $e^\psi$ is one of the five
vielbeins and is globally defined, because $\psi$ exhibits a shift symmetry. Therefore it is the minus term
which is a representative of the second cohomology:
\begin{eqnarray*}
\sin \theta_1 d \theta_1\wedge d\phi_1 - \sin \theta_2 d \theta_2 \wedge d\phi_2.
\end{eqnarray*}

This means that the dual object to the domain wall in the type IIA
picture is something that carries a charge away from the $(x^4,
x^5)$ plane and deposits it on the $(x^8, x^9)$
plane.\footnote{Recall that $\theta_1, \phi_1$ map to $x^4, x^5$,
and $\theta_2, \phi_2$ map to $x^8, x^9$. The $U(1)$ fibre is $\psi$
which maps to $x^6$ and is the direction in which we perform the
T-duality.} A D4-brane with one end on NS5 and the other on NS5'
performs this function. Thus an exotic domain wall maps under this
T-duality into a D4-brane stretched only part of the way along
$x^6$.

Next we can use the brane picture to study the gauge theory on the
D4-branes. The coupling constant of the gauge theory is determined
by the distance between the branes in $x^6$ \cite{9811139,
9904131,9803232}, which is given by the B-field and therefore matches arguments from Sections \ref{ks} and
\ref{vafaembed}. For our product gauge group set-up,
 \begin{eqnarray}
\frac{1}{g_1^2}& = & \frac{l_6 - a}{g_s},\\
\frac{1}{g_2^2}&=&\frac{a}{g_s},
\end{eqnarray}
where $g_1$ and $g_2$ are the couplings of the two theories, $g_s$
is the string coupling, and $l_6$ gives the circumference of the
$x^6$ circle so that $a$ is the separation between the NS5-branes.
We see that it is exactly the presence of the $M$ fractional branes
that breaks conformal invariance: In the IIA description, a non-zero
$\beta$-function arises in the gauge theory when the NS5-branes are
bent. This is because the endpoints of the D4-branes on the
NS5-branes introduce a dependence of $a$ on $\theta_i, \phi_i$. Thus
the positions of the NS5-branes should really be measured far from
the D4-branes. As shown in \textcite{Witten:1997sc}, $a$ will only have
a well-defined limiting value when the NS5-brane has an equal number
of D4-branes on either side. For the KS brane configuration, this is
not the case - in other words the branes are bent in $r$ (the only
available direction, suppressed in the figures). This introduces a
dependence of $a$ on $r$, from which one can rederive the running of
the gauge theory coupling constant(s).

\begin{figure}[htp]
\centering
\includegraphics[scale=0.4]{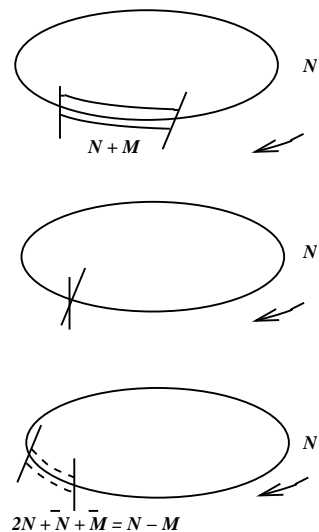}
\caption{A Seiberg duality transformation in the dual brane picture. Note that here $N = 1$ and $M = 2$ so in the final picture two of the branes are actually antibranes (denoted by dashed lines).}
\label{casc}
\end{figure}

In addition, the duality cascade is also observed in the
brane configuration picture. An early reference is \textcite{EGK}; see
also \textcite{uranga}. The bending of the NS5-branes means that, at some
point far from the suspended D4-branes, $a$ will vanish, implying a
divergence of one coupling constant. To avoid it one must move the
NS5-branes, pulling one through the other around the circle. As NS5'
approaches NS5, the $N$ D4-branes occupy the entire $x^6$, and the $N
+M$ branes shrink to zero size. When NS5' is pulled through NS5,
the $N$ branes double up, while the branes that were originally
between NS5 and NS5' regrow with the opposite orientation as
antibranes. These can partially annihilate the 2N branes, leaving $N
- M$ branes on the expanding segment, as shown in Figure \ref{casc}.
The system now has a dual description in terms of a gauge theory
with gauge group $SU(N-M) \times SU(N)$, so that moving the branes
through each other corresponds to performing a Seiberg duality.

\vspace{5mm}

The branes continue to be bent, so one is led to repeat the motion around the circle. As in the duality cascade described in Section \ref{chisb}, the process continues until only the NS5-branes with $M$ D4-branes stretching between them are left and the gauge group is just $SU(M)$.

To see how this configuration maps to the deformed conifold, we have
to lift it to M-theory. This allows us to study the non-perturbative
dynamics of the theory. The first such analysis was performed by
Witten, for the case of ${\cal N}$=2 theories
\cite{Witten:1997sc}, but it was generalised to ${\cal N}$=1 by
\textcite{Witten:1997ep, 9706127} and \textcite{Hori:1997ab} and it is these results
which apply most directly to our case (in particular elliptic ${\cal
N} = 1$ models). Both D4-branes and NS5-branes map to M5-branes in
M-theory, with the D4-branes corresponding to M5-branes 
wrapped on the eleventh dimension $x^{10}$. Define
\begin{eqnarray}
t & = & e^{(-\frac{x^6}{R} + \imath x^{10})},
\end{eqnarray}
where $R$ is the radius of the eleventh dimension.\footnote{For our case we should keep in mind that $t$ is not periodic in $x^6$ so we should only use it for a finite range of values.} In the case that no fractional branes are present, the D4- and NS5-branes lift to a configuration with three separate components. The fully wrapped $N$ D4-branes become M5-branes wrapping $x^6$ and $x^{10}$ or $t$. These $N$ toroidal branes are described by the equations
\begin{eqnarray}
x^N &= & 0,\\ \nonumber
y^N & = & 0,
\end{eqnarray}
which have to be supplemented with the equations describing the
lifted NS5-branes: $y = 0, t= 1$ (NS5) and $x= 0, t =
e^{-\frac{a}{R}}$ (NS5').  For the brane dual picture at the bottom
of the cascade when the D3-branes have cascaded away and the gauge
group is just $SU(M)$, the $M$ suspended (fractional) branes join with
the NS5-branes to become a single object in the M-theory description
\cite{9904131}. This introduces dynamical effects into the model.
This set-up for pure $\mathcal{N}=1$ SYM was studied in
\textcite{Witten:1997ep}, but in the limit where $x^6 \rightarrow
\infty$. %We can nevertheless adopt the results.

The NS5-branes by themselves would lift to $xy = 0$, which describes
the conifold. The Klebanov-Strassler configuration of NS5-branes and
fractional branes lifts to an M5-brane described by \textcite{0110050} as
\begin{eqnarray}
\label{eq-m5}
xy & = & \zeta,\\ \nonumber
t & = & x^M,
\end{eqnarray}
where $\zeta \in \mathbb{C}$ is some complex parameter. If we now
try to continue the cascade by shrinking the distance $x^6$ between
the NS5-branes, we find $t = e^{\imath x^{10}}$, i.e. $t$ now
parametrises a circle. The embedding of the M5-brane into ($x, y,
t$) must be holomorphic in order for susy to be preserved but as
there is no non-constant holomorphic map into $S^1$, $t$ must be
constant. Then (\ref{eq-m5}) become
\begin{eqnarray}
xy & = & \zeta, \\ \nonumber
t & = & t_0.
\end{eqnarray}
This object is exactly the lifted deformed conifold dual described
by (\ref{conlift}). The appropriate RR flux arises from the usual
one-form one obtains by dimensional reduction of the M-theory lift
to IIA on a twisted circle. It is then T-dualised to $F_3$ in IIB.

\subsubsection{The Vafa set-up via brane configurations}

\begin{figure}[htp]
\centering
\includegraphics[scale=0.3]{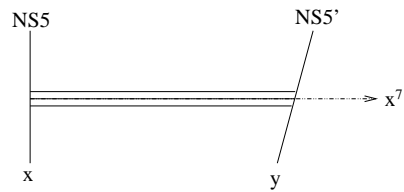}
\caption{The IIA brane configuration dual to the Vafa set-up.}
\label{vafabrane}
\end{figure}

The duality put forward by Vafa was studied from the brane
configuration point of view by \textcite{0105066, 0110050, 0106040}. We
begin with the IIB embedding set-up, in which $N$ D5-branes wrap the
finite $S^2$ of a resolved conifold. Under a T-duality in $x^6$,
this maps to a IIA configuration of  $N$ D4-branes stretching between
two orthogonal NS5-branes, similar to the brane configuration dual
of the KS set-up.\footnote{This should not be confused with the
mirror IIA embedding of Vafa's, in which D6-branes wrap the $S^3$ of
a deformed conifold. In the rest of this section, any reference to
T-duals or T dualities is to those along $\psi$ or $x^6$ which take
us between the IIB embedding set-up described above and the IIA
brane configuration dual shown in Figure \ref{vafabrane}.} There are
two differences between the KS and Vafa brane configuration duals.
Firstly, the NS5-branes here are also separated in $x^7$, with the
separation in $x^7$ given by the size of $S^2$ at the tip of the
conifold, the resolution parameter. More precisely, the two
NS5-branes will be separated along $z = x^6 + \imath x^7$ since
$B_2$ controls the separation in $x^6$ \cite{0105066}. In the IR
limit which interests us, the separation in $x^6$ will be
negligible. In addition, the direction along which the D4-branes
stretch between NS5 and NS5' is non-compact, as shown in Figure
\ref{vafabrane}.

Vafa's geometric transition duality as expressed in the type IIB
embedding can be seen directly in the IIA brane description
\cite{0105066}. The easiest way to track it is ``backwards",
beginning with the final IIB picture of a deformed conifold with
fluxes and no branes and asking what its T-dual (along $x^6$) is.
Then we will shrink the $S^3$ to zero size and blow up $S^2$, both
in the IIA brane configuration, and finally T-dualise back to IIB to
find the resolved conifold with wrapped D5-branes. First note that
the orthogonal NS5-branes described as T-dual to type IIB string
theory on a singular conifold in Section \ref{tdual} are coincident
in $x^6$ and $x^7$. The T-dual of a deformed conifold consists of
two NS5-branes intersecting on a curve $xy = \mu^2$. However, when
there is an RR flux $F_3$ through the $S^3$ in the IIB picture, the
T-dual will be modified. Since $F_3$ has one component in the
direction of the T-duality ($\psi$ or $x^6$), the IIA picture will
have a 2-form flux $F_2$ in the $x, y$ directions.\footnote{This can
be seen from the fact that $F_3 = N \omega_3$, where $\omega_3$ is
given by (\ref{omega3}).} The geometric transition must involve
shrinking the $S^3$ to zero size. Then $x$ and $y$ describe
vanishing spheres. This implies an infinite flux per unit area,
which singularity is resolved by the creation of a D4-brane. $F_2 = dA$
couples to the worldvolume of the NS5-brane through
\begin{eqnarray}
\int A \wedge \star d \phi & = & \int F_2 \wedge C_4,
\end{eqnarray}
where $\phi$ is one of the periodic scalars on the worldvolume of
the NS5. The correct susy-preserving source for $C_4$ in this case
is a D4-brane, which intersects NS5 and NS5' in the requisite 4
dimensions. Since we should complete the geometric transition by
blowing up the two-sphere, it must stretch between the NS5-branes.
Here we see the fundamental connection between the KS and geometric transition
pictures most clearly. The two directions available for the D4-brane
to stretch along are $x^6$ and $x^7$. Growing it in the $x^6$
direction only results in the pre-transition Klebanov-Strassler IIA
brane set-up, while growing it in the $z=x^6+\imath x^7$ direction
gives exactly the brane configuration dual of the resolved conifold
with D5s wrapping the $S^2$. We are able to begin with the deformed
conifold picture in type IIB, T-dualise it to type IIA and find
after the relevant transitions either the T dual of the KS picture
or of the Vafa picture, depending on whether the direction in which
the dual NS5-branes are separated is compact or not.

The geometric transition can also be followed in M-theory
\cite{0105066}, where the argument now runs most easily in the opposite
direction to that of the previous paragraph. The pre-transition
configuration of D4s stretching between orthogonal NS5-branes lifts
to a single M5-brane with a complicated worldvolume structure given
by $R^{1,3} \times \Sigma$ where $\Sigma$ is a complex curve defined
by
\begin{eqnarray}
xy & = & \zeta,\\ \nonumber
t & = & x^N.
\end{eqnarray}
This time we shrink $S^2$ to effect the transition, and find again that $t$ must parametrise a circle and is therefore constant. This implies $x^N y^N  =  \zeta^N$, i.e. $\Sigma \rightarrow \Sigma_k$, where $t = t_0$ and \[ xy  =   \zeta e^{\frac{2 \pi \imath k}{N}}. \] We obtain $N$ degenerate curves which are no longer the M-theory lift of D4-branes, but correspond to a closed string background, the deformed conifold. The main difference between the pre-transition set-ups in M-theory is that the KS M5-brane is wrapped on a torus parametrised by $x^6 + \imath x^{10}$ while the M5-brane of the Vafa set-up is wrapped on a cylinder $x^7 + \imath x^{10}$, at least in the IR.

\setcounter{equation}{0}
\section{Supergravity treatment and non-K\"ahler duality chain}\label{chainour}

\begin{center}
\begin{figure*}[htp]
\begin{center}
\mbox{\begin{minipage}{14.5 cm}
\begin{center}
\includegraphics[scale=1]{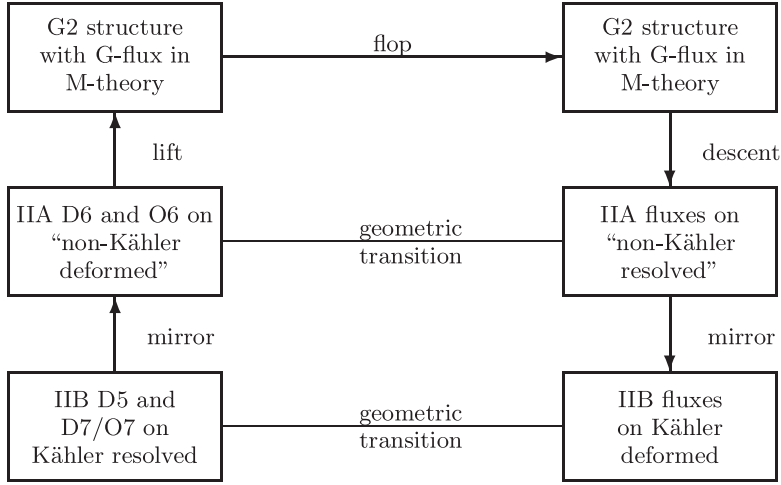}
  \caption{The modified duality. The backgrounds in IIA have to be replaced by non-K\"ahler versions of the deformed and resolved conifolds and the $\mathcal{M}$-theory lift does not possess $G_2$ holonomy anymore.} \label{ourchain}
\end{center}
\end{minipage}
 }
 \end{center}
\end{figure*}
\end{center}

In the last section we reviewed three arguments for geometric transition: one based on gauge theory, one on topological strings and the third on brane constructions and MQCD methods. The ``duality chain'' derived by \textcite{vafa, flop}, see Figure \ref{vafachain}, seems to suggest a straightforward way to verify geometric transitions on the supergravity level. Exploiting the well-known fact that mirror symmetry on Calabi--Yaus can be realized (in a certain limit) by three T-dualities, see Strominger--Yau--Zaslow (SYZ) \cite{syz}, and that a T-duality for a given metric and set of background fields is performed by applying Buscher's rules \cite{buscher, buschertwo}, one should be able to explicitly formulate the supergravity solution corresponding to all the links in the above chain. Proving more subtle than naively expected, this analysis was nevertheless carried out \cite{gtone,realm,gttwo,dipole,thesis} and we will review it in this section.

The issues that make the supergravity treatment non-trivial are the following:
\begin{itemize}
\item Resolved and deformed conifolds are only approximately mirror to each other.
Whereas the resolved conifold does indeed admit a $T^3$ fibre for T-dualising, the deformed conifold
possesses less symmetry. It is therefore only possible to recover a deformed conifold mirror in the
large complex structure limit. This was discussed in great detail by \textcite{thesis} in chapter 2.
\item Taking the back-reaction of RR and NS-NS fluxes (or D-branes) into account changes the IIB supergravity
solution. Instead of D5-branes wrapping a resolved conifold, the manifold is only {\sl conformally} a Calabi--Yau.
It acquires a warp factor $h(r)$, depending on the radial direction. Strictly speaking this spoils the SYZ argument,
but this problem can be overcome by working in the ``local limit'', i.e. restricting the metric to a patch over
which $h(r)$ does not vary too much. This approach is inherent to all references \cite{gtone,realm,gttwo,dipole,thesis}
and we will comment on its shortcomings.
\item We have seen that a supergravity solution \`a la Klebanov--Strassler (with fractional branes) necessarily
contains non-trivial NS-NS flux. Assuming the SYZ argument still
holds and performing T-dualities along the isometry directions of
the resolved conifold, the metric and the B-field get mixed. This
phenomenon can be used to argue that the mirror of a CY in the
presence of NS-NS flux is no longer CY and it was postulated that
these manifolds are half-flat \cite{jan}.\footnote{In contrast to a
CY, which is characterised by a closed fundamental 2-form $J$ and a
closed holomorphic three-form $\Omega=\Omega_++i\Omega_-$, half-flat
manifolds only obey $d(J\wedge J)=0$ and $d\Omega_-=0$, and are
therefore a special class of non-K\"ahler manifolds ($dJ\ne0$).} It
has now been established \cite{minasone, minastwo, minasthree} that the most general
$\mathcal{N}=1$ susy-preserving backgrounds are generalised CYs in
the sense of \textcite{hitch, gualtieri}; they contain half-flat
manifolds as a subclass. Even if we don't start with a simple torus,
we encounter the same phenomenon of twisted fibres and the IIA
solutions in Figure \ref{vafachain} have to be replaced by
non-K\"ahler backgrounds.
\item All that said, there remains another problem: no susy-preserving background for D5-branes on the resolved conifold is known (without other ingredients). The one derived by \textcite{pt} was later on shown by \textcite{cvetgiblupo} to break susy completely; we will review the argument in Appendix \ref{fluxes}. This problem was avoided by \textcite{gttwo, thesis} by constructing a IIB solution from F-theory, which in addition to the D5-branes contains D7 and O7-planes. Note that this subtlety is not visible in the local limit and does not alter the calculations presented in \textcite{gtone, realm} much. Nevertheless, the whole analysis with all fluxes that are consistent with this orientifold construction was repeated by \textcite{thesis} in full detail.
\end{itemize}

To treat all these subtleties simultaneously, we will first review mirror symmetry \`a la SYZ and explain the local limit. After that we will describe the orientifold set-up used by \textcite{gttwo, thesis} and then walk the reader through the whole duality chain, as depicted in Figure \ref{ourchain}.

\subsection{Mirror symmetry between the resolved and deformed conifolds}\label{mirror}

The resolved and deformed conifolds describe asymptotically a cone over $S^2\times S^3$, but the singularity at $r=0$ is smoothed out to an $S^2$ or $S^3$, respectively. The Ricci-flat K\"ahler metric of the resolved conifold was derived by \textcite{candelas, pt}:
\begin{eqnarray}\label{resmetric}\nonumber
ds^2_{\rm res} & = & \widetilde{\gamma}'\,d\widetilde{r}^2\\ \nonumber
& &+\,   \frac{\widetilde{\gamma}'}{4}\,\widetilde{r}^2\big(d\widetilde{\psi}
  +\cos\widetilde{\theta}_1\,d\widetilde{\phi}_1
  +\cos\widetilde{\theta}_2\,d\widetilde{\phi}_2\big)^2 \\
& &+\,   \frac{\widetilde{\gamma}}{4}\,\big(d\widetilde{\theta}_1^2
  +\sin^2\widetilde{\theta}_1\,d\widetilde{\phi}_1^2\big)\\ \nonumber
& &+ \, \frac{\widetilde{\gamma}+4a^2}{4}\,\big(d\widetilde{\theta}_2^2
  + \sin^2\widetilde{\theta}_2\,d\widetilde{\phi}_2^2\big)\,,
\end{eqnarray}
where $(\widetilde{\phi}_i,\widetilde{\theta}_i)$ are the usual
Euler angles on $S^2$, $\widetilde{\psi}=0\ldots 4\pi$ is a $U(1)$
fibre over these two spheres and $\widetilde\gamma$ is a
function\footnote{The function $\widetilde{\gamma}$ is related to
the K\"ahler potential $\widetilde{\mathcal{F}}$ as
$\widetilde{\gamma}=\widetilde{r}^2
(\partial\widetilde{\mathcal{F}}/\partial r^2)$, and similarly for
$\hat{\gamma}$ below, see \eqref{defgamma}.} of $\widetilde{r}$ that
goes to zero as $\widetilde{r}\to 0$. The constant $a$ is called the
resolution parameter, because it produces a finite size prefactor
for the $(\widetilde{\phi}_2,\widetilde{\theta}_2)$-sphere at
$\widetilde{r}=0$. This metric clearly has three isometries related
to shift symmetries in the co-ordinates
$\widetilde{\psi},\,\widetilde{\phi}_1$ and $\widetilde{\phi}_2$.
These are indeed the appropriate Killing directions as the metric
was constructed to be invariant under $SU(2)\times SU(2)\times U(1)$
\cite{candelas}, see Appendix \ref{appsingconi}, \ref{appresconi}
for a brief review.

The deformed conifold metric on the other hand is given by \cite{Ohta:1999we, papatseyt, minasian}
%\begin{widetext}
\begin{eqnarray}\label{defmetric}
ds^2_{\rm def} &=& \hat{\Gamma}\, \left[\frac{4\,d\widetilde{r}^2}{\widetilde{r}^2
  (1-\mu^4/\widetilde{r}^4)}\right.\\ \nonumber
& & \left.\qquad\qquad+ \left(d\widetilde{\psi}+\cos\widetilde{\theta}_1\,
  d\widetilde{\phi}_1+\cos\widetilde{\theta}_2\,d\widetilde{\phi}_2\right)^2\right] \\ \nonumber
 && + \, \frac{\hat{\gamma}}{4}\;\left[\left(\sin\widetilde{\theta}_1^2\,
  d\widetilde{\phi}_1^2+d\widetilde{\theta}_1^2\right)
  +\left(\sin\widetilde{\theta}_2^2\, d\widetilde{\phi}_2^2
  + d\widetilde{\theta}_2^2\right)\right]\\ \nonumber
&& + \, \frac{\hat{\gamma}\mu^2}{2\, \widetilde{r}^2}\,\left[\cos\widetilde{\psi}\,
  \left(d\widetilde{\theta}_1\, d\widetilde{\theta}_2
  -\sin\widetilde{\theta}_1\sin\widetilde{\theta}_2\, d\widetilde{\phi}_1 \,
  d\widetilde{\phi}_2\right) \right.\\ \nonumber
& & \left. \qquad+\sin\widetilde{\psi}\,\left(\sin\widetilde{\theta}_1 \,
d\widetilde{\phi}_1\, d\widetilde{\theta}_2 + \sin\widetilde{\theta}_2\,
d\widetilde{\phi}_2\,d\widetilde{\theta}_1\right)\right]\,,
\end{eqnarray}
%\end{widetext}
with the deformation parameter $\mu$ and a similar function $\hat{\gamma}(\widetilde{r})$. $\hat{\Gamma}$ can be read off from equation \eqref{defmetricapp}.  This metric exhibits the same structure of a $\widetilde{\psi}$-fibration over two spheres, but there are additional cross terms in the last two lines. We see that the $U(1)$ symmetry associated with shifts $\widetilde{\psi}\to\widetilde{\psi}+k$ is absent. This is not a peculiarity of our co-ordinate choice but an inherent property of the deformed conifold. As discussed in Section \ref{ks}, the deformed conifold breaks the $U(1)$ symmetry of the singular conifold (which the resolved conifold also exhibits).

Clearly, the manifolds cannot be mirrors of each other in the naive
SYZ sense: one possesses a $T^3$ fibration and the other one
doesn't.
From the point of view of cohomology, mirror symmetry
implies an exchange of odd and even cohomologies; more precisely
$h^{1,2}\leftrightarrow h^{1,1}$. This is expressed in the exchange
of the blown up two-cycle of the resolved conifold with the blown-up
three-cycle of the deformed conifold.
But \textcite{hori, aganagic} made an attempt to find the mirror
of the resolved conifold, and the resulting manifold was found to
differ from the deformed conifold in that some co-ordinates are
elements of $\mathbb{C}^*$ instead of $\mathbb{C}$. The mirror
manifold can be described by $x_1+x_2+x_1 x_2 e^t+1-uv=0$, where
$x_i\in \mathbb{C}^*$ and $u,v\in \mathbb{C}$ and $t$ is the size of
the blown-up $S^2$ in the original resolved geometry. The mirror
symmetry between the two manifolds only becomes exact in the limit
where the size of the $S^2$ and $S^3$ shrink to zero.

The way to realise mirror symmetry via T-duality even in the absence of isometry directions is to go to the large complex structure limit \cite{syz} that takes us away from the singular fibres. We can still apply SYZ if the base is large compared to the $T^3$ fibre. If we identify $(\widetilde{r},\widetilde{\theta}_1,\widetilde{\theta}_2)$ as the base co-ordinates and $(\widetilde{\psi},\widetilde{\phi}_1,\widetilde{\phi}_2)$ as the co-ordinates of the $T^3$ fibre in the resolved metric \eqref{resmetric}, we can T-dualise along the latter.

It was furthermore shown by \textcite{gtone} that the large complex structure limit has to be imposed ``by hand'', i.e. the co-ordinates $(\widetilde{\theta}_1,\widetilde{\theta}_2)$ receive a large boost. This would be a non-trivial manipulation of the metric that is not guaranteed to preserve the Calabi--Yau property, which is why the following analysis will be presented in local co-ordinates in which this boost is actually nothing but a co-ordinate redefinition. See Chapter 2 of \textcite{thesis} for details; we will just review the results.

We restrict our analysis to a small neighbourhood of $(r_0,\,\z,\,\langle\phi_1\rangle, \,\langle\phi_2\rangle, \,\langle\theta_1\rangle,\,\langle\theta_2\rangle)$ by introducing
\begin{eqnarray}\label{local}\nonumber
\widetilde{r} &=& r_0+\frac{r}{\sqrt{\widetilde{\gamma}_0'}},\qquad\qquad\quad\, \,
  \widetilde{\psi} \,=\, \z + \frac{2 z}{\sqrt{\widetilde{\gamma}_0'}\,r_0}, \\ \nonumber
\widetilde{\phi}_1 &=& \langle\phi_1\rangle + \frac{2x}{\sqrt{\widetilde{\gamma}_0}\,\sin\langle\theta_1\rangle},\quad
  \widetilde{\theta}_1 \,=\, \langle\theta_1\rangle+\frac{2\theta_1}
  {\sqrt{\widetilde{\gamma}_0 }},\\ \nonumber
\widetilde{\phi}_2 &=& \langle\phi_2\rangle + \frac{2y}{\sqrt{(\widetilde
  {\gamma}_0 + 4 a^2)}\,\sin\langle\theta_2\rangle}\,,\\ 
\widetilde{\theta}_2 &=& \langle\theta_2\rangle+\frac{2\theta_2}{\sqrt{
  (\widetilde{\gamma}_0+4 a^2)}},
\end{eqnarray}
where $\widetilde{\gamma}_0$ is constant, namely $\widetilde{\gamma}(\widetilde{r})$ evaluated at $\widetilde{r}=r_0$. The co-ordinates $(r,\,z,\,x,\,y,\,\theta_1,\,\theta_2)$ describe small deviations from these expectation values and we will call them ``local co-ordinates'' henceforth. In these local co-ordinates the metric on the resolved conifold takes a very simple form (in lowest order in local co-ordinates):
\begin{eqnarray}\label{startmetric} \nonumber
ds^2 &=& dr^2+(dz+A\,dx + B\, dy)^2\\
& &+ (dx^2+d\theta_1^2)+(dy^2+d\theta_2^2)\,,
\end{eqnarray}
where we have defined
\begin{eqnarray}\label{defineab}\nonumber
A &=& \sqrt{\frac{\widetilde{\gamma}_0'}{\widetilde{\gamma}_0}}\;r_0\,\cot
\langle\theta_1\rangle \,,\\
B &=& \sqrt{\frac{\widetilde{\gamma}_0'}{(\widetilde{\gamma}_0+4 a^2)}}\;r_0\,
\cot\langle\theta_2\rangle\,.
\end{eqnarray}
Note that at lowest order in local co-ordinates these are simply
constants. In \textcite{dipole} it was shown that at linear order in $r$
the $\theta_i$-dependences can be resummed to give precisely
$\cot\theta_i$ instead of $\cot\langle\theta_i\rangle$, but for
illustrative purposes we will stick to the simplest case of lowest
order in local co-ordinates in this article. For later convenience
we define
\begin{equation}\label{alpha}
\alpha=(1+A^2+B^2)^{-1}.
\end{equation}
The metric \eqref{startmetric} is easily T-dualised along $x,y$ and $z$ (which correspond to the former isometry directions $\widetilde{\psi},\,\widetilde{\phi}_1,\,\widetilde{\phi}_2$) with the help of Buscher's rules \cite{buscher, buschertwo}. To illustrate the large complex structure limit, consider again \eqref{startmetric}, which can be written as
\begin{equation}\label{torimetric}
ds^2 \,=\,dr^2+(dz+A\,dx + B\, dy)^2+ \vert d\chi_1\vert^2+\vert d\chi_2\vert^2\,,
\end{equation}
with the two tori
\begin{equation}\label{deftori}
d\chi_1\,=\,dx+\tau_1\,d\theta_1\,,\qquad d\chi_2\,=\,dy+\tau_2\,d\theta_2\,.
\end{equation}
In \eqref{startmetric} the complex structures are simply $\tau_1=\tau_2=i$. Note that these tori are just local versions of the spheres in \eqref{resmetric}, since locally a sphere resembles a degenerate torus.\footnote{The appearance of tori instead of spheres is also consistent with the dual brane pictures as described in Section \ref{bc}.} The large complex structure limit is then given by letting
\begin{equation}\label{boost}
\tau_1\,\longrightarrow\, i-f_1\,,\qquad\qquad \tau_2\,\longrightarrow\, i-f_2,
\end{equation}
with real and large $f_{1,2}$. With the choice $f_1^2=f_2^2=\alpha/\epsilon$ one recovers the mirror metric in type IIA (taking $\epsilon\to 0$ and rotating the $(y,\,\theta_2)$ torus -- see Section 2.2 of \textcite{thesis})
\begin{eqnarray}\label{defmirror}\nonumber
d\tilde{s}^2 &=& dr^2 \,+\,\alpha^{-1}\,\left(dz-\alpha A\,dx-\alpha B
\,dy\right)^2\\ \nonumber
& &+\, \alpha(1+B^2)\,(dx^2+d\theta_1^2)\,+\,\alpha (1+A^2)\,
(dy^2+d\theta_2^2)\\ \nonumber
 & &+ \, 2\alpha AB \,\left[\cos\langle z\rangle\,(d\theta_1 d\theta_2- dx\,dy)\right.\\
& & \left.\phantom{2\alpha AB}+\sin \langle z\rangle\,(dx\,d\theta_2+dy\,d\theta_1)\right]\,,
\end{eqnarray}
which matches indeed the local limit of a deformed conifold. To see this, simply rewrite \eqref{defmetric} in local co-ordinates similar to those of \eqref{local} (but $A$ and $B$ will differ). There is one subtle difference, though: the two spheres (tori) parameterised by $(x,\theta_1)$ and $(y,\theta_2)$ are not of the same size as in the CY metric \eqref{defmetric}. It was discussed in \textcite{thesis} how the mirror symmetry becomes exact in the limit where the resolution and deformation parameters approach zero, as expected by \textcite{hori}. But since this means having vanishingly small two- or three-cycles (``close to the transition point''), this is a regime where the base cannot be large compared to the $T^3$ fibre, i.e. we cannot expect SYZ to work. This is why the large complex structure boost ``by hand'' became necessary.

\subsection{IIB orientifold and resolved conifold}\label{orifold}

The $SL(2,\mathbb{Z})$ symmetry of IIB string theory has been proposed to have a geometrical interpretation in terms of F-theory \cite{vafaf}. Consider an elliptically fibred Calabi--Yau fourfold $K$ which is a toroidal fibre bundle over a base $B$. Even though $K$ is a smooth manifold, there will be points in the base where the fibre becomes singular and its complex structure parameter $\tau$ can have non-trivial monodromy around these points. An F-theory compactification on $K$ refers to a compactification of type IIB on $B$, where the IIB axion-dilaton $\lambda=\chi+i e^{-\phi}$ is identified with the geometrical parameter $\tau$ \cite{vafaf}. This leads to orientifolds in IIB \cite{senone, sentwo}, see e.g. \textcite{dabholkar} for a detailed review. In our case, the base $B$ is an orientifold of the resolved conifold\footnote{$B$ will not be a Calabi--Yau threefold anymore, since $X$ is a Calabi--Yau, but it is still K\"ahler \cite{gttwo}.}
\begin{equation}
\frac{B}{\Omega(-1)^{F_L}I_2}\,,
\end{equation}
where $F_L$ indicates the left-moving fermion number, $\Omega$ stands for the worldsheet parity operator and $I_2$ is a target space parity that inverts both co-ordinates of the toroidal fibre.

In general, $\tau$ varies over the base resulting in a non-constant field $\lambda$. However, there are possible scenarios that allow for a constant solution of $\lambda$ \cite{senone, sentwo, keshavmukhi}. These solutions are characterised by 24 singularities in the function describing the elliptic fibration. In the special case where these singularities appear at four different locations with a multiplicity of six, $\lambda$ is constant. The singularities are interpreted as 24 seven-branes in F-theory and give rise to 4 orientifold seven-planes with 4 D7-branes on top of each (to cancel their charges), located at the four orientifold fixed points.

If we now wrap D5-branes on the $S^2$ of the resolved geometry, we obtain an intersecting D5/D7-brane scenario on a IIB  orientifold. This preserves supersymmetry, because it can form a bound state \cite{narain}. The boundstate metric was actually derived by \textcite{D5D7} and agrees with the local limit we use here. We will not make explicit use of the boundstate description in order to keep this section illustrative. The metric of the base $B$ has to resemble the resolved conifold locally, but globally it will also contain singularities that correspond to the 7-branes. Adding D5-branes creates warp factors. To incorporate these effects we make the following generic ansatz for the base
\begin{eqnarray}\label{basemetric}\nonumber
ds^2 & = & h_0(\widetilde{r})\,d\widetilde{r}^2 + h_1(\widetilde{r})\,\big
  (d\widetilde{\psi}+\cos\widetilde{\theta}_1\,d\widetilde{\phi}_1
  +\cos\widetilde{\theta}_2\,d\widetilde{\phi}_2\big)^2\\ \nonumber
& &+ h_2(\widetilde{r})\,d\widetilde{\theta}_1^2
  +h_3(\widetilde{r})\,\sin^2\widetilde{\theta}_1\,d\widetilde{\phi}_1^2 \\ 
& & + h_4(\widetilde{r})\,d\widetilde{\theta}_2^2+ h_5(\widetilde{r})\,
  \sin^2\widetilde{\theta}_2\,d\widetilde{\phi}_2^2\,,
\end{eqnarray}
which allows in particular for the two spheres to be asymmetric and squashed. This ansatz is motivated by the idea that in the absence of D-branes and fluxes we should recover the K\"ahler metric. Also, for $\widetilde{r}\to\infty$ the warp factors should approach 1, so we will suppress any $\theta_{1,2}$-dependence in the functions $h_i$ although it would not influence the following local analysis.
We again define local co-ordinates and absorb the prefactors $h_i(\widetilde{r})$, which gives the same simple form of the local metric
\begin{eqnarray}\label{startiib}\nonumber
ds^2 &=& dr^2+(dz+A\,dx + B\, dy)^2\\
& & + (dx^2+d\theta_1^2)+(dy^2+d\theta_2^2)\,,
\end{eqnarray}
where we have with a slight abuse of notation kept the names $A$ and $B$ for the constants, but they are now more generically defined. Apart from this redefinition of $A$ and $B$, the mirror symmetry analysis will be completely unchanged from Section \ref{mirror}. The mirror is then given by \eqref{defmirror}, which is the local limit of a deformed conifold. We will show shortly the consistency of this construction with an orientifold in IIA.

We need our IIB background to be invariant under the orientifold action, which is given by $\Omega\,(-1)^{F_L}\,I_{ij}$. Since the IIB background is invariant under $ \Omega\,(-1)^{F_L}$, we require the metric to be invariant under spacetime parity $I_{ij}$ of the two co-ordinates $x_i$ and $x_j$ over which the fibre degenerates. Furthermore, we require the IIA orientifold metric that results after three T-dualities to resemble the deformed conifold. This severely restricts the possibilities for $x_i$ and $x_j$.

The choice we adopt is that the F-theory torus is fibred non-trivially over the two-torus $(x,\,\theta_1)$. This is actually the only choice that preserves all the symmetries we require \cite{gttwo}. The D5-branes are wrapped on the two-torus (or sphere) given by $(y,\,\theta_2)$ (recall from \eqref{resmetric} that this is the sphere that remains finite as $\widetilde{r}\to 0$). This means that under three T-dualities
\begin{equation}\nonumber
\mbox{IIB on}\quad\frac{B}{\Omega (-1)^{F_L} I_{x\theta_1}}
\,\xrightarrow{T_{xyz}}\, \mbox{IIA on}\quad \frac{B'}{\Omega (-1)^{F_L}
I_{yz\theta_1}}\,.
\end{equation}
We find the following system of D-branes and orientifold planes in type IIB
\begin{equation}\nonumber
\begin{array}{lllllllllll}
D5:\quad & 0&1&2&3&-&-&-&-&y&\theta_2\\
D7/O7:\quad & 0&1&2&3&r&z&-&-&y&\theta_2\,,
\end{array}
\end{equation}
which turns after three T-dualities along $x,\,y$ and $z$ into IIA with
\begin{equation}\nonumber
\begin{array}{lllllllllll}
D6:\quad & 0&1&2&3&-&z&x&-&-&\theta_2\\
D6/O6:\quad & 0&1&2&3&r&-&x&-&-&\theta_2\,.
\end{array}
\end{equation}
It is easy to see that the metric \eqref{startiib} is indeed invariant under $I_{x\theta_1}$ (remember that $A$ contains $\cot\langle \theta_1\rangle$, so it is odd under this parity)
and the mirror will be symmetric under $I_{yz\theta_1}$ after we impose some restrictions on the B-field components (more in the next section).

Note that the D7-branes extend along the non-compact direction $r$. A similar brane configuration on the \emph{singular} conifold was considered by \textcite{ouyang}, but it was not constructed from F-theory\footnote{This analysis has recently been extended to the resolved conifold \cite{dterms}.}. It was shown there how strings stretching between D7 and D5-branes (or D6 and D6) give rise to a global symmetry. It is not a gauge symmetry because of the large volume factor associated with the D7-branes extending along the non-compact direction $r$. We will call the D7 or D6s that originate from F-theory ``flavour branes'' to distinguish them from the D5 or D6s that carry the gauge theory.

\subsection{Mirror symmetry with NS-NS flux and ``non-K\"ahler deformed conifold''}\label{nsmirror}

Having argued that \eqref{startmetric} or equivalently \eqref{startiib} is the correct local metric for D5s and D7/O7s on the resolved conifold, we now turn to the first link in the duality chain. As already mentioned, the mirror symmetry analysis from Section \ref{mirror} is changed once we take fluxes into account. The full RR spectrum that is consistent with the IIB orientifold was studied in \textcite{thesis}. We only focus on the NS-NS sector here, as it alone is relevant for the geometry. It should also have become clear that we can only present a local analysis here, for two reasons: 1) The mirror symmetry argument between resolved and deformed conifold fails globally and 2) We don't know the full F-theory solution, in other words the functions $h_i(\widetilde r)$ in \eqref{basemetric} remain unknown.

For the NS-NS flux we make the most generic ansatz that is consistent with our orientifold, with one exception: we only allow for electric NS-NS flux (magnetic NS-NS flux leads in general to non-geometrical solutions \cite{wecht, wechtaylor, simeon, nongeomet,hull}), i.e. B-field components that have only one leg along T-duality directions:\footnote{Without loss of generality we do not include components involving $dr$ since components of the 3-form field strength like $dr\wedge dx\wedge d\theta_i$ can easily be obtained from $\partial_r b_{xi}(r)\,dr\wedge dx\wedge d\theta_i$.}
\begin{equation}\label{bstartiib}
B_2^{IIB} \,=\, b_{z\theta_1}\,dz\wedge d\theta_1 + b_{x\theta_2}\,dx\wedge d\theta_2
+ b_{y\theta_1}\,dy\wedge d\theta_1\,.
\end{equation}
In general, the coefficients $b_{z\theta_1},\,b_{x\theta_2}$ and $b_{y\theta_1}$ can depend on all base co-ordinates $(r,\,\theta_1,\,\theta_2)$ to preserve the background's isometries.

This B-field has non-trivial consequences when we perform T-dualities along $x$, $y$ and $z$. We will not merely find a local version of the deformed conifold, but a manifold with {\it twisted fibres} that is clearly the local limit of a non-K\"ahler version of the deformed conifold.

The reason why mirror symmetry with NS-NS field gives rise to a non-K\"ahler manifold is actually very easy to illustrate in the SYZ picture. T-duality mixes B-field and metric. In the presence of NS-NS flux, Buscher's rules \cite{buscher, buschertwo} read
\begin{eqnarray}\label{busch}\nonumber
\widetilde{G}_{yy} &=& \frac{1}{G_{yy}},\qquad\qquad\qquad
  \widetilde{G}_{\mu y}\,=\,\frac{B_{\mu y}}{G_{yy}},\\
\widetilde{G}_{\mu\nu} &=& G_{\mu\nu}-\frac{G_{\mu y}G_{\nu y}-B_{\mu y},
  B_{\nu y}}{G_{yy}},\\ \nonumber
\widetilde{B}_{\mu\nu} &=& B_{\mu\nu}-\frac{B_{\mu y}G_{\nu y}-G_{\mu y}B_{\nu y}}{G_{yy}}, \\ \nonumber
\widetilde {B}_{\mu y} &=& \frac{G_{\mu y}}{G_{yy}}\,,
\end{eqnarray}
so cross terms in the metric are traded against the corresponding
components in $B_2$ and vice versa. Therefore, the $T^3$ fibres
acquire a twisting by $B_2$-dependent one-forms under T-duality that
we denote by
\begin{eqnarray}\label{fibration}\nonumber
dx\;\longrightarrow\;d\hat{x} &=& dx-b_{x\theta_2}\,d\theta_2,\\
dy\;\longrightarrow\;d\hat{y} &=& dy-b_{y\theta_1}\,d\theta_1,\\ \nonumber
dz\;\longrightarrow\;d\hat{z} &=& dz-b_{z\theta_1}\,d\theta_1.
\end{eqnarray}
This does not mean that $d\hat x$ etc. are exact; in general the
B-field is non-constant. Apart from this modification, we perform
the same steps as in Section \ref{mirror}: we boost the complex
structure as in \eqref{boost} and take the limit $\epsilon\to 0$.

Then we find the mirror in IIA to be
%\begin{widetext}
\begin{eqnarray}\label{nkdef}\nonumber
d\widetilde{s}^2 &=& dr^2 \,+\,\alpha^{-1}\,\left[d\hat z-\alpha A\,d\hat x-\alpha B\,d\hat y\right]^2\\ \nonumber
&& + \,\alpha(1+B^2)\left[d\theta_1^2+d\hat x^2\right] + \alpha (1+A^2)\left[d\theta_2^2 + d\hat y^2\right]\\ \nonumber
&& + \, 2\alpha AB\,\cos \langle z\rangle \,\big[d\theta_1 d\theta_2 \,-\, d\hat x\,d\hat y\big] \\ 
& &+\,  2\alpha AB\,\sin\langle z\rangle\,\big[d\hat x\,d\theta_2
+d\hat y\,d\theta_1 \big]\,,
\end{eqnarray}
%\end{widetext}
with $\alpha$ defined in \eqref{alpha}. We therefore conjecture the local resolved conifold to be mirror dual to a local ``non-K\"ahler deformed conifold'' with twisted fibres that make this metric inherently non-K\"ahler.

In the absence of a B-field this is clearly a K\"ahler background, since in this local version all coefficients in the metric are constants. With a B-field dependent fibration the fundamental two-form will in general no longer be closed, because it will depend on derivatives of $b_{ij}$. A more thorough analysis of this geometry was attempted in \textcite{thesis}, but it remains somewhat incomplete because we lack the knowledge of a global background. Strictly speaking, we only know the metric in a small patch and have no global information about the manifold. We can, however, make some predictions on what we expect for the global solution, since supersymmetry imposes restrictions on allowed non-K\"ahler manifolds.

We were able to show that this metric admits a symplectic structure, but we were not able to find a half-flat structure (with quite a generic ansatz for the almost complex structure, see Chapter 4 in \textcite{thesis}). This is not in contradiction with the results from \textcite{jan}, where the mirror of a torus was found to be half-flat. Our IIB starting background is not a Calabi--Yau; it is at best a conformal Calabi--Yau. Conformal Calabi--Yaus are complex manifolds,\footnote{This is obvious from their $SU(3)$ torsion classes, for example. See e.g. \textcite{salamon, lust}.} and there is ample evidence in the literature, see e.g. \textcite{kachrutomas, jeschek}, that the mirror of a complex manifold is symplectic (which can be half-flat at the same time, but does not have to be). Furthermore, our background includes RR flux and therefore has to fulfill different susy requirements than a purely geometrical background, i.e. it does not lift to a manifold with $G_2$-holonomy, as half-flat manifolds do, but only to one with $G_2$ structure.
%\vspace*{0.8cm}

\subsection{Completing the duality chain}

With the metric \eqref{nkdef} we are now ready to follow the duality chain through the M-theory flop and back to type IIB.

\subsubsection*{M-theory lift}

Note that there is no longer any NS-NS flux in our IIA background since it was completely ``used up" under T-duality and became part of the metric, but there is RR flux, dual to the RR three- and five-forms of the original IIB background. This flux means that we have to lift the ten-dimensional solution on a twisted fibre instead of a circle and that there will be G-flux in M-theory. It was shown in \textcite{gtone, thesis} that the RR flux does not change during the flop, so let us only consider the RR one-form potential that enters into the eleven-dimensional metric. It can be written as
\begin{equation}\label{oneform}
C_1\,=\,\Delta_1\,d\hat{x} - \Delta_2\,d\hat{y}\,,
\end{equation}
where $\Delta_1$ and $\Delta_2$ depend on the assumptions made about the IIB RR forms in the very beginning. (Recall the definition of the twisted fibres $d\hat{x}, d\hat{y}, d\hat{z}$ from \eqref{fibration}.) As usual in the presence of a gauge field $C_1$ and dilaton $\phi$ (which remains unchanged under three T-dualities \cite{gtone, thesis}), type IIA on a manifold $X$ is lifted to $\mathcal{M}$-theory on a twisted circle via
\begin{equation}
ds_{\mathcal{M}}^2 \,=\, e^{-2\phi/3}\,ds_X^2 + e^{4\phi/3}\,(dx_{11}+C_1)^2,
\end{equation}
with $x_{11}$ parametrising the extra dimension with radius $R$: $x_{11} \in (0,  2\pi R)$. In the limit $R\to 0$ we recover 10-dimensional IIA theory. The gauge fields in our case enter into the metric so that it becomes
%\begin{widetext}
\begin{eqnarray}\label{liftbefore}\nonumber
ds_{\mathcal{M}}^2 &=& e^{-2\phi/3}\,dr^2 + e^{-2\phi/3}\,\alpha^{-1}\,
\big(dz-\alpha A\,d\hat{x} -\alpha B\, d\hat{y}\big)^2\\ 
&& + \, e^{4\phi/3}\,\big(dx_{11}+\Delta_1\,d\hat{x}-\Delta_2\,d\hat{y}\big)^2\\
\nonumber
&& +\, e^{-2\phi/3}\,\left[\alpha(1+B^2)\,(d\theta_1^2+d\hat{x}^2)\right.\\ \nonumber
& & \left.\qquad\quad \;+\,
\alpha(1+A^2)\,(d\theta_2^2+d\hat{y}^2)\right]\\ \nonumber
&& + \, e^{-2\phi/3}\,2\alpha AB\,\left[\cos\langle z\rangle\, (d\theta_1\,
d\theta_2-d\hat{x}\,d\hat{y})\right.\\ \nonumber
& & \left.\qquad\qquad\quad\;\:\:\, + \, \sin\langle z\rangle \,
(d\hat{x}\,d\theta_2 + d\hat{y}\,d\theta_1)\right]\,.
\end{eqnarray}
%\end{widetext}

The two fibration terms in the first line are of special interest. They are very similar in structure, even more so if one introduces new co-ordinates $\psi_1$ and $\psi_2$ via
\begin{equation}\label{xeleven}
dz\,=\,d\psi_1-d\psi_2\qquad\mbox{and}\qquad dx_{11}\,=\,d\psi_1+d\psi_2.
\end{equation}
This choice is particularly convenient for performing the flop, as we will explain now.

\subsubsection*{Flop}

The metric of all three conifold geometries can be written in terms of two sets of $SU(2)$ left-invariant one-forms, see Appendix \ref{cs}. In terms of Euler angles on two $S^3$s these left-invariant one-forms are given as\footnote{We are essentially following the conventions of \textcite{cvetlupope, cvetlupo}, but our notation differs from theirs by $\phi_2\to -\phi_2$.}
\begin{eqnarray}\label{leftforms}\nonumber
\sigma_1 &=& \cos\psi_1\,d\theta_1+\sin\psi_1\,\sin\theta_1\,d\phi_1, \\ \nonumber
\sigma_2 &=& -\sin\psi_1\,d\theta_1+\cos\psi_1\,\sin\theta_1\,d\phi_1, \\
\sigma_3 &=& d\psi_1+\cos\theta_1\,d\phi_1,\\ \nonumber
\Sigma_1 &=& \cos\psi_2\,d\theta_2-\sin\psi_2\,\sin\theta_2\,d\phi_2, \\ \nonumber
\Sigma_2 &=& -\sin\psi_2\,d\theta_2-\cos\psi_2\,\sin\theta_2\,d\phi_2,\\ \nonumber
\Sigma_3 &=& d\psi_2-\cos\theta_2\,d\phi_2.
\end{eqnarray}
The Calabi--Yau metrics for resolved and deformed conifolds are written in these vielbeins as (with the definition $\psi=\psi_1-\psi_2$)
\begin{eqnarray}\nonumber
ds_{\rm def}^2 &=& A^2\,\sum_{i=1}^2(\sigma_i-\Sigma_i)^2 + B^2\,\sum_{i=1}^2 (\sigma_i+\Sigma_i)^2\\ \nonumber
 & & + C^2 \,(\sigma_3-\Sigma_3)^2 +D^2\,dr^2;\\
ds_{\rm res}^2 &=& \widetilde{A}^2\,\sum_{i=1}^2(\sigma_i)^2 + \widetilde{B}^2
\,\sum_{i=1}^2(\Sigma_i)^2\\ \nonumber
 & & + \widetilde{C}^2\,(\sigma_3-\Sigma_3)^2 +
\widetilde{D}^2\,dr^2,
\end{eqnarray}
with the coefficients $A,B$ etc. determined by K\"ahler and Ricci flatness conditions, see \eqref{resmetricapp} and \eqref{defmetricapp}. This clearly shows that the deformed conifold is completely symmetric under a $\mathbb{Z}_2$ that acts as $\sigma_i\leftrightarrow \Sigma_i$, whereas the resolved conifold does not have this symmetry since $\widetilde{A}\ne \widetilde{B}$.

Both geometries can be lifted to a $G_2$ holonomy manifold, a cone over $S^3\times \widetilde{S}^3$, where $S^3$ describes a sphere with vanishing size at the tip of the cone, whereas $\widetilde{S}^3$ remains finite. As described in Section \ref{chainvafa}, the flop corresponds to an exchange  $S^3\;\leftrightarrow\;\widetilde{S}^3$. In terms of vielbeins, the flop simply amounts to an exchange $\sigma_i\leftrightarrow \Sigma_i$, since each $S^3$ is described by a set of $SU(2)$ left invariant one-forms. But note that this also implies that the $U(1)$ fibre along which one reduces to d=6 is contained either in $\sigma_3$ or $\Sigma_3$, i.e. it is given either by $\psi_1$ or $\psi_2$, but not by $x_{11}=\psi_1+\psi_2$ as we defined it in \eqref{xeleven}.

This discussion was for Calabi--Yau metrics. The ``non-K\"ahler deformed conifold'' we found in Section \ref{nsmirror} does {\it not} have two $S^2$s of the same size. We therefore need to use a more general ansatz. In  \textcite{cvetlupope, cvetlupo} it was shown that there exists a one-parameter family of $G_2$--holonomy metrics (that includes the lift of the resolved and deformed conifolds\footnote{In particular, \textcite{cvetlupo} solved the differential equations for the $r$--dependent coefficients $a,b,c,f,g_3$ and $\xi$ and showed that the resulting K\"ahler form looks like that for the resolved conifold. It was not considered how a flop between resolved and deformed conifolds can be performed.}) of the form
\begin{eqnarray}\label{gtwoansatz}
ds^2 &=& dr^2 + a^2\,\big[(\Sigma_1+\xi \sigma_1)^2 + (\Sigma_2+\xi \sigma_2)^2\big]\\ \nonumber
&&+ \,b^2(\sigma_1^2+\sigma_2^2) + c^2 (\Sigma_3-\sigma_3) + f^2 (\Sigma_3+g_3\,\sigma_3)^2\,,
\end{eqnarray}
where $\psi_1-\psi_2$ was identified as the eleventh direction in \textcite{cvetlupo}, i.e. the limit $c\to 0$ corresponds to a reduction to ten dimensions. This metric has less symmetry than the metric in \textcite{flop}, for which the flop was discussed. Note that the parameter $\xi$ describes an asymmetry between the two $S^2$ in a deformed metric. It seems therefore appropriate to adopt this ansatz for our purposes.

Of course our metric \eqref{liftbefore} does not describe $S^3\times S^3$ principal orbits. Recall that our co-ordinates $x,y,z$ are non-trivially fibred due to the B-field components which entered into the metric. We can nevertheless adopt the ansatz \eqref{gtwoansatz} with a different definition of vielbeins:
\begin{eqnarray}\label{sigmas} \nonumber
\sigma_1 &=& \cos\psi_1\,d\theta_1 +\sin\psi_1\,d\hat{x},\\ \nonumber
\sigma_2 &=& -\sin\psi_1\,d\theta_1 +\cos\psi_1\,d\hat{x},\\ 
\sigma_3 &=& d\psi_1-\alpha A\,d\hat{x}, \\ \nonumber
\Sigma_1 &=& \cos\psi_2\,d\theta_2 -\sin\psi_2\,d\hat{y}, \\ \nonumber
\Sigma_2 &=& -\sin\psi_2\,d\theta_2 -\cos\psi_2\,d\hat{y}, \\ \nonumber
\Sigma_3 &=& d\psi_2+\alpha B\,d\hat{y}\,.
\end{eqnarray}

The flop has to be different from the case considered in \textcite{flop}, since we do not want to exchange the roles of $\psi_1$ and $\psi_2$, but we want to exchange $x_{11}$ and $z$ as these are the naturally fibred co-ordinates in \eqref{liftbefore}. Furthermore, we have the asymmetry factor $\xi$, so that our metric does not exhibit the $\mathbb{Z}_2$ symmetry $\sigma_i\leftrightarrow\Sigma_i$ as the lift of the Calabi--Yau deformed conifold does. As explained in Section 3.3 of \textcite{thesis}, the flop in our conventions corresponds to
\begin{eqnarray}\label{flipflop}\nonumber
\sigma_3-\Sigma_3 &\leftrightarrow&\sigma_3+\Sigma_3,\\
\sigma_i &\to& \Sigma_i,\\ \nonumber
\Sigma_i &\to& \xi\,(\sigma_i-\Sigma_i),
\end{eqnarray}
with $i=1,2$.
This results in the following metric after the flop\footnote{Here we used an explicit gauge choice for the RR one-form corresponding to $C_1\,=\,-\alpha A \,d\hat{x}+\alpha B\,d\hat{y}$.}
\begin{eqnarray}\label{liftafter} \nonumber
ds^2 &=& e^{-2\phi/3}\,dr^2+e^{-2\phi/3}\,\frac{\alpha A^2B^2}{1+A^2}\,\big
  (d\theta_1^2+d\hat{x}^2\big)\\
&&+\, e^{-2\phi/3}\,\frac{1}{1+A^2}\,\big(d\theta_2^2+d\hat{y}^2\big) \\ \nonumber
&&+ \,e^{-2\phi/3}\,\alpha^{-1}\,\big(dx_{11}-\alpha A\,d\hat{x}\alpha B\,d\hat{y}\big)^2\\ \nonumber
&&+\, e^{4\phi/3}\,\big(dz-\alpha A\,d\hat{x}-\alpha B\,
  d\hat{y}\big)^2\,,
\end{eqnarray}
which can now be reduced along the same $x_{11}$ to the IIA background after transition.

\subsubsection*{M-theory reduction}

Dimensional reduction on the same $x_{11}$ clearly does not give the same metric as before the flop. Instead, we find
%\begin{widetext}
\begin{eqnarray}\label{iiaafter}\nonumber
ds^2_{IIA} &=& dr^2 +e^{2\phi}\,\big[(dz-b_{z\theta_1}\,d\theta_1)\\ \nonumber
& & - \, \alpha A\,(dx-b_{x\theta_2}\,d\theta_2)-\alpha B\,(dy-b_{y\theta_1}\,d\theta_1)\big]^2\\ \nonumber
& &+\,  C\,\big[d\theta_1^2+(dx-b_{x\theta_2}\,d\theta_2)^2 \big]\\ 
& &+ \, D\,\big [d\theta_2^2+(dy-b_{y\theta_1}\,d\theta_1)^2\big] \,,
\end{eqnarray}
%\end{widetext}
where we have written the fibration structure explicitly as a reminder that the original IIB B-field is contained in this metric. We test the reader's patience by introducing another set of symbols for the metric components giving the spheres:
\begin{eqnarray}\label{candd} \nonumber
C &=& \frac{\alpha A^2B^2}{1+A^2}\,,\qquad D\,=\,\frac{1}{1+A^2}\,,\\ 
\mbox{and}\quad
\alpha_0^{-1} &=& CD+\alpha^2\,e^{2\phi}\,(CB^2+DA^2)
\end{eqnarray}
in analogy with the definition of $A,\,B$ and $\alpha$ in \eqref{defineab} and \eqref{alpha}.

This manifold is non-K\"ahler in precisely the same spirit as the ``non-K\"ahler deformed conifold'' before the flop \eqref{nkdef}. Comparing it to \eqref{startiib} shows that it also possesses the characteristic metric of a resolved conifold (locally). We therefore baptise this manifold a ``non-K\"ahler resolved conifold'' and claim it to be transition dual to the metric \eqref{nkdef}. The latter is a manifold with D6-branes wrapping a 3-cycle, whereas the former describes a blown-up 2-cycle with fluxes on it.

\subsubsection*{The final mirror}

We can now ``close the duality chain'' by performing another mirror which takes us back to IIB. We should recover a K\"ahler background similar to the Klebanov--Strassler model \cite{ks}, since we started with a K\"ahler manifold in IIB. In principle the analysis follows the same steps as laid out when T-dualising the resolved conifold from IIB to IIA without NS-NS flux in Section \ref{mirror}, only now our starting metric is the non-K\"ahler version of the resolved conifold \eqref{iiaafter}.

T-dualising this background along $x,\,y$ and $z$ is tedious but nevertheless straightforward. See Section 3.4 of \textcite{thesis} for the details of the calculation. We make the fascinating observation that the same mechanism that converted the B-field into metric cross terms now serves to restore $b_{x\theta_2}$, $b_{y\theta_1}$ and $b_{z\theta_1}$ as the B-field and the metric is completely free of any B-field dependent fibration. The final IIB metric after transition is found to be
%\begin{widetext}
\begin{eqnarray}\label{finaliibmetric}\nonumber
d\tilde{s}_{IIB}^2 &=& dr^2\\ \nonumber
& &+   \frac{e^{-2\phi}}{\alpha_0 CD}\big[dz+\alpha_0\alpha AD e^{2\phi}dx+\alpha_0\alpha BC e^{2\phi}dy\big]^2\\
& &+\,  \alpha_0\left(D+\alpha^2B^2e^{2\phi}\right)\,(dx^2 + \zeta\, d\theta_1^2)\\ \nonumber
& & +\,  \alpha_0\left(C+\alpha^2A^2e^{2\phi}\right)\,(dy^2 + d\theta_2^2)\\\nonumber
& &+ \, 2\alpha_0\alpha^2 AB e^{2\phi}\,\big[\cos\langle z\rangle(d\theta_1\,d\theta_2-dx\,dy)\\ \nonumber
& & \qquad\qquad\quad\quad \;\!+\,\sin\langle z\rangle (dx\,d\theta_2+dy\,d\theta_1)
 \big]\,,
\end{eqnarray}
%\end{widetext}
where we have introduced the ``squashing factor''
\begin{equation}\label{squash}
\zeta \,=\,\frac{C-\alpha^2 A^2\tilde{\beta}_1^2}{\alpha_0\,(D+\alpha^2 B^2
e^{2\phi})}\,.
\end{equation}

We therefore find that the final IIB metric after the flop \eqref{finaliibmetric} is not quite a deformed conifold due to the asymmetry in the $(x,\,\theta_1)$ sphere/torus. In the local version presented above it is of course K\"ahler (all coefficients are constant), but we cannot make any statement about the global behaviour. Remember that we do not have the global metric for our starting background with D7/O7 and D5-branes.

The cross terms in the metric \eqref{iiaafter} are now converted into B-field components by the same mechanism of Buscher's rules \eqref{busch}. One recovers the B-field \eqref{bstartiib} we started with in IIB before the transition. The same holds true for the RR flux. The flux does not change under geometric transition \cite{gtone, thesis}, confirming the picture advocated by \textcite{vafa}.

In conclusion, we have shown that we can construct a new pair of IIA string theory backgrounds that are non-K\"ahler and deviate from the deformed and resolved conifolds in a very precise manner: the $T^3$ fibres are twisted by the B-field. They are related by a geometric transition, because their respective lifts to $\mathcal{M}$-theory are related by a flop. The IIB backgrounds \eqref{startiib} and \eqref{finaliibmetric}, on the other hand, are K\"ahler and are also transition dual, based on mirror symmetry.

\setcounter{equation}{0}
\section{Discussion}
We have presented a supergravity analysis confirming Vafa's
``duality chain'', see Figure \ref{vafachain}, with the inclusion of
non-K\"ahler manifolds in type IIA. These manifolds are non-K\"ahler
due to a twisting of their fibres by the B-field that is introduced
via T-duality. Thus, they should fall into the classification of
T-folds \cite{hull}, where the transition functions of a manifold
are allowed to take values in the T-duality group
$O(d,d;\mathbb{Z})$ or into generalised complex geometries
\cite{hitch, gualtieri}. They are only trivial examples  though, as
we have only considered T-duality with a B-field of (1,1)-type (also
called ``electric''). Thus our backgrounds are still geometric, i.e.
true manifolds rather than T-folds, and their generalised complex
structure is not of a mixed type, but purely symplectic (as the IIB
background we started with was complex and it is by now well
established that mirror symmetry with ``electric'' NS-NS flux
connects complex and symplectic manifolds \cite{kachrutomas,
jeschek}). A symplectic structure was indeed found in terms of $SU(3)$
torsion classes; see \textcite{gttwo, thesis} for details.

We still lack a global description for these manifolds, as mirror symmetry between the resolved and the deformed conifolds forced us to adopt a local limit. On top of that, there is no known global description for our IIB starting background: D5-branes wrapped on the resolution of the conifold. The Pando Zayas--Tseytlin solution \cite{pt} suggested for this case explicitly breaks supersymmetry, as explained in Appendix \ref{fluxes}. We circumvented this problem by viewing the IIB background as an orientifold stemming from F-theory. This background contains additional D7-branes and O7-planes, but allows for a supersymmetric background with D5-branes. We left the ansatz for the fluxes generic, as long as they are invariant under the orientifold operation. It would of course be more satisfying to find a global background with all these properties and explicitly confirm its supersymmetry.

\begin{center}
\begin{figure*}[htp]
\begin{center}
\mbox{\begin{minipage}{14.5 cm}
\begin{center}
\includegraphics[scale=1]{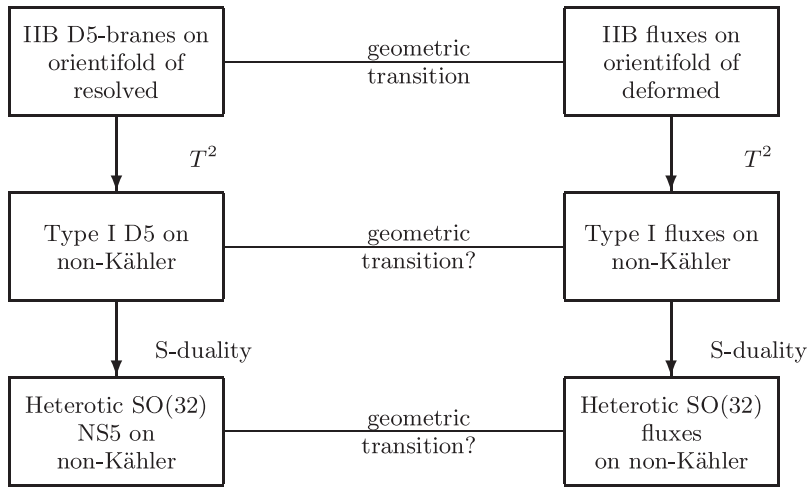}
 \caption{The heterotic duality chain. Following the arrows we can construct non-K\"ahler backgrounds in type I and heterotic theory that are dual to the type IIB backgrounds before and after transition. This implies that the new backgrounds are also transition duals in some sense.} \label{hetenchain}
\end{center}
\end{minipage}
 }
 \end{center}
\end{figure*}
\end{center}

Once we introduce additional D7-branes, we might immediately ask two
questions: Can the D5s and D7s form a supersymmetric bound state?
And do these branes introduce additional symmetry into the dual
gauge theory? Both questions have been answered affirmatively; see
\textcite{D5D7} and \textcite{gttwo}, respectively. In \textcite{D5D7} the
metric of a D5/D7 bound state on a resolved conifold geometry was
found exactly, in the sense that all backreactions neglected in
earlier papers were taken into account. Difficulties inherent in
solving the equations of motion were circumvented by a U-duality
chain which took as its starting point a D1/F1 bound state or (m,n)
string. The result matches earlier conjectures in certain limits,
completing the supergravity description of the Vafa duality chain
starting point as found from the F-theory set-up. The 7-branes from
F-theory will lead to a global symmetry group depending on the
special degeneration of the F-theory torus over the base. In
\textcite{gttwo} the symmetry group was argued to be $SU(2)^{16}$. This
was due to the fact that in IIB every orientifold fixed point
contributes four D7-branes giving rise to an $SO(8)$ that is broken
by Wilson lines to $SO(4)\times SO(4)\simeq SU(2)^4$. This is
consistent with the IIA orientifold that contains eight fixed
points, each accompanied by two D6-branes. The symmetry group
generated by eight stacks of D6s is therefore $SO(4)^8\simeq
SU(2)^{16}$. In \textcite{keshavmukhi} it was shown that one can even
construct the exceptional gauge groups $E_6$, $E_7$ and $E_8$, which
are of particular interest for Grand Unified Theories (GUTs). The
D7s don't give rise to a gauge symmetry because they extend along
the non-compact radial direction and therefore suffer from a
large-volume suppression. A similar set-up was suggested by
\textcite{ouyang}, but there the flavour branes did not have an F-theory
origin and were treated in a probe approximation. 
%Note that the Klebanov--Strassler model starts out with flavour branes (the regular D3-branes along non-compact external space), but these ``cascade away", until only fractional D3-branes (and therefore a pure SYM gauge theory dual) is left at the bottom of the cascade.

The superpotential in our flux backgrounds also remains to be calculated. One of the remarkable results from \textcite{vafa}  was to show that the flux generated superpotential does indeed agree (at lowest order) with the Veneziano--Yankielowicz superpotential for Super--Yang--Mills (SYM) theory. This superpotential actually receives corrections from field theory \cite{schwetz, ckl} as well as from string theory considerations \cite{dv}. One question we would like to address is whether a generalised superpotential (taking the non-K\"ahler structure of the target manifold into account) might be better suited to reproduce these corrections. Furthermore, we would like to address the additional global symmetry. The field theory analogue to Veneziano--Yankielowicz for an $SU(N)$ theory with matter is given by the Affleck--Dine--Seiberg superpotential \cite{affleck}. It would be interesting to see if we could reproduce this superpotential (as in the case of a Calabi--Yau orbifold $\mathbb{C}^3/\mathbb{Z}_2\times\mathbb{Z}_2$ \cite{lerda}) or if we would find an extension to it when including the flux due to D7-branes. We would need the precise supergravity solution to see which fluxes are actually turned on. In our set-up, the charge of the D7-branes is immediately cancelled by the orientifold planes. We would have to move the orientifold planes away from the flavour branes to observe their effect. This would lead to non-perturbative corrections.

Another generalisation of our duality chain was suggested by \textcite{realm}. One can exploit the idea of a IIB orientifold to go to the orientifold limit i.e. type I. Another S-duality takes us to heterotic SO(32) and we find two non-K\"ahler backgrounds that must in a certain sense be dual to each other, since they are individually U-dual to the IIB backgrounds for which we confirmed the geometric transition picture, see Figure \ref{hetenchain}. The orientifold operator we have to choose here is different from the one used in Section \ref{orifold} and the heterotic backgrounds will therefore not resemble non-K\"ahler versions\footnote{These manifolds are non-K\"ahler because we perform 2 T-dualities before S-dualising to heterotic. The aim is to ``use up'' all the NS-NS field of the IIB theory, so that only RR flux is left, which becomes NS-NS flux in the heterotic theory and to convert the D5/D7 system into D5/D9s.} of conifolds anymore. See \textcite{realm} or \textcite{thesis} for details.

The interpretation of this duality is still mysterious. Since heterotic string theory does not contain any open strings, the interpretation as an open/closed duality fails. We think this is a case where the geometric transition changes the vector bundle in a way that it requires the introduction of NS5-branes as localised sources for anomaly cancellation (before transition). It would also be interesting to study the underlying topological (0,2) theory. Results from \textcite{zerotwoKatz, zerotwoSharpe, zerotwoKapu} and \textcite{zerotwoWitten} should prove useful here.

We would also like to gain a better understanding of how our IIA
non-K\"ahler backgrounds fit into generalised complex geometry. We
have not explicitly shown that our manifolds are (twisted)
generalised Calabi--Yau (i.e. possess a (twisted) closed pure spinor
\cite{minasone, minastwo, minasthree} or have $SU(3)\times SU(3)$ structure, see
\textcite{granlouwald} and references therein), which is the most general condition for all
non-K\"ahler backgrounds with fluxes to preserve supersymmetry
(there is a second pure spinor, which is not closed, but its
derivative is proportional to the RR field strengths; NS-NS flux and
dilaton enter into the (twisted) $d$-operator). With much progress
having been made in the field of generalised topological sigma
models \cite{lindstrom, kapuli} and topological string theory
\cite{pestun, kapustin, kapustinli} one could hope to repeat the analysis of
\textcite{bcov} on these kinds of backgrounds and show agreement of the
open and closed \em generalised \em topological partition functions.
This is complicated by the fact that we also have RR flux in our
model, whose role in topological string theory is still not well
understood.

To make contact with phenomenology, one would need to compactify the
six-dimensional manifolds suggested here. Since our analysis was
performed in a local limit anyway, it would still hold if the
conifold bulk was cut off and replaced with a compact Calabi--Yau.
This is similar in spirit to all the cosmological models working
with the ``warped Klebanov--Strassler throat''. Indeed, once we
compactify we would also be forced to introduce extra objects for
charge cancellation. If these were anti-D-branes, we'd find
ourselves in the realm of non-supersymmetric, potentially viable
cosmological models. Another phenomenologically appealing direction
is the study of more realistic gauge groups, like the standard model
or simply QCD, in terms of geometric transitions. For the future one
might hope that open-closed duality can teach us something about the
strong coupling behaviour of confining gauge theories.

\begin{acknowledgments}
We would like to thank our collaborators on the various projects
that entered into this review, Stephon Alexander, Katrin Becker,
Melanie Becker, Keshav Dasgupta, Marc Grisaru, Sheldon Katz and Radu
Tatar, and we are very grateful to Keshav Dasgupta, Andrew Frey and Igor Klebanov
for many helpful discussions. We would also like to thank Arvind
Murugan for an illuminating correspondence. R.G. is supported in
part by a Chalk--Rowles fellowship. The work of A.K. is supported by
an NSERC grant. A.K. would also like to thank the Galileo Galilei
Institute for Theoretical Physics in Florence, where this work was
completed, and the organizers of the workshop ``String-- and
M--theory approaches to particle physics and cosmology'' for creating
an enjoyable and stimulating atmosphere, as well as the INFN for
partial financial support.
\end{acknowledgments}

\appendix
\def\theequation{\Alph{section}.\arabic{equation}}
\section{Conifolds}\label{coni}

\setcounter{equation}{0}

The (singular) conifold is a cone over a five-dimensional base and is a Calabi--Yau threefold. There are two ``relatives'' of the conifold, in which the singularity has been smoothed out in two different ways: one is called a ``resolved conifold'', with a blown-up $S^2$ at the tip of the cone; the other is the ``deformed conifold'', in which the singularity is blown up into an $S^3$. All three manifolds look asymptotically the same, like a cone over $S^2\times S^3$. Their metrics then take the form
\begin{eqnarray}\label{conemetric}
ds^2 & = & dr^2 + r^2 ds_{\mathrm{base}}^2\,.
\end{eqnarray}
It was shown in \textcite{candelas} that all three possess a K\"ahler metric and are Ricci flat and that one can pass continuously from one geometry to another. This is despite the fact that they are topologically different, which is seen immediately from the Euler numbers: $\chi(S^3) = 0, \chi(\mathrm{point}) = 1$ and $\chi(S^2) = 2$. This transition is called a ``conifold transition'' and can be pictured as follows:

\begin{center}
\includegraphics[scale=0.4]{gwyn_fig03.eps}
\end{center}
Here, the deformed conifold on the left approaches the singular conifold as the $S^3$ shrinks to zero size and the resolved conifold is obtained by blowing up the orthogonal $S^2$.

We will in the following review the symmetry properties and
Ricci-flat K\"ahler metrics on all three manifolds, as well as
discuss their complex structures. Useful references are \textcite{
candelas,minasian, pt} and \textcite{papatseyt}.

\subsection{The singular conifold}\label{appsingconi}

Just as a two-dimensional cone is embedded in real three-dimensional space as
$x^2 + y^2 - z^2 =0$, a real six-dimensional conifold can be expressed in terms of three complex co-ordinates, and is therefore embedded in $\mathbb{C}^4$ as
\begin{eqnarray}\label{eq-con}
\sum_{i=1}^4 z_i^2 = 0\,.
\end{eqnarray}
This describes a surface which is smooth apart from a singularity at $z_i =0$. The space has an $SO(4) \approx SU(2) \times SU(2)$ symmetry by which the $z_i$ are rotated into each other, and a $U(1)$ which rotates all the $z_i$ by the same phase. There is also a scaling symmetry given by the transformation $z_i \rightarrow tz_i, t \in \mathbb{C}^\star$.
By choosing $z_i = x_i + \imath y_i$, we can rewrite \eqref{eq-con} as
 \begin{eqnarray}\label{splitparts}
 \sum_{i=1}^4 x_iy_i &=& 0\,,\qquad \sum_{i=1}^4 (x_i^2 - y_i^2) \, =\, 0\,.
 \end{eqnarray}
The $x_i$ describe a three-sphere for any $y_i$, with vanishing radius at $y_i=0$, and the co-ordinates $y_i$ are orthogonally fibred to them. Therefore the space is given by $T^*S^3$.

To find the base of the conifold we take its intersection with a three-sphere of radius r:
\begin{eqnarray}\label{eq-int}
\sum_{i=1}^4 |z_i|^2 & =& \sum_{i=1}^4 (x_i^2+y_i^2) \,=\, r^2,
\end{eqnarray}
which removes the scaling symmetry $z_i \rightarrow tz_i$. The resulting five-dimensional space is a Sasaki--Einstein manifold\footnote{The base of a K\"ahler cone is a Sasakian manifold and
the base of a Ricci-flat cone is an Einstein manifold, so the base of a Calabi-Yau cone is a Sasaki-Einstein manifold.} called $T^{1,1}$. Together with \eqref{splitparts} we see that \eqref{eq-int} gives a three-sphere of radius $r/\sqrt{2}$ parametrised by $x_i$, whereas the $y_i$ describe a two-sphere fibred over the $S^3$. Since all such fibrations are trivial, the topology of the base $T^{1,1}$ is $S^2\times S^3$ \cite{candelas}.

$T^{1,1}$ also has a coset space description as $(SU(2)\times SU(2))/U(1)$. To see this, define
\begin{eqnarray}\label{defW}
W & = &  \frac{1}{\sqrt{2}} \sum_n \sigma^n  z_n,
\end{eqnarray}
with $\sigma^n$ the Pauli matrices for $n=1, 2, 3$ and $\sigma^4 = \imath \mathbf{1}$ so that
\begin{eqnarray*}
W & = &\frac{1}{\sqrt{2}} \left ( \begin{array}{cc} z_3 + \imath z_4 & z_1 - \imath z_2 \\ z_1 + \imath z_2 & - z_3 + \imath z_4  \end{array} \right ).
\end{eqnarray*}
Then the defining equation for the conifold (\ref{eq-con}) and the base (\ref{eq-int}) can be written as
\begin{eqnarray}
\det W & = & 0\,,\\ \label{defr}
\mathrm{tr} W^\dagger W & = & r^2\,.
\end{eqnarray}
By a rescaling $ Z = \frac{W}{r}$ these become
\begin{eqnarray*}
\det Z & = & 0,\\
\mathrm{tr} Z^\dagger Z & = & 1.
\end{eqnarray*}
Given a particular solution $Z_0$, say $Z_0 = \frac{1}{2}( \sigma_1 + \imath \sigma_2)$, the general solution can be written as
\begin{eqnarray}
Z & = & L Z_0 R^\dagger,
\end{eqnarray}
where
\begin{eqnarray}\label{defLR}
L = \left ( \begin{array} {cc} a & - \bar b \\ b & \bar a  \end{array} \right )\,,\quad & R = \left ( \begin{array} {cc}k & - \bar l \\ l & \bar k  \end{array} \right) .
\end{eqnarray}
$L,R \in SU(2)$ so \( |a|^2 + |b|^2 = |k|^2 + |l|^2 = 1 \). Thus we have shown that $SU(2) \times SU(2)$ acts transitively on the base. When $(L,R) = ( \Theta, \Theta^ \dagger)$ with
\begin{eqnarray*}
\Theta & = & \left ( \begin{array}{cc} e^{\imath \theta} & 0 \\ 0 & e^{-\imath \theta } \end{array} \right ),
\end{eqnarray*}
$Z_0$ is left fixed. This means that we can identify $(L,R)$ and $(L \Theta, R \Theta^\dagger)$, i.e. the base is the coset space $\frac{SU(2) \times SU(2)}{U(1)} = \frac{S^3 \times S^3 }{U(1)}$ with topology $S^2 \times S^3$ and symmetry group $SU(2) \times SU(2) \times U(1)$.

We now turn to the discussion of the K\"ahler metric on the singular conifold. The metric on a complex manifold is K\"ahler if and only if it can be written as
\begin{eqnarray*}
g_{\mu \bar \nu} & = & \partial_{\mu} \partial_{\nu} {\cal F},
\end{eqnarray*}
where $\cal F$ is the K\"ahler potential. If this potential is to be invariant under the action of $SU(2) \times SU(2)$ it can only be a function of $r^2$, so
\begin{eqnarray*}
g_{\mu \bar \nu} & = & (\partial_\mu \partial_{\bar \nu} r^2) {\cal F'} + (\partial_\mu r^2)(\partial_ {\bar \nu} r^2) {\cal F''},
\end{eqnarray*}
where the prime indicates a derivative with respect to $r^2$.  In terms of $W$
\begin{equation}\label{kaehlermetric}
  ds^2\,=\,\mathcal{F}'\,Tr(dW^\dagger dW)+\mathcal{F}''\,|Tr\,W^\dagger dW|^2\,.
\end{equation}
To find the condition that our metric be in addition Ricci flat, we need the Ricci tensor, which takes the form
$R_{\mu \bar \nu}  =  \partial_\mu \partial_{\bar \nu} \,\ln (\det g_{\mu \bar \nu})$ on a K\"ahler manifold.
Define a function
\begin{equation}\label{defgamma}
\gamma(r)\,=\,r^2\,\mathcal{F}'\,,
\end{equation}
then requiring Ricci flatness leads to
\begin{equation}
\gamma(r)\,=\,r^{4/3}\,.
\end{equation}
After a rescaling $r\to\tilde{r}=\sqrt{3/2}\,r^{2/3}$ one recovers indeed a metric of the form \eqref{conemetric} from \eqref{kaehlermetric}.
The metric of the base has an especially useful description in terms of Euler angles. Choosing the following parametrisation of $L$ and $R$ in \eqref{defLR}
\begin{eqnarray}\label{eulerangles}\nonumber
a &=& \cos\frac{\theta_1}{2}\,e^{\frac{i}{2}(\psi_1+\phi_1)}\,,\qquad
k\,=\,\cos\frac{\theta_2}{2}\,e^{\frac{i}{2}(\psi_2+\phi_2)},\\
b &=& \sin\frac{\theta_1}{2}\,e^{\frac{i}{2}(\psi_1-\phi_1)}\,,\qquad
 l\,=\,\sin\frac{\theta_2}{2}\,e^{\frac{i}{2}(\psi_2-\phi_2)},
\end{eqnarray}
where $\psi_i,\phi_i,\theta_i$ are the Euler angles of each $SU(2)$, one obtains from \eqref{kaehlermetric}
\begin{equation}\label{singmetric}
ds_{T^{1,1}}^2 \, = \, \frac{1}{9}(d\psi + \sum_{i =1 }^2 \cos \theta_i d \phi_i)^2 + \frac{1}{6}\sum_{i=1}^2 (d \theta_1^2 +
\sin^2\theta_i d \phi_i^2)\,.
\end{equation}
This form of the metric clearly shows the two spheres $(\theta_i,\phi_i)$ and the $U(1)$ fibre over them, parametrised by $\psi=\psi_1+\psi_2$. We also observe that the $U(1)$ symmetry (discussed after \eqref{eq-con}) manifests itself as a shift symmetry in $\psi$.

\subsection{The deformed conifold}

One way to repair the singularity of a conifold is by deformation  in which the defining equation (\ref{eq-con}) near $r=0$  is replaced by
\begin{eqnarray}\label{eq-def}
\sum_{i=1}^4 z_i^2 = \mu^2.
\end{eqnarray}
By taking again the intersection with the three-sphere to find the base, one finds $2x_i^2 = \mu^2 + r^2$, i.e. a finite $S^3$ remains at $r=0$. This is called a \em deformed conifold\em. Note that the $U(1)$ symmetry of the singular conifold (corresponding to a rotation $z_i\to e^{i\alpha} z_i$ with constant phase $\alpha$ for all $i$) is broken to a $\mathbb{Z}_2$ that sends $z_i \to -z_i$.

In terms of the matrix $W$ as defined in \eqref{defW} the deformed conifold is given by
\begin{equation*}
det\,W\,=\,-\mu^2/2
\end{equation*}
and as in \eqref{defr} we define a radial co-ordinate via the relation $r^2=Tr\,(W^\dagger W)$. Splitting the $z_i$ into real and imaginary parts we obtain
\begin{equation}
r^2\,=\,\sum_{i=1}^4 (x_i^2+y_i^2)\,,\qquad \mu^2\,=\,\sum_{i=1}^4(x_i^2-y_i^2)\,,
\end{equation}
which implies that $r$ ranges from $\mu$ to $\infty$. But it is also clear that the deformed conifold is still the cotangent bundle over a three-sphere $T^*S^3$, only that the $S^3$ has a minimal size; it never shrinks to zero.
A particular solution is found to be
\begin{equation}
W_\mu \,=\, \begin{pmatrix} \frac{\mu}{\sqrt{2}} & \sqrt{r^2-\mu^2}\\
0 & -\frac{\mu}{\sqrt{2}}\end{pmatrix}
\end{equation}
and the general solution is obtained by setting
$  W\,=\,L\,W_\mu\,R^\dagger$.
For $r\ne\mu$ the stability group is again $U(1)$. So for each $r\ne\mu$ the surfaces $r=$constant are again $S^2\times S^3$. Note however, that for $r=\mu$ the matrix $W_\mu$ is proportional to $\sigma_3$ and is invariant under an entire $SU(2)$. Thus, the ``origin'' of co-ordinates $r=\mu$ is in fact an $SU(2)=S^3$.

Again we define a K\"ahler potential $\hat{\mathcal{F}}$ and $\hat{\gamma}=r^2
\hat{\mathcal{F}}$. Then the metric is given by \eqref{kaehlermetric} and the condition for Ricci flatness becomes \cite{candelas}
\begin{equation}\label{deformedgamma}
r^2(r^4-\mu^4)(\hat{\gamma}^3)'+3\mu^4\hat{\gamma}^3\,=\,2 r^8\,.
\end{equation}
This can be integrated and one finds that for $r\to\infty$ the function $\hat{\gamma}$ approaches $r^{4/3}$, which agrees with the singular conifold solution. So asymptotically (for large $r$) the two spaces look the same.
In terms of Euler angles \eqref{eulerangles} the metric is explicitly given as \cite{papatseyt, minasian} (see also \textcite{Ohta:1999we})
\begin{widetext}
\begin{eqnarray}\label{defmetricapp}\nonumber
ds^2_{\rm def} &=& \left[\left(r^2 \hat{\gamma}'-\hat{\gamma}\right)
\,\left(1-\frac{\mu^4}{r^4}\right)+\hat{\gamma}\right]\,\left(\frac{dr^2}{r^2(1-\mu^4/r^4)}
+ \frac{1}{4}\,(d\psi+\cos\theta_1\,d\phi_1+\cos\theta_2\,d\phi_2)^2\right)\\
&& +\, \frac{\hat{\gamma}}{4}\,\left[(\sin\theta_1^2\,d\phi_1^2+d\theta_1^2)
+(\sin\theta_2^2\,d\phi_2^2+d\theta_2^2)\right]\\ \nonumber
&&+\,\frac{\hat{\gamma}\mu^2}{2 r^2}\,\left[\cos\psi(d\theta_1
d\theta_2-\sin\theta_1\sin\theta_2 d\phi_1
d\phi_2)+\sin\psi(\sin\theta_1 d\phi_1 d\theta_2+\sin\theta_2 d\phi_2
d\theta_1)\right]\,,
\end{eqnarray}
\end{widetext}
where we would still need to rescale $r$ to ensure that $\hat\gamma$ has dimension $r^2$.
Note that even the metric now shows the absence of the $U(1)$ symmetry formerly associated with shifts in $\psi$. As stressed at the beginning of this section, this is not an accident of the parametrisation we chose, but inherent to the defining equation \eqref{eq-def} of the deformed conifold.

\subsection{The resolved conifold}\label{appresconi}

Another way to repair the conifold singularity is to \em resolve \em it by blowing up a two-sphere. Upon defining new variables
\begin{eqnarray}\label{coordchange}
x & = & z_1 + \imath z_2,\\
y & = & z_2 + \imath z_1,\\
u & = & z_3 - \imath z_4,\\
v & = & z_4 - \imath z_3,
\end{eqnarray}
the conifold equation (\ref{eq-con}) becomes
\begin{eqnarray}\label{eq-res}
xy-uv\, =\, 0\,.
\end{eqnarray}
This is equivalent to requiring non-trivial solutions to
\begin{equation}
\begin{pmatrix} x & u \\ v & y\end{pmatrix}
\begin{pmatrix} \xi_1 \\ \xi_2 \end{pmatrix}
\,=\, 0\,,
\end{equation}
in which $\xi_1, \xi_2$ are not both zero. So, for $(u,v,x,y)\ne 0$ (away from the tip), they describe again a conifold. But at $(u,v,x,y)=0$ this is solved by any pair $(\xi_1,\xi_2)$. Due to the overall scaling freedom $(\xi_1,\xi_2)\sim (\lambda\xi_1,\lambda\xi_2)$ we can mod out by this equivalence class and $(\xi_1,\xi_2)$ actually describe a $\mathbb{CP}^1\sim S^2$ at the tip of the cone. Therefore, the resolved conifold is depicted as $\mathcal{O}(-1)\oplus\mathcal{O}(-1)\to \mathbb{CP}^1$.
We will work in a patch where $\xi_2/\xi_1=\lambda$ is a good inhomogeneous co-ordinate on $\mathbb{CP}^1$. Hence
\begin{equation}
W\,=\,\begin{pmatrix} -u\lambda & u \\ -y\lambda & y \end{pmatrix}\,.
\end{equation}
The radial co-ordinate is defined as in \eqref{eq-int} and becomes
\begin{equation}
r^2\,=\,Tr\,W^\dagger W \,=\,\sigma\Lambda,
\end{equation}
with $\sigma = |u|^2+|y|^2$ and $\Lambda = 1+|\lambda|^2$.
The K\"ahler potential $\mathcal{K}$ in this case is not only a function of $r^2$, but
\begin{equation}
 \mathcal{K}\,=\,\widetilde{\mathcal{F}}+4 a^2\ln\Lambda,
\end{equation}
with $\widetilde{\mathcal{F}}$ a function of $r^2$ and $a$ a constant, the resolution parameter. This gives the metric on the resolved conifold
\begin{equation}\nonumber
ds^2\,=\,\widetilde{\mathcal{F}}'\,Tr(dW^\dagger dW)+\widetilde{\mathcal{F}}''\,
|Tr\,W^\dagger dW|^2+4a^2\,\frac{|d\lambda|^2}{\Lambda^2}\,.
\end{equation}
This reduces to the singular conifold metric when $a\to 0$. We again define $\widetilde{\gamma}=r^2 \widetilde{\mathcal{F}}$. Then Ricci flatness requires
\begin{equation} \label{resgamma}
\widetilde{\gamma}'\widetilde{\gamma}(\widetilde{\gamma}+4a^2)\,=\,2r^2/3\,,
\end{equation}
which can be solved for $\widetilde{\gamma}(r)$.
In terms of the Euler angles \eqref{eulerangles} with $\psi=\psi_1+\psi_2$, this metric was derived by \textcite{pt} to be\footnote{Again we have to rescale the radial co-ordinate such that $\widetilde\gamma$ has dimension $r^2$.}
\begin{eqnarray}\label{resmetricapp}\nonumber
ds^2_{\rm res} & = & \widetilde{\gamma}'\,dr^2 + \frac{\widetilde{\gamma}'}{4}\,r^2\big(d\psi+\cos\theta_1\,d\phi_1
+\cos\theta_2\,d\phi_2\big)^2 \\
& &+ \frac{\widetilde{\gamma}}{4}\,\big(d\theta_1^2+\sin^2\theta_1\,d\phi_1^2\big) \\ &&\nonumber +
\frac{\widetilde{\gamma}+4a^2}{4}\,\big(d\theta_2^2+\sin^2\theta_2\,d\phi_2^2\big)\,,
\end{eqnarray}
with $\widetilde{\gamma}=\widetilde{\gamma}(r)$ going to zero like $r^2$ and $\widetilde{\gamma}'=\partial\widetilde{\gamma}/\partial r^2$. $a$ is called the resolution parameter because it determines the size of the blown up $S^2$ at $r=0$. This shows very nicely that the $(\theta_2,\phi_2)$ sphere is the only part of the metric that remains finite as we approach the tip at $r=0$.

It is convenient to define a new radial co-ordinate via $\rho^2=3/2\,\gamma$. Using \eqref{resgamma}, the Ricci-flat metric with appropriate dimensions can be written as
\begin{eqnarray}\label{resmetricrho}\nonumber
ds^2_{\rm res} & = & \frac{\kappa(\rho)}{9}\,\rho^2\big(d\psi+\cos\theta_1\,d\phi_1
+\cos\theta_2\,d\phi_2\big)^2 \\
& &+ \frac{\rho^3}{6}\,\big(d\theta_1^2+\sin^2\theta_1\,d\phi_1^2\big)\\&& \nonumber+
\frac{\rho^2+6a^2}{6}\,\big(d\theta_2^2+\sin^2\theta_2\,d\phi_2^2\big) +  \kappa(\rho)^{-1}\,d\rho^2\,,
\end{eqnarray}
with $\kappa(\rho)=(\rho^2+9a^2)/(\rho^2+6a^2)$.
It is interesting that there is another K\"ahler metric on the resolved conifold which is related to this one by a flop, basically corresponding to the exchange of the two $S^2$.

\subsection{Complex structures of conifolds}\label{cs}

In this section we will explore a set of vielbeins that does not only give rise to the Ricci flat K\"ahler metric on all three conifold geometries, but also makes the closed K\"ahler and holomorphic three-form apparent. We follow the convention of \textcite{cvetlupo}, which is very similar to \textcite{papatseyt}. But note that the simpler set of vielbeins advocated by \textcite{ks} and \textcite{minasian} does not produce a closed holomorphic three-form for the deformed conifold. We assume that the reader is familiar with the basic concepts of complex differential geometry. A good introduction is found for example in \textcite{nak}.

The vielbeins are deduced from the symmetry group $SU(2)\times SU(2)$. In terms of Euler angles on the corresponding two $S^3$s, we choose left-invariant one-forms on the conifold base:
\begin{eqnarray}\label{forms}\nonumber
\sigma_1 &=& \cos\psi_1\,d\theta_1+\sin\psi_1\,\sin\theta_1\,d\phi_1,\\ \nonumber
\sigma_2 &=& -\sin\psi_1\,d\theta_1+\cos\psi_1\,\sin\theta_1\,d\phi_1,\\
\sigma_3 &=& d\psi_1+\cos\theta_1\,d\phi_1,\\ \nonumber
\Sigma_1 &=& \cos\psi_2\,d\theta_2+\sin\psi_2\,\sin\theta_2\,d\phi_2, \\  \nonumber
\Sigma_2 &=& -\sin\psi_2\,d\theta_2+\cos\psi_2\,\sin\theta_2\,d\phi_2,\\ \nonumber
\Sigma_3 &=& d\psi_2+\cos\theta_2\,d\phi_2.
\end{eqnarray}
They satisfy a Maurer--Cartan equation $d\sigma_i=-i/2\,\epsilon_i^{\phantom{i}jk}\,\sigma_j\wedge\sigma_k$ and similarly for $\Sigma_i$. \textcite{papatseyt} used only five angles and $\psi_1=\psi_2=\psi/2$. This is sufficient for the six-dimensional conifolds, but \textcite{cvetlupo} lifted these geometries to a unified solution in M-theory. It was shown here that all three conifold geometries give rise to one $G_2$ holonomy metric. The eleventh direction is identified with $\psi_1-\psi_2$ and therefore the co-ordinate choice $\psi_1+\psi_2=\psi$ and $\psi_1-\psi_2=0$ can indeed be viewed as a dimensional reduction from 7 to 6 dimensions.

\subsubsection{Singular conifold}

The one-forms \eqref{forms} give rise to vielbeins on the six-dimensional conifold:
\begin{eqnarray}\label{singvielb}\nonumber
e_1 &=&\frac{r}{\sqrt{6}}\,\sigma_1, \qquad\quad
e_2 \,=\,\frac{r}{\sqrt{6}}\,\sigma_2, \\
e_3 &=&\frac{r}{\sqrt{6}}\,\Sigma_1,\qquad \quad
e_4 \,=\,\frac{r}{\sqrt{6}}\,\Sigma_2,\\ \nonumber
e_5 &=&\frac{r}{3}\,(\sigma_3+\Sigma_3), \quad
e_6 \,=\,dr,
\end{eqnarray}
and the metric is diagonal in these vielbeins:
\begin{eqnarray}\nonumber
ds^2 &=& \sum_{i=1}^6 e_i^2\, = \, dr^2+ \frac{r^2}{9}(d\psi + \sum_{i =1 }^2 \cos \theta_i d \phi_i)^2\\
& &  + \frac{r^2}{6}\sum_{i=1}^2 (d\theta_1^2 + \sin^2\theta_i d \phi_i^2)\,.
\end{eqnarray}
Here we identify $\psi=\psi_1+\psi_2$. Note that the odd combination of $\psi_1-\psi_2$ does not appear and we recover the by-now-familiar structure of the base -- an $S^2$ fibred over an $S^3$ -- although we started out with co-ordinates for two $S^3$s.

An (almost) complex structure on this real six-dimensional manifold is defined by choosing complex vielbeins
\begin{equation}\label{defcs}
 E_1 = e_1+\imath\,e_2,\;  E_2 = e_3+\imath\,e_4,\;  E_3 = e_5+\imath\,e_6.
\end{equation}
In terms of these complex vielbeins, the fundamental two-form $J$ and holomorphic three form $\Omega$ are defined as
\begin{eqnarray}\label{defJ}
J^{(1,1)} &=& \frac{\imath}{2}(E_1\wedge\bar E_1+E_2\wedge\bar E_2+E_3\wedge\bar E_3)\\ \label{defOmega}
\Omega^{(3,0)} &=& E_1\wedge E_2 \wedge E_3\,.
\end{eqnarray}
For the singular conifold their co-ordinate expressions are still fairly simple (again, only the even combination of $\psi_1$ and $\psi_2$ appears):
\begin{widetext}
\begin{eqnarray}\nonumber
J &=& -\frac{r}{3}\,dr\wedge(d\psi+\cos\theta_1\,d\phi_1+\cos\theta_2\,d\phi_2)
  -\frac{r^2}{6}\,(\sin\theta_1\,d\phi_1\wedge d\theta_1+
  \sin\theta_2\,d\phi_2\wedge d\theta_2)\,,\\ \nonumber
\Omega &=& \frac{r^2}{6}\,(\cos\psi-\imath\sin\psi)\,dr\wedge\Big[\sin\theta_1\, d\theta_2\wedge d\phi_1 - \sin\theta_2\,d\theta_1\wedge d\phi_2 +\imath(d\theta_1\wedge d\theta_2-\sin\theta_1\sin\theta_2\,d\phi_1\wedge d\phi_2)\Big]\\ \nonumber
& &+\, \frac{r^3}{18}\,(\cos\psi-\imath\sin\psi)\Big[d\theta_1\wedge d\theta_2\wedge(d\psi+\cos\theta_1\,d\phi_1+\cos\theta_2\,d\phi_2) -\sin\theta_1\sin\theta_2\,d\phi_1\wedge d\phi_2\wedge d\psi \\ \nonumber
& &  \qquad\!\!
-\imath(\sin\theta_1\,d\theta_2\wedge d\phi_1 - \sin\theta_2\,d\theta_1\wedge d\phi_2)\wedge d\psi %\\ \nonumber
%& &  \qquad\qquad
-\imath(\sin\theta_1\cos\theta_2\,d\theta_2+\cos\theta_1\sin\theta_2\,d\theta_1)\wedge d\phi_1\wedge d\phi_2\Big],
\end{eqnarray}
\end{widetext}
and one can easily show that
\begin{equation}
dJ\,=\,0\,\qquad\mbox{and}\qquad d\Omega \,=\,0\,.
\end{equation}
Together these relations imply that the almost complex structure is actually integrable, so the closure of the fundamental two-form means that this manifold is K\"ahler. For a K\"ahler manifold the closure of $\Omega$ means furthermore that it is Calabi--Yau (see e.g. \textcite{salamon}).

The complex structure induced by these vielbeins is of course identical with the one coming from the holomorphic coordinates $z_i$ used in \eqref{eq-con} to define the singular conifold. One finds that, up to a numerical factor, the holomorphic three--form can also be expressed as
\begin{equation}\label{omegainz}
  \Omega \,=\, \frac{dz_1\wedge dz_2\wedge dz_3}{z_4}\,,
\end{equation}
which agrees with above coordinate expression if the holomorphic coordinates are parameterized as
\begin{eqnarray}\label{holocoord}\nonumber
  x & =& r^{3/2} e^{\imath/2(\psi-\phi_1-\phi_2)}\,\sin\frac{\theta_1}{2}\,\sin\frac{\theta_2}{2} \\ \nonumber
  y & =& r^{3/2} e^{\imath/2(\psi+\phi_1+\phi_2)}\,\cos\frac{\theta_1}{2}\,\cos\frac{\theta_2}{2} \\
  u & =& r^{3/2} e^{\imath/2(\psi+\phi_1-\phi_2)}\,\cos\frac{\theta_1}{2}\,\sin\frac{\theta_2}{2} \\ \nonumber
  v & =& r^{3/2} e^{\imath/2(\psi-\phi_1+\phi_2)}\,\sin\frac{\theta_1}{2}\,\cos\frac{\theta_2}{2}\,.
\end{eqnarray}
We also make use of the coordinate redefinition \eqref{coordchange} to relate these coordinates to $z_i$
\begin{eqnarray}\nonumber
  z_1 &=& \frac{1}{2}\,(x-\imath\,y)\\ \nonumber
  z_2 &=& \frac{1}{2\imath}\,(x+\imath\,y)\\ \nonumber
  z_3 &=& \frac{1}{2}\,(u+\imath\,v)\\ \nonumber
  z_4 &=& \frac{-1}{2\imath}\,(u-\imath\,v)\,.
\end{eqnarray}
For practical computations these coordinates are not very useful, as they are the homogeneous ones. The real coordinates make the structure of the six dimensional manifold much more transparent and the vielbeins serve as a convenient basis for all sorts of differential forms, like fluxes.

\subsubsection{Resolved conifold}\label{csres}

The same complex structure \eqref{defcs} can be used for the resolved conifold. We only have to scale the vielbeins according to the metric:
\begin{eqnarray}\label{resvielb}
e_1 &=& \frac{\rho}{\sqrt{6}}\,\sigma_1, \qquad\qquad\qquad\qquad\,
e_2 \,=\,\frac{\rho}{\sqrt{6}}\,\sigma_2, \\ \nonumber
e_3 &=& \frac{\sqrt{\rho^2+6a^2}}{\sqrt{6}}\,\Sigma_1,\qquad\qquad\quad\!\!\!\;
e_4 \,=\,\frac{\sqrt{\rho^2+6a^2}}{\sqrt{6}}\,\Sigma_2,\\ \nonumber
e_5 &=&\frac{\rho}{3}\sqrt{\frac{\rho^2+9a^2}{\rho^2+6a^2}}\,(\sigma_3+\Sigma_3), \quad
e_6 \,=\,\sqrt{\frac{\rho^2+6a^2}{\rho^2+9a^2}}\,d\rho\,,
\end{eqnarray}
then the metric remains diagonal and we recover \eqref{resmetricrho} with $\psi=\psi_1+\psi_2$. The fundamental two-form \eqref{defJ} is found to be
\begin{eqnarray}
J &=& -\frac{\rho}{3}\,d\rho\wedge(d\psi+\cos\theta_1\,d\phi_1+\cos\theta_2\,d\phi_2)\\ \nonumber
& & -\frac{\rho^2}{6}\,\sin\theta_1\,d\phi_1\wedge
d\theta_1- \frac{\rho^2+6a^2}{6}\, \sin\theta_2\,d\phi_2\wedge d\theta_2
\end{eqnarray}
and is closed, as is the holomorphic three-form one obtains from \eqref{defOmega}:
\begin{widetext}
\begin{eqnarray}\label{Omegares}\nonumber
\Omega &=& \frac{\rho(\rho^2+6a^2)}{6\sqrt{\rho^2+9a^2}}\,(\cos\psi-i\sin\psi)\,d\rho\wedge\Big[\sin\theta_1\,    
  d\theta_2\wedge d\phi_1 - \sin\theta_2\,d\theta_1\wedge d\phi_2
  +i(d\theta_1\wedge d\theta_2-\sin\theta_1\sin\theta_2\,d\phi_1\wedge d\phi_2)\Big]\\ \nonumber
&& + \, \frac{\rho^2}{18}\,\sqrt{\rho^2+9a^2}\,(\cos\psi-i\sin\psi)\Big[d\theta_1\wedge   
  d\theta_2\wedge(d\psi+\cos\theta_1\,d\phi_1+\cos\theta_2\,d\phi_2) -\sin\theta_1\sin\theta_2\,d\phi_1\wedge 
  d\phi_2\wedge d\psi \\ 
& &  \qquad -i(\sin\theta_1\,d\theta_2\wedge d\phi_1 - \sin\theta_2\,d\theta_1\wedge d\phi_2)\wedge d\psi 
  -i(\sin\theta_1\cos\theta_2\,d\theta_2+\cos\theta_1\sin\theta_2\,d\theta_1)\wedge d\phi_1\wedge d\phi_2\Big]\,.
\end{eqnarray}
\end{widetext}
So this complex structure also fulfills the Calabi--Yau conditions. The corresponding homogeneous holomorphic coordinates in tis case read 
\begin{eqnarray}\nonumber
  x & =&  \left ( 9 a^2 \rho^4 + \rho ^6 \right ) ^{1/4} e^{i/2(\psi-\phi_1-\phi_2)}\,\sin\frac{\theta_1}{2}\,\sin\frac{\theta_2}{2} \\ \nonumber
  y & =&  \left ( 9 a^2 \rho^4 + \rho ^6 \right ) ^{1/4} e^{i/2(\psi+\phi_1+\phi_2)}\,\cos\frac{\theta_1}{2}\,\cos\frac{\theta_2}{2} \\ \nonumber
  u & =&  \left ( 9 a^2 \rho^4 + \rho ^6 \right ) ^{1/4} e^{i/2(\psi+\phi_1-\phi_2)}\,\cos\frac{\theta_1}{2}\,\sin\frac{\theta_2}{2} \\ \nonumber
  v & =&  \left ( 9 a^2 \rho^4 + \rho ^6 \right ) ^{1/4} e^{i/2(\psi-\phi_1+\phi_2)}\,\sin\frac{\theta_1}{2}\,\cos\frac{\theta_2}{2}\,.
\end{eqnarray}
They lead to the same holomorphic three--form with the definition \eqref{omegainz}.

\subsubsection{Deformed conifold}

For the deformed conifold the story is more complicated. The metric is not diagonal in the vielbeins \eqref{forms} and we have to define linear combinations of them such that
\begin{eqnarray}\nonumber
e_1 &=& \frac{\sqrt{\hat\gamma}}{2}\,(\alpha\sigma_1-\beta\Sigma_1),\quad\;
e_2 \,=\, \frac{\sqrt{\hat\gamma}}{2}\,(\alpha\sigma_2+\beta\Sigma_2),\\ \nonumber
e_3 &=& \frac{\sqrt{\hat\gamma}}{2}\,(-\beta\sigma_1+\alpha\Sigma_1),\;\;
e_4 \,=\, \frac{\sqrt{\hat\gamma}}{2}\,(\beta\sigma_2+\alpha\Sigma_2),\\ \nonumber
e_5 &=& \frac{1}{2}\sqrt{(r^2\hat\gamma'-\hat\gamma)\left(1-\mu^4/r^4\right)+\hat\gamma}\,(\sigma_3+\Sigma_3),\\ 
e_6 &=& \frac{\sqrt{(r^2\hat\gamma'-\hat\gamma)\left(1-\mu^4/r^4\right)+\hat\gamma}}{r\sqrt{1-\mu^4/r^4}}\,dr,
\end{eqnarray}
where $\alpha^2+\beta^2=1$ has to hold for the metric to turn out correctly. With these linear combinations one recovers \eqref{defmetricapp} from
\begin{equation*}
ds^2 \,=\, \sum_{i=1}^6 e_i^2\,.
\end{equation*}
For the metric to also be Ricci flat and K\"ahler, the coefficients $\alpha$ and $\beta$ are determined to be
\begin{eqnarray}\label{defalphabeta}\nonumber
\alpha &=& \frac{1}{2}\,\sqrt{1+\mu^2/r^2} + \frac{1}{2}\,\sqrt{1-\mu^2/r^2},\\
\beta &=& \frac{\mu^2}{2r^2\alpha}\,.
\end{eqnarray}
The complex structure is defined as in \eqref{defcs} and again gives rise to a Calabi--Yau.

With the choice \eqref{defalphabeta} the K\"ahler form amounts to
\begin{eqnarray}\nonumber
J &=& -\frac{r^6\hat\gamma'+\mu^4\hat\gamma-r^2\mu^4\hat\gamma'}{2r^5\sqrt{1-\mu^4/r^4}}\, dr\wedge (d\psi+\cos\theta_1\, d\phi_1\\ 
& & \phantom{-r^6\hat\gamma'+\mu^4\hat\gamma-r^2\mu^4\hat\gamma'\quad}
  +\cos\theta_2\ d\phi_2)\\ \nonumber
& &+\,\frac{\hat\gamma}{4}\,\sqrt{1-\frac{\mu^4}{r^4}}\,(\sin\theta_1\,d\theta_1\wedge d\phi_1 +
\sin\theta_2\,d\theta_2\wedge d\phi_2)\,,
\end{eqnarray}
which is easily shown to be closed (recall that the prime indicates derivative w.r.t. $r^2$). The expression for the holomorphic 3-form is
\begin{widetext}
\begin{eqnarray}\nonumber
\Omega &=& 2\mathcal{T}\,\frac{\mathcal{S}\cos\psi-i\sin\psi}{r\mathcal{S}}\,dr\wedge(\sin\theta_1\,d\theta_2\wedge d\phi_1
  - \sin\theta_2\,d\theta_1\wedge d\phi_2)\\ \nonumber
& &+ \frac{2\imath\mathcal{T}}{r\mathcal{S}}\,(\cos\psi-i\mathcal{S}\sin\psi)\,dr\wedge(d\theta_1\wedge d\theta_2
  -\sin\theta_1\sin\theta_2\,d\phi_1\wedge d\phi_2) 
  -\frac{2\mu^2\mathcal{T}}{r^3\mathcal{S}}\,dr\wedge (\sin\theta_1\,d\theta_1\wedge d\phi_1-\sin\theta_2\,
  d\theta_2\wedge d\phi_2)\\
& &+  \frac{\imath\mu^2\mathcal{T}}{r^2}\big[\sin\theta_1\,d\theta_1\wedge d\phi_1\wedge(d\psi+\cos\theta_2\,d\phi_2)
  - \sin\theta_2\,d\theta_2\wedge d\phi_2\wedge(d\psi+\cos\theta_1\,d\phi_1)\Big]\\  \nonumber
& &+  \mathcal{T}(\imath\cos\psi+\mathcal{S}\sin\psi)\Big[\sin\theta_2\,d\theta_1\wedge d\phi_2\wedge
  (d\psi+\cos\theta_1\,d\phi_1)-\sin\theta_1\,d\theta_2\wedge d\phi_1\wedge(d\psi+\cos\theta_2\,d\phi_2)\Big]\\ \nonumber
& &+  \mathcal{T}(\mathcal{S}\cos\psi-\imath\sin\psi)\Big[d\theta_1\wedge d\theta_2\wedge
  (d\psi+\cos\theta_1\,d\phi_1+\cos\theta_2\,d\phi_2)-\sin\theta_1\sin\theta_2\,d\phi_1\wedge d\phi_2\wedge d\psi\Big]\,,
\end{eqnarray}
\end{widetext}
where we have introduced the symbols $\mathcal{S}=\sqrt{1-\mu^4/r^4}$ and $\mathcal{T}=\hat\gamma\sqrt{\hat\gamma+(r^2\hat\gamma'-\hat\gamma)(1-\mu^4/r^4)}/8$.
To show that it is indeed closed one needs to make use of \eqref{deformedgamma}.

For the deformed conifold we can use the same holomorphic coordinates as for the singular conifold \eqref{holocoord}, but the three--form \eqref{omegainz} now reads
\begin{equation}
   \Omega \,=\, \frac{dz_1\wedge dz_2\wedge dz_3}{\sqrt{\mu^2-z_1^2-z_2^2-z_3^2}}\,.
\end{equation}

As a side remark, let us note that the much simpler vielbeins from \textcite{ks}:
\begin{eqnarray}\label{ksvielb} \nonumber
g_1 &=& -\sin\theta_1\,d\phi_1, \\
g_2 &=& d\theta_1, \\ \nonumber 
g_3 &=& -\sin\psi\,d\theta_2+\cos\psi\,\sin\theta_2\,d\phi_2,\\ \nonumber
g_4 &=& \cos\psi\,d\theta_2+\sin\psi\,\sin\theta_2\,d\phi_2,\\ \nonumber
g_5 &=& d\psi+\cos\theta_1\,d\phi_1 +\cos\theta_2\,d\phi_2,\quad \\ \nonumber
g_6 &=& dr
\end{eqnarray}
will never give a closed holomorphic three-form on the deformed conifold, even with a generic ansatz for a linear combination of these vielbeins.\footnote{This statement was confirmed with {\em Mathematica} for arbitrary $r$-dependent coefficients.} In other words, they are not compatible with the holomorphic coordinates \eqref{holocoord}. They do work for the singular and resolved conifold, because they happen to give the same 2 and 3-forms. So there is more than one choice of vielbeins that allows for a Calabi--Yau metric. However, if we wish to pass from one geometry to the other, we prefer to employ a complex structure that allows for all three of them to be Calabi--Yau. The set of vielbeins \eqref{ksvielb} was also used by \textcite{minasian}. Caution should therefore be used with their solutions, in particular from the viewpoint of supersymmetry.

\subsection{Fluxes on conifolds}\label{fluxes}

Having studied the complex structure of conifold geometries, we can
now turn to the question of what types of fluxes are allowed on
them. In type IIB compactifications the background three-form flux
$G_3=F_3+\tau H_3$ ($\tau=C_0+ie^{-\phi}$ is the axion-dilaton) has
to obey a self-duality condition \cite{gkp}
\begin{equation}
*_6\,G_3 \,=\, iG_3,
\end{equation}
where $*_6$ indicates the Hodge dual in 6 dimensions. Supersymmetry
requires $G_3$ to be of type (2,1) and primitive \cite{granapol, gubs},
i.e. that it satisfy $G_3\wedge J=0$. In \textcite{cvetgiblupo} it was
shown that the solution of \textcite{ks} for D5-branes
on the singular conifold fulfills these requirements, whereas the
\textcite{pt} (PT) solution for D5-branes on the
\em resolved \em conifold does not. The latter has a (1,2) part in
addition to the allowed (2,1).

Although we agree with the result obtained by \textcite{cvetgiblupo}, we question the complex structure they use. Following PT they take the simplest set of vielbeins that would give the right resolved metric \eqref{resmetricapp} \footnote{Note that there is a typo in equation (6.5) of \textcite{cvetgiblupo}.}, i.e.
\begin{eqnarray}\label{cglpvielb}\nonumber
\epsilon_1 &=& \frac{\rho}{\sqrt{6}}\,d\theta_1,
  \qquad\qquad\!\;\!\;\!\, \epsilon_2\,=\,\frac{\rho}{\sqrt{6}}\,\sin\theta_1\,d\phi_1,\\  \nonumber
\epsilon_3&=&\frac{\sqrt{\rho^2+6a^2}}{\sqrt{6}}\,d\theta_2,\quad
  \epsilon_4\,=\,\frac{\sqrt{\rho^2+6a^2}}{\sqrt{6}}\,\sin\theta_2\,d\phi_2,\\
\epsilon_5&=&-\sqrt{\frac{\rho^2+6a^2}{\rho^2+9a^2}}\,d\rho,\\ \nonumber
\epsilon_6&=&\frac{\rho}{3}\sqrt{\frac{\rho^2+9a^2}{\rho^2+6a^2}}\,(d\psi+\cos\theta_1\,d\phi_1
+\cos\theta_2\,d\phi_2),\quad
\end{eqnarray}
and then show that the fluxes from \textcite{pt} have not only a (2,1) but also a (1,2) part w.r.t. the complex structure:
\begin{equation}\nonumber
E_1 = \epsilon_1+\imath\,\epsilon_2\,,\;  E_2 = \epsilon_3+\imath\,\epsilon_4\,,\; E_3 = \epsilon_5+\imath\,\epsilon_6\,.
\end{equation}
Note, however, that this choice is not the right one to observe the Calabi--Yau property. It leads to a closed fundamental 2-form, but the holomorphic 3-form
%\begin{widetext}
\begin{eqnarray}\nonumber
\Omega &=& -\frac{\rho(\rho^2+6a^2)}{6\sqrt{\rho^2+9a^2}}\,d\rho\wedge\Big[\imath(\sin\theta_2\, d\theta_1\wedge d\phi_2 \\ \nonumber
& &  \qquad\qquad\qquad\qquad\qquad -\sin\theta_1\,d\theta_2\wedge d\phi_1)\\ \nonumber
& &  \qquad\qquad+(d\theta_1\wedge d\theta_2-\sin\theta_1\sin\theta_2\,d\phi_1
  \wedge d\phi_2)\Big]\\ \nonumber
&  & + \frac{\rho^2}{18}\,\sqrt{\rho^2+9a^2}\,\Big[\imath d\theta_1\wedge  d\theta_2\wedge\\ \nonumber
& & \qquad\qquad\qquad\qquad  (d\psi+\cos\theta_1\,d\phi_1+\cos\theta_2\,d\phi_2)\\ \nonumber 
& & -\imath\sin\theta_1\sin\theta_2\,d\phi_1\wedge d\phi_2\wedge d\psi \\ \nonumber
& & +(\sin\theta_1\,d\theta_2\wedge d\phi_1 - \sin\theta_2\,d\theta_1\wedge d\phi_2)\wedge d\psi\\ \nonumber 
& & +(\sin\theta_1\cos\theta_2\,d\theta_2+\cos\theta_1\sin\theta_2\,d\theta_1)\wedge d\phi_1\wedge d\phi_2\Big]
\end{eqnarray}
%\end{widetext}
lacks the $(\cos\psi-\imath\sin\psi)$ terms compared to \eqref{Omegares}. It is therefore not closed but instead
%\begin{widetext}
\begin{eqnarray}\nonumber
d\Omega &=& \frac{\rho(\rho^2+6a^2)}{6\sqrt{\rho^2+9a^2}}\,d\rho\wedge d\psi\wedge\Big[\imath(d\theta_1\wedge
  d\theta_2\\ \nonumber
& & \qquad\qquad\qquad\qquad-\sin\theta_1\sin\theta_2\,d\phi_1\wedge d\phi_2)\\ \nonumber
& & \qquad\qquad-(\sin\theta_2\, d\theta_1\wedge d\phi_2 
  - \sin\theta_1\,d\theta_2\wedge d\phi_1)\Big]\\ \nonumber
& &+  \frac{\rho^2}{18}\,\sqrt{\rho^2+9a^2}\,\Big[-d\theta_1\wedge d\theta_2\wedge d\psi \wedge\\ \nonumber
& & \qquad\qquad\qquad\qquad\qquad (\cos\theta_1 d\phi_1 +\cos\theta_2   d\phi_2)\\ \nonumber
& & \qquad + \imath(\sin\theta_1\cos\theta_2\,d\theta_2 +\cos\theta_1\sin\theta_2\,d\theta_1)\wedge\\ \nonumber
& & \qquad\qquad\qquad\qquad\qquad\qquad\quad d\phi_1\wedge d\phi_2\wedge d\psi\Big],
\end{eqnarray}
%\end{widetext}
which will never vanish. This statement remains true for the singular conifold, so this complex structure should not be used for analysing the KS flux either. We therefore believe that the analysis in \textcite{cvetgiblupo} should be taken with a grain of salt.

We can use our knowledge from Section \ref{cs} to find out to which cohomology group the flux from \textcite{pt} belongs. The imaginary self-dual 3-form flux $G_3=F_3+iH_3$ found in \textcite{pt} is given by\footnote{This is a solution with constant dilaton, which can therefore be set to zero, and vanishing axion. Furthermore, there is a typo in equation (4.3) in \cite{pt}, concerning the sign of $F_3$.}
\begin{eqnarray}\nonumber
H_3 &=& d\rho\wedge(F_1'(\rho)\,\sin\theta_1\,d\theta_1\wedge d\phi_1\\ 
 & & \qquad +F_2'(\rho)\,\sin\theta_2\,d\theta_2\wedge d\phi_2)
\end{eqnarray}
and
\begin{eqnarray}\nonumber
F_3 &=& P\,(d\psi+\cos\theta_1\,d\phi_1 +\cos\theta_2\,d\phi_2)\wedge\\ 
 & & (\sin\theta_1\,d\theta_1\wedge d\phi_1-\sin\theta_2\,d\theta_2\wedge d\phi_2)\,,
\end{eqnarray}
where $P$ is a constant (to ensure $dF_3=0$) and $F_1'(\rho)$ and $F_2'(\rho)$ were determined from the equations of motion to be
\begin{equation}\nonumber
F_1'(\rho) \,=\, 3P\,\frac{\rho}{\rho^2+9a^2}\,,\qquad F_2'(\rho) \,=\, -3P\,\frac{(\rho^2+6a^2)^2}{\rho^3(\rho^2+9a^2)}\,.
\end{equation}
We now use the vielbeins from Section \ref{csres} and invert \eqref{resvielb} to solve for the co-ordinate differentials. We then find the flux in terms of vielbeins
\begin{eqnarray}\nonumber
G_3 &=& \frac{18 P\sqrt{\rho^2+6a^2}}{\rho^3\sqrt{\rho^2+9a^2}}\,(e_1\wedge e_2\wedge e_5-\imath\, e_3\wedge e_4\wedge 
  e_6)\\
& &  -\frac{18P\,(e_3\wedge e_4\wedge e_5-\imath\, e_1\wedge e_2\wedge e_6)}{\rho\sqrt{\rho^2+6a^2}\sqrt{\rho^2+9a^2}}\,.
\end{eqnarray}
The vielbein notation is extremely convenient to see that this flux is indeed imaginary self-dual\footnote{The self--duality should really be checked w.r.t. the  \em warped \em resolved conifold, but since we consider a 3--form flux on a 6--dimensional manifold, the appropriate factors of the warp factor drop out when taking the Hodge dual.} (remarkable since PT also used the wrong set of vielbeins). The Hodge dual is simply found by
\begin{equation}\nonumber
*_6 (e_{i_1}\wedge e_{i_2}\wedge\ldots\wedge e_{i_k}) = \epsilon_{i_1 i_2\ldots i_k}^{\phantom{i_1 i_2\ldots i_k}i_{k+1}\ldots i_6}\,e_{i_{k+1}}\wedge \ldots \wedge e_{i_6}
\end{equation}
and does not involve any factors of $\sqrt{g}$. We use the convention that $\epsilon_{123456}=\epsilon_{123}^{\phantom{123}456}=1$.
With the usual complex structure \eqref{defcs} the PT flux becomes
\begin{eqnarray}\nonumber
G_3 &=& -\frac{9 \imath P}{\rho^3\sqrt{\rho^2+9a^2}\sqrt{\rho^2+6a^2}}\times\\ \nonumber
& & \Big[(\rho^2+3a^2)\,(E_1\wedge E_3\wedge \overline E_1 - E_2\wedge E_3\wedge\overline E_2)\\ 
& & \;-3a^2\,(E_1\wedge\overline E_1\wedge\overline E_3 + E_2\wedge\overline E_2\wedge\overline E_3)\Big].
\end{eqnarray}
We make several observations: This flux is neither primitive\footnote{Since $J=\frac{\imath}{2}\,\sum_i(E_i\wedge\overline E_i)$ it
follows immediately that $J\wedge G_3$ has a nonvanishing $E_1\wedge E_2\wedge\overline{E_1}\wedge\overline{E_2}\wedge\overline{E_3}$ part that is proportional to $a^2$.} nor is it of type (2,1). It has a (1,2) \em
and \em a (2,1) part. With just a (1,2) part present we could have
made this flux supersymmetric by a different choice of complex
structure. But as it stands, this flux indeed
breaks supersymmetry, as claimed by \textcite{cvetgiblupo}. Apart from
that, in the limit $a\to 0$ the (1,2) part vanishes, the flux becomes primitive and we recover
the singular conifold. This seems to indicate that the
resolution parameter forbids a supersymmetric supergravity solution
for wrapped D5-branes on the resolution in the presence of flux.

Let us close this section by repeating the flux analysis for the KS model, which in the region away from the tip agrees with the singular conifold solution first advocated by \textcite{kt} (KT). Again, we agree with the result of \textcite{cvetgiblupo}, but we maintain that a complex structure that allows for a closed holomorphic 3-form on the singular conifold should have been used. We use the set \eqref{singvielb} with the same complex structure as in \eqref{defcs}. The 3-form flux $G_3=F_3+\imath\,H_3$ with
\begin{eqnarray}\nonumber
F_3 &=& \frac{M}{2}\,(d\psi+\cos\theta_1\,d\phi_1 +\cos\theta_2\,d\phi_2)\wedge\\ \nonumber
& & (\sin\theta_1\,d\theta_1\wedge d\phi_1-\sin\theta_2\,d\theta_2\wedge d\phi_2),\\ \nonumber
H_3 &=& \frac{3}{2Mr}\,dr\wedge(\sin\theta_1\,d\theta_1\wedge d\phi_1
-\sin\theta_2\,d\theta_2\wedge d\phi_2),
\end{eqnarray}
becomes
\begin{equation}
G_3 \,=\, -\frac{9\imath M}{2r^3}\,(E_1\wedge E_3\wedge \overline E_1 - E_2\wedge E_3\wedge\overline E_2)\,,
\end{equation}
where $M$ indicates the number of fractional D3-branes in the KT model; see Section \ref{ks}. It is also easy to check that this flux is indeed primitive $(J\wedge G_3=0)$. Also, the resulting 5--form flux $F_5=dC_4+B_2\wedge F_3$ can be made self--dual by choosing 
\begin{equation}
  dC_4=d(h^{-1}(r))\wedge dx_0\wedge ...\wedge dx_3=*_{10}(B_2\wedge F_3)\,,
\end{equation}
where the 10--dimensional Hodge dual is to be taken w.r.t. to the warped metric 
\begin{equation}\nonumber
  ds^2=h^{-1/2}(r)\eta^{\mu\nu}dx_\mu dx_\nu+h^{1/2}(r)(dr^2+r^2ds^2_{T^{1,1}})\,.
\end{equation}
Thus, we have confirmed that the KT model preserves supersymmetry in the correct complex structure. We also see that in the limit where the 2-cycle in the resolved conifold shrinks to zero, the flux in the PT solution agrees with the singular conifold solution of KT.

%%%%%%%%%%%%%%%%%%%%%%%%%%%%%%%%%%%%%%%%%%%%%%%%%%%%%%%%%%%%%%%%%%%%%%%%%%

\bibliographystyle{apsrmp}
\bibliography{conifolds}
\printfigures
\end{document}